\documentclass[12pt]{article}
\usepackage{graphicx}
\usepackage{epsfig}
\title{{\bf Spin Effects in Long Range Gravitational Scattering}}
\author{Barry R. Holstein$^a$\footnote{\tt holstein@physics.umass.edu} \hspace*{1pt} and Andreas
Ross$^{a,b}$\footnote{\tt
andreas.ross@yale.edu} \\ \\
$^a$ Department of Physics -- LGRT\\
University of Massachusetts\\
Amherst, MA  01003, USA\\  \\
$^b$ Department of Physics \\
 Yale University \\
New Haven, CT 06520, USA \\ \\}

\begin{document}
\maketitle
\thispagestyle{empty}

\begin{abstract}
We study the gravitational scattering of massive particles with and
without spin in the effective theory of gravity at one loop level.
Our focus is on long distance effects arising from nonanalytic
components of the scattering amplitude and we show that the
spin-independent and the spin-dependent long range components
exhibit a universal form. Both classical and quantum corrections are
obtained, and the definition of a proper second order potential is
discussed.
\end{abstract}

\vspace{0.2 in}
\setcounter{page}{0}
\newpage

\section{Introduction}

The gravitational interaction of two massive particles is described
by Newton's law
\begin{equation}
 V(r)= - \frac{G \hspace*{0.3pt} m_am_b}{r}
\end{equation}
which is a good approximation for the motion of nonrelativistic
particles at large separations. Einstein's theory of general
relativity, however, predicts important corrections that have been
verified experimentally, {\it e.g.}, in measurements of the
precession of the perihelion of Mercury \cite{wei}.  This effect can
be calculated from the famous Einstein-Infeld-Hoffmann
Lagrangian \cite{Einstein:1938yz} where the terms beyond Newtonian
physics stem from the relativistic $\mathcal O(v^2)$ corrections to
the kinetic energy, from relativistic $\mathcal O(v^2)$ corrections
to the Newtonian potential and from a new piece of the potential
proportional to $G^2M^3/r^2$ which may be regarded as an $\mathcal
O(GM/r)$ correction to the leading Newtonian potential. The two
small expansion parameters here are $v^2$ and $GM/r$ which for bound
states are related due to the virial theorem. We, however, will
examine scattering processes wherein the classical expansion
parameters $v^2$ and $GM/r$ are independent, and our focus will be
on the components that vanish in the nonrelativistic limit $v
\rightarrow 0$.

In the weak field limit gravitational dynamics can be described in
terms of a quantum field theory which is based on expanding around
flat Minkowski spacetime and quantizing the metric fluctuation in
terms of a massless spin-2 field---the graviton. The resulting
theory is an effective field theory whose predictions are
trustworthy only for energies much smaller than the Planck scale.
Nevertheless, one can in this picture calculate well-defined quantum
predictions as well as classical corrections to Newtonian physics.
Donoghue has pioneered the use of the effective field theory of
gravity to extract the leading long distance effects in the
nonrelativistic limit \cite{don1, don2} (see \cite{Donoghue:1995cz} and \cite{Burgess:2003jk} for reviews) 
which yield corrections of the form
\begin{equation}
 V(r) = - \frac{G \hspace*{0.3pt} m_am_b}{r} \left(1 + A_C \, \frac{G M}{r} + A_Q
 \, \frac{G\hbar}{r^2} \right) \label{eq_pot_cor}
\end{equation}
where $A_C$ and $A_Q$ are the coefficients of the classical and
quantum corrections respectively and are evaluated below. Note that
while the classical expansion parameter $GM/r \sim M/M_{Pl} \,
\times \, \ell_{Pl}/r$ can give measurable effects for macroscopic
objects (where a large factor multiplies a tiny factor), the quantum
corrections which scale as $G\hbar / r^2 \sim (\ell_{Pl}/r)^2$ are clearly
tiny and phenomenologically irrelevant for macroscopic systems.
The calculations of these leading long distance corrections are
performed by evaluating the leading nonanalytic components of the
scattering amplitude at one loop level, where $\mathcal O(G^2)$
effects first arise, and by defining a potential in terms of the
Fourier transform of this amplitude.

Once spin is introduced into the calculation, additional interaction
structures arise, such as a spin-orbit coupling and a spin-spin
coupling.  The leading effects of order $G$ stemming from the tree
level one-graviton exchange process give rise to the familiar
geodetic precession and to the Lense-Thirring effect \cite{wei}.  At
the two-graviton exchange level, classical $\mathcal O(GM/r)$ and
quantum $\mathcal O(G\hbar/r^2)$ corrections to these leading
spin-dependent interactions arise and are calculated as part of our
work.

In earlier work on loop corrections to the form factors of graviton
couplings it was found that the long distance corrections to the
spin-independent interactions have the same form for scalars,
spin-1/2 fermions and for spin-1 bosons \cite{Bjerrum-Bohr:2002ks,
Holstein:FF}.  These NLO form factor interactions constitute one
component of the calculation of the full scattering amplitude at the
two-graviton exchange level, so the question arises as to whether
the {\it full} scattering amplitude also exhibits such
universalities, whereby the form of the corrections to the
spin-independent Newtonian potential is independent of the spins of
the scattered particles --- in other words, are the coefficients $A_C$
and $A_Q$ in Eq. (\ref{eq_pot_cor}) independent of spin?

This then is the goal of the present work --- to evaluate the
gravitational scattering of two massive particles having various
spins, in order to check previous work and to verify the
universality hypothesis.  In the next section then we review our
calculational methods and reproduce the long range gravitational
scattering amplitude in the case of a pair of spinless particles. In
the following sections, we generalize these methods to the cases of
spin-0 -- spin-1/2, spin-0 -- spin-1 and spin-1/2 -- spin-1/2
scattering and demonstrate both the expected universality as well as
novel spin-dependent interactions.  We summarize our work and draw
general implications in a brief concluding section. Appendix
\ref{app_iter} encapsulates parts of our calculational methods while
we refer to our companion paper on long distance effects in
electromagnetic scattering \cite{hrem} for many of the details such
as the required Feynman integrals, the Fourier transformations etc.
In Appendix \ref{app_eom} we demonstrate how the classical equations
of motion in the form of the Einstein-Infeld-Hoffmann Lagrangian can
be extracted from our results.

\section{Spin-Independent Scattering}

\begin{figure}
\begin{center}
\epsfig{file=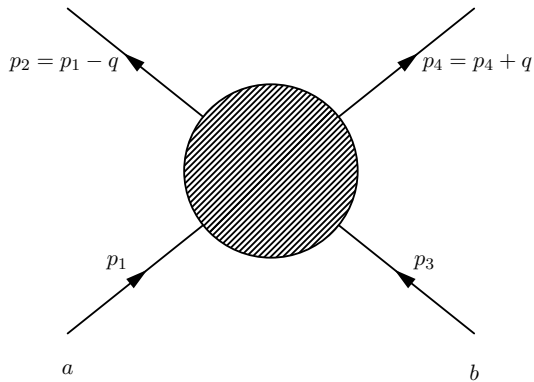,width=7cm} \caption{Basic kinematics of
gravitational scattering. } \label{fig_kinem}
\end{center}
\end{figure}

We first set the kinematic framework for our study.  We consider the
gravitational scattering of two non-identical particles---particle
$a$ with mass $m_a$, and incoming four-momentum $p_1$ and particle
$b$ with mass $m_b$, and incoming four-momentum $p_3$.  After
interacting the final four-momentum of particle $a$ is $p_2=p_1-q$
and that of particle $b$ is $p_4=p_3+q$---{\it cf.} Fig.
\ref{fig_kinem}. Now we need to be more specific.

\subsection{Spin-0 -- Spin-0 Scattering}

We begin by examining the gravitational scattering of two spinless
particles. The gravitational coupling of a spin-0 particle is found by expanding
the minimally coupled scalar field matter Lagrangian
\begin{equation}
\sqrt{-g}{\cal L}_m=\sqrt{-g}\left({1\over 2} g^{\mu \nu} \partial_\mu \phi
\partial_\nu \phi  -{1\over 2}m^2\phi^2 \right)
\end{equation}
in terms of the gravitational field $h_{\mu\nu}$ which is a small
fluctuation of the metric about flat Minkowski space defined as
\begin{equation}
g_{\mu\nu}=\eta_{\mu\nu} + \kappa h_{\mu\nu}
\end{equation}
with $\kappa = \sqrt{32 \pi G} \propto 1/M_P$. The inclusion of this
factor $\kappa$ in the definition of the graviton field $h_{\mu
\nu}$ gives this field  a mass-dimension of unity and thus yields a
kinetic term of standard normalization without a dimensionful
parameter. For matter interactions, this choice is convenient since
the order of $\kappa$ keeps track of the number of gravitons
involved in an interaction. Once the action is written in terms of
the expansion of the graviton field, all indices are understood to
be lowered or raised using the Minkowski metric $\eta_{\mu \nu}$. We
also require the expansion of the inverse metric and square root of
the determinant of the metric tensor---
\begin{eqnarray}
g^{\mu\nu}&=&\eta^{\mu\nu} - \kappa h^{\mu\nu} + \kappa^2 h^{\mu \alpha} h^\nu_\alpha
 + \mathcal O(\kappa^3)\nonumber \\
\sqrt{-g}&=& 1 + \frac{\kappa}{2} h + \frac{\kappa^2}{8} \left(h^2 - 2 h_{\mu \nu}
h^{\mu \nu} \right) + \mathcal O(\kappa^3).
\end{eqnarray}
Then, expanding in powers of $\kappa$, we find
\begin{eqnarray}
\sqrt{-g}{\cal L}_m^{(0)}&=&{1\over
2} \partial_\mu\phi\partial^\mu\phi - \frac{1}{2} m^2\phi^2 \nonumber\\
\sqrt{-g}{\cal L}_m^{(1)}&=&{\kappa\over 2} \, h^{\mu\nu}\left[\eta_{\mu\nu}
\left(\frac{1}{2} \partial_\alpha\phi\partial^\alpha\phi - \frac{1}{2} m^2\phi^2\right)-
\partial_\mu\phi\partial_\nu\phi\right]\nonumber\\
\sqrt{-g}{\cal L}_m^{(2)}&=&{\kappa^2\over 2} \hspace*{0.5pt} \Bigg[\frac{1}{4} \Big(h^2
- 2 h_{\mu \nu} h^{\mu \nu}\Big) \bigg(\frac{1}{2} \partial_\alpha \phi \partial^\alpha \phi
- \frac{1}{2} m^2 \phi^2\bigg)\nonumber\\
&& \hspace*{13pt} + \bigg(h^{\mu \alpha} h_\alpha^\nu - \frac{1}{2} h h^{\mu \nu}\bigg)
\partial_\mu \phi \partial_\nu \phi \Bigg]
\end{eqnarray}
where $h\equiv \eta^{\alpha\beta}h_{\alpha\beta}$ represents the
trace and the one- and two-graviton vertices are identified as
\begin{figure}[h]
  \centering
  \includegraphics{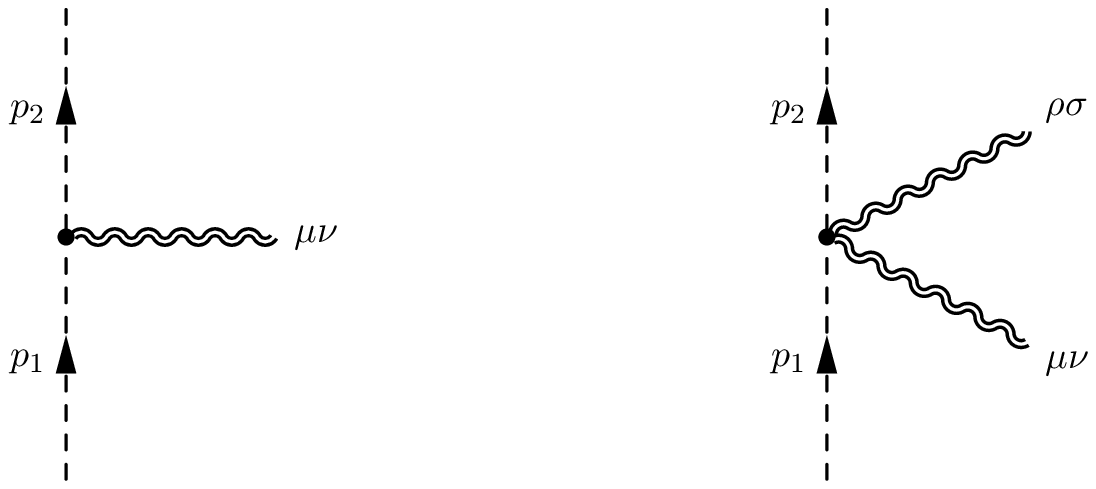}
\end{figure}
\vspace*{-20pt}
\begin{eqnarray}
{}^{0} \tau_{\mu\nu}^{(1)}(p_2, p_1, m) \hspace*{-7.5pt}&=\hspace*{-7.5pt}& {-i \kappa\over
2}\left[p_{1\mu}p_{2\nu}+p_{1\nu}p_{2\mu}-\eta_{\mu\nu}(p_1\cdot
p_2-m^2)\right]\nonumber\\
{}^{0} \tau^{(2)}_{\mu\nu,\rho\sigma}(p_2, p_1, m)\hspace*{-7.5pt}&=\hspace*{-7.5pt}&
\frac{i\kappa^2}{2}\bigg[2 I_{\mu\nu,\kappa\zeta}{I^\zeta}_{\lambda,\rho\sigma}(p_1^\kappa
p_2^\lambda+p_1^\lambda p_2^\kappa) \nonumber\\
&& {}\hspace*{14pt} - \hspace*{-1pt} (\hspace*{-0.5pt} \eta_{\mu\nu}I_{\kappa\lambda,\rho\sigma} \hspace*{-1.5pt} + \hspace*{-1.5pt} \eta_{\rho\sigma}I_{\kappa\lambda,\mu\nu}\hspace*{-0.5pt} )p_1^\kappa p_2^\lambda \hspace*{-1pt}
- \hspace*{-1pt} P_{\mu\nu, \rho\sigma} (p_1 \hspace*{-1.5pt} \cdot \hspace*{-1.3pt} p_2 \hspace*{-1.8pt} - \hspace*{-1.5pt} m^2) \hspace*{-1pt} \bigg]  \label{eq:tp}
\end{eqnarray}
where we have defined
\begin{eqnarray}
I_{\alpha\beta,\gamma\delta}&\equiv&{1\over
2}(\eta_{\alpha\gamma}\eta_{\beta\delta}+\eta_{\alpha\delta}\eta_{\beta\gamma}) \nonumber \\
P_{\alpha\beta,\gamma\delta}&\equiv&{1\over
2}(\eta_{\alpha\gamma}\eta_{\beta\delta}+\eta_{\alpha\delta}\eta_{\beta\gamma} - \eta_{\alpha\beta}\eta_{\gamma\delta}).
\end{eqnarray}

The purely gravitational dynamics are derived from the
Einstein-Hilbert action
\begin{equation}
 S_{GR} = \frac{1}{16 \pi G} \int d^4x \hspace*{1pt} \sqrt{-g} \, R \label{eq_actionGR}
\end{equation}
where higher dimensional operators such as $R^2$, which are needed
when performing renormalization, are not relevant
here \cite{don1,don2}.  Since general relativity is invariant under
{\it local} coordinate transformations, it is a gauge theory and one
must deal with the subtleties that arise in the quantization of
gauge theories---we have to perform gauge fixing.  The procedure
works analogously to gauge fixing in Yang-Mills theories.
Furthermore, we will make use of the background field method, which
provides a powerful organizational scheme for the quantization of
the effective field theory of gravity: Keeping the background metric
general instead of restricting to the flat Minkowski metric, we use
the expansion
\begin{equation}
 g_{\mu \nu} = \bar g_{\mu \nu} + \kappa h_{\mu \nu} \label{eq_metric-exp-bfm}.
\end{equation}
where $\bar g_{\mu \nu}$ is the classical background metric (or
field) and $h_{\mu \nu}$ is the quantum field.  A gauge fixing
condition is only imposed on the quantum field $h_{\mu \nu}$,
leaving the general covariance of the background unaffected. This
procedure has the advantage of ensuring that the resulting theory
can be renormalized, since the loop expansion then has the same
symmetry properties as the action.

Moreover, the background field method greatly simplifies
calculations involving graviton loops.  Gravitons running in loops
must be derived from from an expansion involving the quantum field
$h_{\mu \nu}$ whereas gravitons that are {\it not} within a loop may
be derived from expanding the background field
\begin{equation}
 \bar g_{\mu \nu} = \eta_{\mu \nu} + \kappa H_{\mu \nu}
\end{equation}
where $H_{\mu \nu}$ denotes an ``external'' graviton, {\it i.e.}, a
graviton that is not inside a loop. At the one loop level, at most
two gravitons involved in a vertex are propagating within a loop,
which allows us to use a gauge fixing condition {\it linear} in the
quantum field $h_{\mu \nu}$ and greatly simplifies the derivation of
both the triple graviton vertex and the vertex that couples one
graviton to the ghost fields.

In order to fix the gauge, we will use the harmonic gauge---
\begin{equation}
 \bar D^\nu h_{\mu \nu} - \frac{1}{2} \bar D_\mu h = 0 \label{eq_harmonic_gauge_condition}
\end{equation}
where $\bar D_\mu$ denotes the covariant derivative on the
background metric. This condition leads to an additional gauge
fixing piece of the action
\begin{equation}
 S_{GF} = \int d^4x \sqrt{- \bar g} \hspace*{1pt} \left(\bar D^\nu h_{\mu \nu} - \frac{1}{2} \bar D_\mu h\right) \! \left(\bar D_\rho h^{\mu \rho} - \frac{1}{2} \bar D^\mu h\right)
  \label{eq_actionEFTGR+GF}
\end{equation}
as well as the ghost action
\begin{equation}
 S_{Ghost} = \int d^4x \sqrt{- \bar g} \, \hspace*{1pt} \bar \eta^\mu \left(\bar D_\mu \bar D_\nu - R_{\mu \nu}\right) \eta^\nu
  \label{eq_actionEFTGR+ghost}
\end{equation}
where $\eta^\mu$ is the ghost field that annihilates a ghost
particle while $\bar \eta^\mu$ creates a ghost particle.\footnote{Note that
the ghost fields anticommute since they obey Fermi-Dirac statistics
and we have to include a factor of $(-1)$ for each closed ghost
loop.}

Now we are in the position to derive the Feynman rules for the
effective field theory of gravity. The complete quantum
gravitational action consists of three components
\begin{equation}
 S_{Grav} = S_{GR} + S_{GF} + S_{Ghost}, \label{eq_full_action_gr}
\end{equation}
and we will illustrate the use of the background field method during
the derivations of the Feynman rules. When deriving the graviton
propagator we expand the action to second order in the quantum
fields $h_{\mu \nu}$ and to zeroth order in the external gravitons
$H_{\mu \nu}$ and the ghosts, yielding
\begin{equation}
D_{\alpha\beta,\gamma\delta}(q)={iP_{\alpha\beta,\gamma\delta}\over q^2} .
\end{equation}
We obtain the ghost propagator expanding the action to second order
in the ghost fields and to zeroth order in the gravitons, whereby
\begin{equation}
D_{\mu\nu}(q)= \frac{i\eta_{\mu \nu}}{q^2}.
\end{equation}
Our triple graviton vertex is obtained by expanding to second order
in the quantum fields $h_{\mu \nu}$, to first order in the
background gravitons $H_{\mu \nu}$, and to zeroth order in the ghost
fields and reads
\begin{figure}[h]
  \centering
  \includegraphics{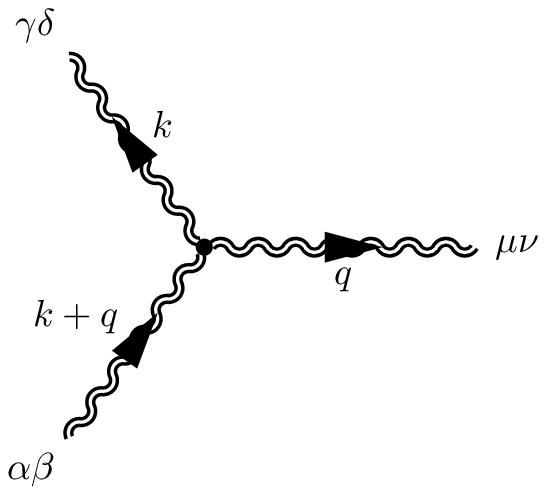}
 \end{figure}
\vspace*{-25pt}
\begin{eqnarray}
\tau^{\mu\nu}_{\alpha\beta,\gamma\delta}(k,q) \hspace*{-7pt}&=&\hspace*{-7pt} -{i\kappa\over 2}
\Bigg\{ P_{\alpha\beta,\gamma\delta} \hspace*{-1pt}
\left[k^\mu k^\nu + (k+q)^\mu (k+q)^\nu + q^\mu q^\nu - {3\over 2} \eta^{\mu\nu} q^2\right]\nonumber\\
&& \hspace*{17pt} + 2q_\lambda q_\sigma \hspace*{-1pt}
 \bigg[I^{\lambda\sigma,}{}_{\alpha\beta}I^{\mu\nu,}{}_{\gamma\delta}
  +I^{\lambda\sigma,}{}_{\gamma\delta}I^{\mu\nu,}{}_{\alpha\beta} \nonumber \\
&& \hspace*{54pt} -I^{\lambda\mu,}{}_{\alpha\beta}I^{\sigma\nu,}{}_{\gamma\delta}
 -I^{\sigma\nu,}{}_{\alpha\beta}I^{\lambda\mu,}{}_{\gamma\delta}\bigg]\nonumber\\
&& \hspace*{17pt} + \hspace*{-1pt} \bigg[
 q_\lambda q^\mu(\eta_{\alpha\beta}I^{\lambda\nu,}{}_{\gamma\delta} \hspace*{-1pt} + \hspace*{-1pt} \eta_{\gamma\delta}I^{\lambda\nu,}{}_{\alpha\beta}) \hspace*{-2pt}
 + \hspace*{-1pt} q_\lambda q^\nu(\eta_{\alpha\beta}I^{\lambda\mu,}{}_{\gamma\delta} \hspace*{-1pt} + \hspace*{-1pt} \eta_{\gamma\delta}I^{\lambda\mu,}{}_{\alpha\beta})\nonumber\\
&& \hspace*{24pt} - q^2 \hspace*{-0.6pt} (\hspace*{-0.5pt} \eta_{\alpha\beta}I^{\mu\nu,}{}_{\gamma\delta} \hspace*{-1.5pt} + \hspace*{-1.5pt} \eta_{\gamma\delta} I^{\mu\nu,}{}_{\alpha\beta} \hspace*{-0.5pt} ) \hspace*{-2.3pt}
 - \hspace*{-1.5pt} \eta^{\mu\nu} q^\lambda q^\sigma \hspace*{-0.6pt} (\hspace*{-0.5pt} \eta_{\alpha\beta} I_{\gamma\delta,\lambda\sigma} \hspace*{-2pt} + \hspace*{-1.5pt} \eta_{\gamma\delta} I_{\alpha\beta,\lambda\sigma}\hspace*{-0.5pt} ) \hspace*{-1.5pt} \bigg]\nonumber\\
&& \hspace*{17pt} + \hspace*{-1pt} \bigg[2q^\lambda \Big(I^{\sigma\nu,}{}_{\gamma\delta} I_{\alpha\beta,\lambda\sigma}k^\mu + I^{\sigma\mu,}{}_{\gamma\delta}I_{\alpha\beta,\lambda\sigma}k^\nu \nonumber\\
&& \hspace*{50pt} -I^{\sigma\nu,}{}_{\alpha\beta}I_{\gamma\delta,\lambda\sigma}(k+q)^\mu
 - I^{\sigma\mu,}{}_{\alpha\beta}I_{\gamma\delta,\lambda\sigma}(k+q)^\nu \Big) \nonumber\\
&& \hspace*{24pt} + q^2(I^{\sigma\mu,}{}_{\alpha\beta}I_{\gamma\delta,\sigma}{}^\nu + I_{\alpha\beta,\sigma}{}^\nu I^{\sigma\mu,}{}_{\gamma\delta}) \nonumber \\
&& \hspace*{24pt} +\eta^{\mu\nu}q^\lambda q_\sigma (I_{\alpha\beta,\lambda\rho}I^{\rho\sigma,}{}_{\gamma\delta} + I_{\gamma\delta,\lambda\rho}I^{\rho\sigma,}{}_{\alpha\beta})\bigg]\nonumber\\
&& \hspace*{17pt} + \hspace*{-1pt} \bigg[\hspace*{-0.5pt} \big(k^2 \hspace*{-1pt} + \hspace*{-1pt} (k \hspace*{-1pt} + \hspace*{-1pt} q)^2 \big) \hspace*{-3pt} \left( \hspace*{-2pt} I^{\sigma\mu,}{}_{\alpha\beta}I_{\gamma\delta,\sigma}{}^\nu \hspace*{-1pt}
 + \hspace*{-1pt} I^{\sigma\nu,}{}_{\alpha\beta}I_{\gamma\delta,\sigma}{}^\mu \hspace*{-1pt} - \hspace*{-1pt} {1\over 2}\eta^{\mu\nu}P_{\alpha\beta,\gamma\delta} \hspace*{-2pt} \right) \nonumber\\
&& \hspace*{24pt} - \big((k+q)^2\eta_{\alpha\beta}I^{\mu\nu,}{}_{\gamma\delta}+k^2\eta_{\gamma\delta} I^{\mu\nu,}{}_{\alpha\beta}\big)\bigg]\Bigg\}
\end{eqnarray}
where the graviton with Lorentz indices $\mu \nu$ represents a
background graviton, and therefore, it is {\it not} to be used
within any loop!\footnote{For each closed graviton bubble loop, we also have
to include a symmetry factor of $1/2!$ in the amplitude.} The final
missing Feynman rule from the purely gravitational action is the
vertex coupling one graviton to two ghost fields which we derive by
expanding to second order in the ghost fields and to first order in
$H_{\mu \nu}$---
\begin{figure}[h]
  \centering
  \includegraphics{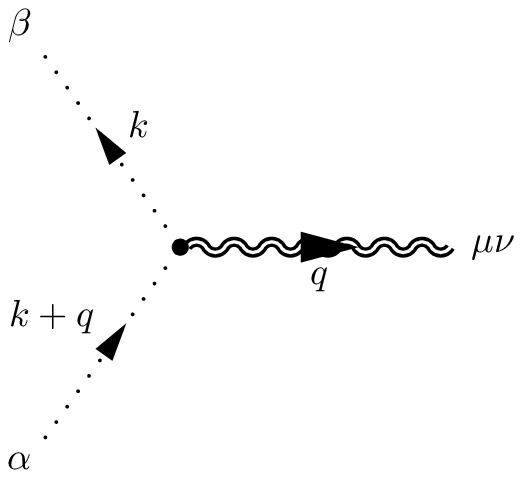}
 \end{figure}
\vspace*{-10pt}
\begin{eqnarray}
 \tau_{\alpha,\beta}^{\mu \nu}(k,q) \hspace*{-5pt} & = & \hspace*{-5pt} \frac{i \kappa}{2}
  \bigg[\big(k^2+(k+q)^2+q^2\big) I_{\alpha \beta,}{}^{\mu \nu}
   + 2 \hspace*{1pt} \eta_{\alpha \beta} \hspace*{1pt} k^\lambda (k+q)^\sigma P_{\lambda \sigma,}{}^{\mu \nu} \nonumber \\
 && \hspace*{10pt} + \hspace*{1pt} 2 \hspace*{1pt} q_\alpha k^\lambda \hspace*{1pt} I_{\beta \lambda,}{}^{\mu \nu} - 2 \hspace*{1pt} q_\beta (k+q)^\lambda \hspace*{1pt}I_{\alpha \lambda,}{}^{\mu \nu} + q_\alpha q_\beta \eta^{\mu \nu}\bigg]
\end{eqnarray}
where the graviton with Lorentz indices $\mu \nu$ is again a background graviton and not to be used as part of any loop.

We begin our calculation with the simplest tree level
single-graviton exchange which leads to the amplitude\footnote{Our
normalization of the amplitude is a nonrelativistic one such that
after applying all Feynman rules we divide the amplitude by a factor
of $\sqrt{2E_12E_22E_32E_4}$.}
\begin{eqnarray}
{}^0{\cal M}^{(1)}(q)&=&{-i\over
\sqrt{2E_12E_22E_32E_4}} \, \tau^{(1a)}_{\mu\nu}(p_2,p_1) \hspace*{1pt} {iP^{\mu\nu,\alpha\beta}\over
q^2} \, \tau^{(1b)}_{\alpha\beta}(p_4,p_3)\nonumber\\
&=&{- 8\pi G\over \sqrt{2E_12E_22E_32E_4}} \left[{(s-m_a^2-m_b^2+{1\over
2} q^2)^2-2m_a^2m_b^2 - \frac{1}{4}q^4 \over q^2}\right]. \nonumber\\
\quad
\end{eqnarray}
A convenient way to define the nonrelativistic potential is as the
Fourier transform of the nonrelativistic center of mass scattering
amplitude.  We utilize a symmetric center of mass
frame\footnote{These symmetric momentum labels of the center of mass
frame are chosen so that the leading order coordinate space
potential is real in the calculation of spin-0 -- spin-1 scattering
presented below.} with incoming momenta $\vec p_1 = \vec p + \vec q
/ 2$ and $\vec p_3 = - \vec p_1 = - \vec p - \vec q / 2$ and with
outgoing momenta $\vec p_2 = \vec p - \vec q / 2$ and $\vec p_4 = -
\vec p + \vec q / 2$. Conservation of energy then requires $\vec p
\cdot \vec q = 0$ so that $\vec p_i^{\hspace*{1.4pt} 2} = \vec
p^{\hspace*{1.4pt} 2} + \vec q^{\hspace*{1.4pt} 2} / 4$ for $i = 1,
2, 3, 4$ and $q^2 = - \vec q^{\hspace*{1.4pt} 2}$. In the
nonrelativistic limit--- $\vec q^{\hspace*{1.4pt} 2}, \vec
p^{\hspace*{1.4pt} 2} \ll m^2$ ---the lowest order amplitude reads
\begin{eqnarray}
{}^0 \! {\cal M}^{(1)}(\vec q) & \simeq & \frac{4 \pi G m_a m_b} {\vec q^{\hspace*{1.4pt} 2}}
\left[1 + \frac{\vec p^{\hspace*{1.4pt} 2}}{m_a m_b} \left(1+\frac{3(m_a + m_b)^2}{2 m_a m_b}\right)
+ \ldots\right] \nonumber \\
&+& G \pi \left[\frac{3(m_a^2 + m_b^2)}{2 m_a m_b} + \frac{\vec p^{\hspace*{1.4pt} 2}}{m_a m_b}
\left(3 - \frac{5(m_a^2 + m_b^2)^2}{4 m_a^2 m_b^2}\right) + \ldots \right] + \ldots \nonumber \\ \label{eq_ampLO_00}
\end{eqnarray}
yielding the potential
\begin{eqnarray}
{}^0V^{(1)}_G(\vec{r})&=&-\int {d^3q\over (2\pi)^3} \,
{}^0 \!{\cal M}^{(1)}(\vec q) \, e^{-i\vec{q}\cdot\vec{r}} \nonumber\\
&=&- \hspace*{1pt} \frac{G m_a m_b}{r} \left[1 + \frac{\vec p^{\hspace*{1.4pt} 2}}{m_a m_b}
\left(1+\frac{3(m_a + m_b)^2}{2 m_a m_b}\right) + \ldots\right] \nonumber \\
&& + \, G \pi \delta^3(\vec r) \left[\frac{3(m_a^2 + m_b^2)}{2 m_a m_b} +
\frac{\vec p^{\hspace*{1.4pt} 2}}{m_a m_b}\left(3 - \frac{5(m_a^2 + m_b^2)^2}{4 m_a^2 m_b^2}\right) + \ldots \right]
\nonumber\\
\label{eq:po}
\end{eqnarray}
The leading component of Eq. (\ref{eq:po}) is recognized as the
usual Newtonian potential (accompanied by a small kinematic
correction) while the second piece is a short range modification.

\begin{figure}
\begin{center}
\epsfig{file=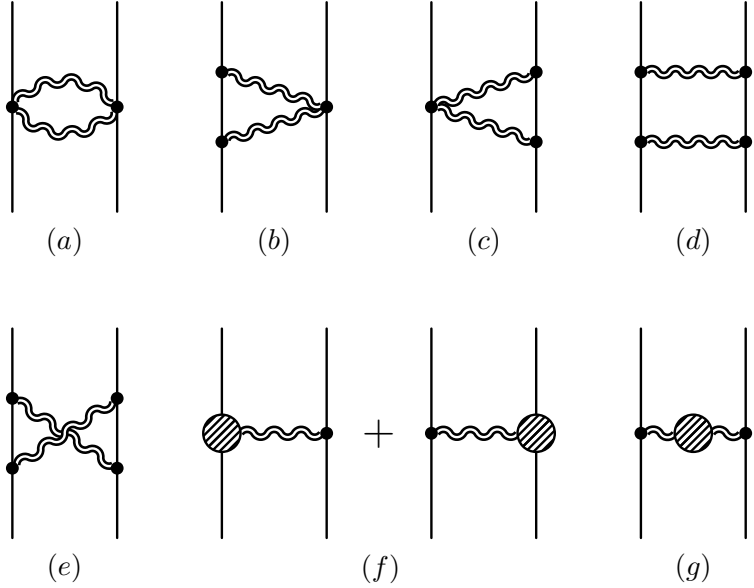,width=10cm} \caption{One loop diagrams of
gravitational scattering. } \label{fig_loops}
\end{center}
\end{figure}

Our purpose in this paper is to study the long distance corrections
to this lowest order potential which arise from the two-graviton
exchange diagrams shown in Fig. \ref{fig_loops}.  This problem has been previously
studied by Iwasaki using noncovariant perturbation theory
\cite{iwa}, and by Khriplovich and Kirilin \cite{kk, Khriplovich:2004cx} and by
Bjerrum-Bohr, Donoghue, and Holstein \cite{bdh} using conventional
Feynman diagrams.  Our approach will be similar to that used in
\cite{kk, Khriplovich:2004cx} and \cite{bdh}.  That is, using the ideas of effective
field theory, we evaluate these Feynman diagrams by keeping only the
leading nonanalytic structure in $q^2$, since it is these pieces
that lead to the long range corrections to the potential while
components analytic in $q^2$ only yield short range contributions,
{\it i.e.}, delta functions or derivatives of delta functions. The
leading nonanalytic behavior is of two basic forms
\begin{itemize}
\item [i)] terms in $1/\sqrt{-q^2}$ which are $\hbar$-independent
and therefore classical
\item [ii)] terms in $\log -q^2$ which are $\hbar$-dependent and
therefore quantum mechanical.
\end{itemize}
The former terms, when Fourier transformed lead to corrections to
the nonrelativistic potential of the form $V_{classical}(r)\sim
1/r^2$ while the latter lead to $V_{quantum}(r)\sim \hbar/mr^3$
corrections.  For typical masses and separations the quantum
mechanical forms are themselves numerically insignificant.
However, they are intriguing in that their origin appears to be
associated with zitterbewegung.  That is, classically we can
define the potential by measuring the energy when two objects are
separated by distance $r$.  However, in the quantum mechanical
case the distance between two objects is uncertain by an amount of
order the Compton wavelength due to zero point motion---$\delta
r\sim \hbar/ m$. This leads to the replacement
$$V(r)\sim {1\over r^2}\longrightarrow {1\over (r\pm\delta r)^2}
\sim {1\over r^2}\mp 2{\hbar\over mr^3}$$ which is the form found
in our calculations.

\begin{figure}
\begin{center}
\epsfig{file=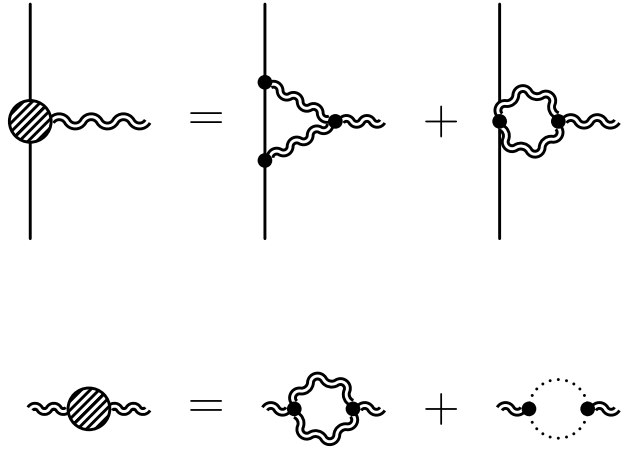,width=0.6\textwidth} \caption{Loop
corrections subsumed in vertex and in vacuum polarization functions.} \label{fig_blobs}
\end{center}
\end{figure}

The diagrams to be evaluated are shown in Fig. \ref{fig_loops}
where the ``blobs'' shown in diagrams (f) and (g) are explained in
Fig. \ref{fig_blobs}. The calculational details including the Feynman integrals
are described in an appendix of our companion paper on electromagnetic scattering
\cite{hrem}. Here we present only the results. Defining\footnote{Note that our definition of $S$ differs slightly from previous publications \cite{don1, don2, bdh, Bjerrum-Bohr:2002ks} in that it does not include a factor of mass in the numerator.}
$$S=\frac{\pi^2}{\sqrt{-q^2}} \quad \quad{\rm and} \quad \quad L=\log -q^2$$
we have, from diagrams \ref{fig_loops}(a)-(g) respectively
\begin{eqnarray}
{}^0{\cal M}_{\ref{fig_loops}a}^{(2)}(q)&=&G^2 m_a m_b \left[-44 L\right]\nonumber\\
{}^0{\cal M}_{\ref{fig_loops}b}^{(2)}(q)&=&G^2m_am_b\left[28L + 8 m_a S\right]\nonumber\\
{}^0{\cal M}_{\ref{fig_loops}c}^{(2)}(q)&=&G^2m_am_b\left[28L + 8 m_b S\right]\nonumber\\
{}^0{\cal M}_{\ref{fig_loops}d}^{(2)}(q)&=&G^2 m_a m_b \left[\hspace*{-0.5pt} \left(\frac{4 m_a m_b}{q^2}+\frac{8(m_a^2 + m_b^2)}{m_a m_b} - 8\right) \hspace*{-1.5pt} L + \hspace*{-0.5pt}4 (m_a + m_b) S \right]\nonumber\\
&-&i 4\pi G^2 m_a^2 m_b^2 \hspace*{1pt} {L\over q^2} \sqrt{m_a m_b \over s-s_0}\nonumber\\
{}^0{\cal M}_{\ref{fig_loops}e}^{(2)}(q)&=&G^2 m_a m_b \left[\hspace*{-0.5pt} \left(\hspace*{-1.5pt} - \frac{4 m_a m_b}{q^2}- \frac{8(m_a^2 + m_b^2)}{m_a m_b} - \frac{70}{3}\right) \hspace*{-1.5pt} L - \hspace*{-0.5pt}4 (m_a \hspace*{-0.5pt} + \hspace*{-0.5pt} m_b) S \right]\nonumber\\
{}^0{\cal M}_{\ref{fig_loops}f}(q)&=&G^2m_am_b\left[14 L - 2 (m_a + m_b) S\right]\nonumber\\
{}^0{\cal M}_{\ref{fig_loops}g}(q)&=&G^2m_am_b\left[-{43\over 15} L \right]\quad
\end{eqnarray}
where $s=(p_1+p_3)^2$ is the square of the center of mass energy and
$s_0=(m_a+m_b)^2$ is its threshold value. The contributions from the
form factor diagrams in Fig. \ref{fig_loops}(f) have been calculated
both directly and using the results from \cite{Bjerrum-Bohr:2002ks,
Holstein:FF}, and the contribution from vacuum polarization diagrams
in Fig. \ref{fig_loops}(g) have been obtained both by direct
evaluation and by using previous results of 't Hooft and Veltman
\cite{'tHooft:1974bx} for the divergences of the graviton
self-energy that allow us to infer its nonanalytic
contributions\footnote{In a massless theory the divergences of
dimensional regularization are accompanied by logarithms of the
momentum transfer.}. Here we should note that {\it any} massless
particle would contribute to the nonanalytic terms of the graviton
vacuum energy since gravitons couple to anything. We choose here not
to include other massless particles besides the graviton---in
particular we do not include the contribution of the photon in the
quantum piece of our potential.

Summing, we find the total result
\begin{equation}
{}^0{\cal M}^{(2)}_{tot}(q)=G^2m_am_b\left[6(m_a+m_b)S - {41\over
5}L \right]-i4\pi G^2m_a^2m_b^2 \hspace*{1pt} {L\over
q^2}\sqrt{m_am_b\over s-s_0}\label{eq:kp}
\end{equation}
and we observe that, in addition to the expected terms involving $L$
and $S$, there arises a piece of the second order amplitude which is
{\it imaginary}.  The origin of this imaginary piece is, of course,
from the second Born approximation to the Newtonian potential, and
reminds us that in order to define a proper correction to the first
order Newtonian potential we must subtract off such terms. For this
purpose we will work in the nonrelativistic limit and the center of
mass frame---$\vec{p}_1+\vec{p}_3=0$ as defined above.  We have then
\begin{equation}
s-s_0=2\sqrt{m_a^2+ \vec{p}_1^{\hspace*{1.4pt} 2}} \sqrt{m_b^2+\vec{p}_1^{\hspace*{1.4pt} 2}}
+2 \vec{p}_1^{\hspace*{1.4pt} 2} -2m_am_b
\end{equation}
and
\begin{equation}
\sqrt{m_am_b\over s-s_0}\simeq {m_r\over p_0}
\end{equation}
where $m_r=m_am_b/(m_a+m_b)$ is the reduced mass and $p_0 \equiv
|\vec{p}_i|, \  i=1,2,3,4$.  The transition amplitude Eq.
(\ref{eq:kp}) then assumes the form
\begin{equation}
{}^0{\cal M}^{(2)}_{tot}(\vec q) \simeq G^2m_am_b\left[6(m_a+m_b)S -{41\over 5}L\right]-i4\pi G^2m_a^2m_b^2 \hspace*{1pt} {L\over q^2}{m_r\over p_0} .
\end{equation}
For the iteration we shall use the simple potential
\begin{equation}
{}^0V^{(1)}_G(\vec r)=-{Gm_am_b\over r}\label{eq:co}
\end{equation}
which reproduces the long distance behavior of the lowest order
amplitude for spin-0 -- spin-0 gravitational scattering---Eq.
(\ref{eq:po})---in the nonrelativistic limit. We will employ the
momentum space representation
\begin{equation}
 {}^0V^{(1)}_G(\vec q) \equiv \left<\vec p_f \left| {}^0 \hat V^{(1)}_G \right|
 \vec p_i \right> = - \frac{4 \pi G m_a m_b}{\vec q^{\hspace*{1.4pt}2}} =
 - \frac{4 \pi G m_a m_b}{(\vec p_i - \vec p_f)^2} \label{eq_lopot0momsp}
\end{equation}
where we identify $\vec p_i = \vec p_1$ and $\vec p_f = \vec p_2$,
and the second Born term is then
\begin{eqnarray} \label{eq:iteration00a}
{}^0{\rm Amp}_G^{(2)}(\vec q)&=&- \int{d^3\ell\over
(2\pi)^3} \, \frac{\left<\vec p_f \left| {}^0 \hat V^{(1)}_G \right| \vec \ell \,
\right> \left<\vec \ell \left| {}^0 \hat V^{(1)}_G \right|
\vec p_i \right>}{E(p_0) - E(\ell) + i \epsilon}\nonumber\\
&=&i\int{d^3\ell\over
(2\pi)^3}{}^0V_G^{(1)}(\vec \ell - \vec p_f) \, G^{(0)}(\vec{\ell})
\, {}^0V_G^{(1)}(\vec{p}_i-\vec{\ell} \, )
\end{eqnarray}
where
\begin{equation}
G^{(0)}(\ell)={i\over {p_0^2\over 2m_r}-{\ell^2\over
2m_r}+i\epsilon}
\end{equation}
is the nonrelativistic propagator. Note that in Eq.
(\ref{eq:iteration00a}) we take both the leading order potential as
well as the total energies $E(p_0)$ and $E(\ell)$ in the
nonrelativistic limit. The remaining integration can be performed
exactly, yielding\footnote{Note that the iteration integrals are
listed in Appendix \ref{app_iter}}
\begin{eqnarray} \label{eq:iteration00b}
{}^0{\rm Amp}_G^{(2)}(\vec q) \hspace*{-5pt} &= \hspace*{-5pt} &i\int {d^3\ell\over (2\pi)^3} \hspace*{1pt} {- 4 \pi G m_a m_b\over
|\vec{\ell} - \vec{p}_2|^2+\lambda^2} \hspace*{1pt} {i\over {p_0^2\over
2m_r}-{\ell^2\over 2m_r}+i\epsilon} \hspace*{1pt} {- 4 \pi G m_a m_b\over
|\vec{p}_1 - \vec{\ell}|^2+\lambda^2}\nonumber\\
&= \hspace*{-5pt} &H =-i \hspace*{0.4pt} 4\pi G^2 m_a^2 m_b^2 \hspace*{1pt} \frac{L}{q^2} \frac{m_r}{p_0}
\end{eqnarray}
which precisely reproduces the imaginary component of ${}^0{\cal
M}_{tot}^{(2)}(\vec q)$, as expected.  In order to produce a properly
defined second order potential ${}^0V^{(2)}_G(\vec r)$ we must then
subtract this second order Born term from the second order
transition amplitude, yielding the result
\begin{eqnarray}
{}^0V_G^{(2)}(\vec r)&=&-\int{d^3q\over
(2\pi)^3}e^{-i\vec{q}\cdot\vec{r}}\left[{}^0{\cal
M}_{tot}^{(2)}(q)-{}^0{\rm Amp}_G^{(2)}(q)\right]\nonumber\\
&=&\int{d^3q\over (2\pi)^3}e^{-i\vec{q}\cdot\vec{r}}
\hspace*{1pt} G^2 m_a m_b\left[-6S(m_a+m_b) + {41\over 5}L\right]\nonumber\\
&=&-{3G^2m_am_b(m_a+m_b)\over r^2
}-{41G^2m_am_b\hbar\over 10\pi r^3}\nonumber\\
\quad\label{eq:so}
\end{eqnarray}

The quantum mechanical---$\sim \hbar/r^3$---component of the second
order potential given in Eq. (\ref{eq:so}) agrees with that
previously given by Bjerrum-Bohr, Donoghue, and Holstein \cite{bdh}
and by Kirilin and Khriplovich \cite{Khriplovich:2004cx}.
However, the classical---$\sim 1/r^2$---contribution quoted by
Iwasaki
\begin{equation}
 {}^0V_{IW}^{(2)}(\vec r)= {G^2m_am_b(m_a+m_b)\over 2 r^2}.
\end{equation}
differs from that quoted above in Eq. (\ref{eq:so}) and by
Bjerrum-Bohr et al. in \cite{bdh}.  The resolution of this issue was
given by Sucher, who pointed out that the classical term depends
upon the precise definition of the first order potential used in the
iteration \cite{js}. Moreover, it depends and on whether one uses
relativistic forms of the leading order potentials and the
propagator $G^{(0)}(\ell)$ in the iteration. In modern terms, the
potential depends on how one performs the matching---{\it e.g.}, Iwasaki
\cite{iwa} performs an off-shell matching while we match on-shell.
In our companion paper on electromagnetic scattering \cite{hrem} we
have provided a more detailed discussion of these
ambiguities\footnote{Besides the dependence on the forms used in the
iteration, the classical piece also depends on the coordinates used.
The quantum piece however depends neither on the choice of
coordinates \cite{bdh} nor on the iteration forms \cite{hrem}.}. Use
of the simple lowest order form Eq. (\ref{eq_lopot0momsp}) within a
nonrelativistic iteration yields our result for the iteration
amplitude given in Eq. (\ref{eq:iteration00b}) and is sufficient to
remove the offending imaginary piece of the scattering amplitude. In
Appendix \ref{app_eom} we derive an alternative form of the
$\mathcal O(G^2)$ classical potential which results from an
iteration that includes the leading relativistic corrections and
which reproduces the classical equations of motion.

Therefore, a unique definition of the potential does not exist. Of
course, the ambiguities in the form of the second order classical
potential should not be a concern, since the potential is {\it not}
an observable.  What {\it is} an observable is the on-shell
transition amplitude, which is uniquely defined in each case as
\begin{equation}
{}^0 \!{\cal M}_{tot}(\vec q)=-\int
d^3re^{i\vec{q}\cdot\vec{r}}\left[{}^0V_i^{(1)}(\vec
r)+{}^0V_i^{(2)}(\vec r)\right]+{}^0{\rm Amp}_i(\vec{q})
\end{equation}
where the index $i$ denotes differing possible definitions of the
potentials and the iteration. Thus we regard our potential as a nice
way to display our resulting scattering amplitudes in coordinate space,
but we emphasize that our main results are the long distance components
of the scattering amplitude.

\section{Spin-Dependent Scattering: Spin-Orbit Interaction}
\subsection{Spin-0 -- Spin-1/2}

Having determined the form of the potential for the
spinless scattering case we move on to the case of scattering of
particles carrying spin.  We begin with the scattering of a
spinless particle $a$ from a spin-1/2 particle $b$.

For the case of spin-1/2 we require some additional formalism in
order to extract the gravitational couplings, which is necessary
because the Dirac algebra $\big\{\gamma^a , \gamma^b \big\} = 2 \eta^{a b}$
is defined with respect to the Minkoswki flat space metric.
In this case the Dirac matter Lagrangian coupled to gravity reads
\begin{equation}
\sqrt{-g}{\cal L}_m = \sqrt {-g} \, \bar \psi \left[\frac{i}{2} \hspace{1pt} {e^{\mu}}_a \{ \gamma^a, D_\mu \} - m\right]\psi
\end{equation}
and involves the vierbein ${e^{\mu}}_a$ which links global
coordinates with those in a locally flat space.  The vierbein is
in some sense the ``square root'' of the metric tensor
$g_{\mu\nu}$ and satisfies the relations
\begin{eqnarray}
{e_{\mu}}^a \hspace*{1pt} {e_{\nu}}^b \, \eta_{ab} &=& g_{\mu \nu} \quad \quad \quad \quad {e^{\mu}}_a \hspace*{1pt} {e^{\nu}}_b \, \eta^{ab} \ \, = \, \ g^{\mu \nu} \nonumber\\
{e_{\mu}}^a \hspace*{1pt} {e_{\nu}}^b \, g^{\mu \nu} &=& \eta_{ab} \hspace*{0.5pt} \quad \quad \quad \quad {e^{\mu}}_a \hspace*{1pt} {e^{\nu}}_b \, g_{\mu \nu} \ \, = \ \, \eta^{ab} .
\end{eqnarray}
The covariant derivative is
\begin{equation}
 D_\mu = \frac{1}{2} \hspace{1pt} \partial^{LR}_\mu + \frac{i}{4} \, {{\omega_{\mu}}^a}_b \, \eta_{a c} \ \sigma^{c b}
\end{equation}
with $\sigma^{c b} = \frac{i}{2} \Big[\gamma^c, \gamma^d\Big]$ and the partial derivative $\partial^{LR}_\mu$ acts only on spinors and in such a way that
\begin{equation}
 \bar \psi \partial^{LR}_\mu \psi = \bar \psi \, \partial_\mu \psi - \left(\partial_\mu \bar \psi\right) \psi.
\end{equation}
Putting everything together, we find then
\begin{equation} \label{lagrangian12}
 \sqrt {-g} \, \mathcal L_m = \sqrt {-g} \, \bar \psi \left[\frac{i}{2} \hspace{1pt} \gamma^a {e^{\mu}}_a \partial^{LR}_\mu - \frac{1} {8} \hspace{1pt} {e^{\mu}}_{a'} \ {{\omega_{\mu}}^a}_b \, \eta_{a c} \, \{\gamma^{a'} , \sigma^{c b}\} - m\right]\psi.
\end{equation}
The spin connection ${{\omega_{\mu}}^a}_b \, \eta_{a c}$ can be
derived in terms of vierbeins by requiring $D_\mu {e_\nu}^a = 0$ and
by antisymmetrization in $\mu \leftrightarrow \nu$ in order to get
rid of Christoffel symbols\footnote{For our purposes we shall use
only the symmetric component of the vierbein matrices, since these
are physical and can be connected to the metric tensor, while their
antisymmetric components are associated with freedom of homogeneous
transformations of the local Lorentz frames and do not contribute to
nonanalyticity \cite{Deser:1974cy}.}. The result is:
\begin{equation}
 {{\omega_{\mu}}^a}_b \, \eta_{a c} = \left(\frac{\eta_{ab}} {2} \, {e^\nu}_c \left(\partial_\mu {e_\nu}^a - \partial_\nu {e_\mu}^a \right) + \frac{\eta_{af}}{2} \, {e^\nu}_c \hspace{1pt} {e^\rho}_b \hspace{1pt} {e_\mu}^f \hspace{1pt} \partial_\rho {e_\nu}^a \right) - \Big(b \leftrightarrow c\Big)
\end{equation}
In order to derive the Feynman rules we expand the ingredients in
Eq. (\ref{lagrangian12}) that contain graviton couplings, that is we
need ${e^{\mu}}_a$ and ${{\omega_{\mu}}^a}_b \, \eta_{a c}$ expanded
up to $\mathcal O(\kappa^2)$
\begin{eqnarray}
 {e_\mu}^a \hspace*{-3pt}& = \hspace*{-3pt}&\delta_\mu^a + \frac{\kappa}{2} \hspace{1pt} h_\mu^a
  - \frac{\kappa^2}{8} \hspace{1pt} h_{\mu \rho} h^{a \rho}+\ldots \nonumber \\
 {e^\mu}_a \hspace*{-3pt}& = \hspace*{-3pt}& \delta_a^\mu - \frac{\kappa}{2} \hspace{1pt} h_a^\mu
  + \frac{3 \kappa^2}{8} \hspace{1pt} h_{a \rho} h^{\mu \rho}+\ldots \nonumber \\
 {{\omega_{\mu}}^a}_b \, \eta_{a c} \hspace*{-3pt}& = \hspace*{-3pt}&\frac{\kappa}{2}
 \hspace{1pt} \partial_b h_{\mu c} + \frac{\kappa^2}{8} \hspace{1pt} h_b^\rho \partial_\mu h_{c \rho}
 - \frac{\kappa^2}{4} \hspace{1pt} h_b^\rho \partial_\rho h_{\mu c} + \frac{\kappa^2}{4} \hspace{1pt} h_b^\rho \partial_c h_{\mu \rho} - \Big(b \leftrightarrow c\Big) \nonumber \\
\end{eqnarray}
After these expansions are employed, we no longer need to
distinguish between Latin Lorentz indeces and Greek covariant 
indices and can use the Minkowski metric to lower and raise all indices.

The matter Lagrangian then has the expansion---(note here that our
conventions are $\gamma_5=-i\gamma^0\gamma^1\gamma^2\gamma^3$ and
$\epsilon^{0123} = + 1$)
\begin{eqnarray}
\sqrt{-g}{\cal L}_m^{(0)}&=&\bar{\psi} \hspace*{-1pt} \left({i\over 2} \hspace*{-2pt} \not\!{\partial}^{LR}-m \right) \hspace*{-1.3pt} \psi\nonumber\\
\sqrt{-g}{\cal L}_m^{(1)}&=&{\kappa\over 2} \hspace*{1pt} h \, \bar{\psi} \hspace*{-1pt}\left({i\over 2} \hspace*{-2pt} \not\!{\partial}^{LR}-m\right) \hspace*{-1.3pt}\psi
  -{\kappa\over 2} \hspace*{1pt} h^{\mu \nu} \, \bar{\psi} \, \frac{i}{2} \partial_\mu^{LR} \gamma_\nu \hspace*{1pt} \psi\nonumber\\
\sqrt{-g}{\cal L}_m^{(2)}&=& {\kappa^2 \over 8} \left(h^2 - 2 h_{\alpha \beta} h^{\alpha \beta} \right) \bar{\psi} \hspace*{-1pt} \left({i\over 2} \hspace*{-2pt} \not\!{\partial}^{LR}-m\right) \hspace*{-1.3pt} \psi \nonumber \\
&+& \frac{\kappa^2}{8} \left(3 h^{\mu \rho} h_\rho^\nu - 2 h h^{\mu \nu} \right) \bar{\psi} \, \frac{i}{2} \partial_\mu^{LR} \gamma_\nu \hspace*{1pt} \psi\nonumber\\
&+& \frac{i \kappa^2}{16}  \epsilon^{\alpha \beta \gamma \delta} \, h^\rho_\alpha (i\partial_\beta h_{\rho \gamma}) \, \bar \psi \gamma_\delta \gamma_5 \psi
\end{eqnarray}
and the corresponding one- and two-graviton vertices are found to be
\begin{figure}[h]
  \centering
  \includegraphics{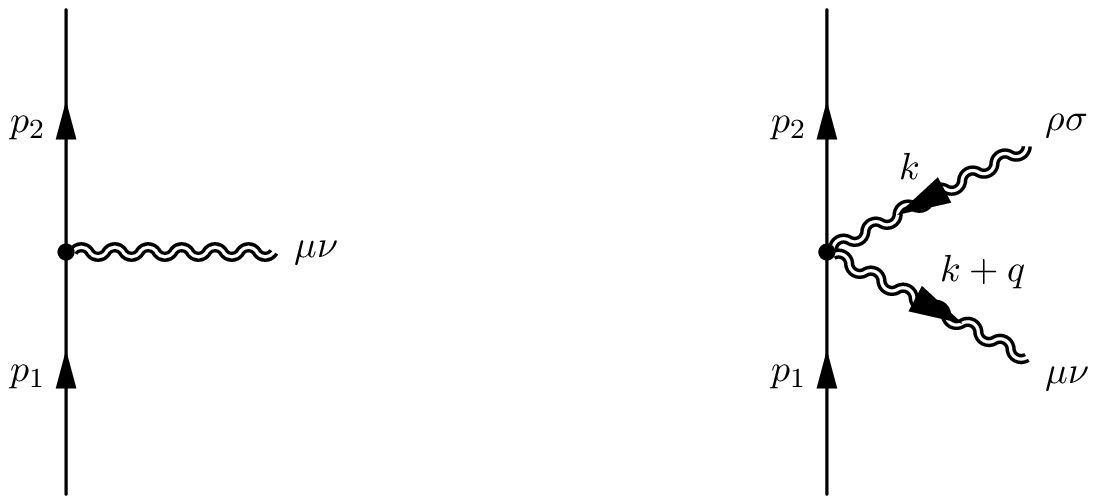}
\end{figure}
\vspace*{-20pt}
\begin{eqnarray}
{}^{\frac{1}{2}}\tau_{\mu\nu}^{(1)}(p_2, p_1, m)\hspace*{-8pt} & = \hspace*{-7.5pt}&{-i\kappa\over 2}\hspace*{-0.5pt}\Bigg[\hspace*{-0.5pt}{1\over 4}\hspace*{-0.7pt}\Big(\hspace*{-1pt}\gamma_\mu(p_1 \hspace*{-1.8pt} + \hspace*{-1.5pt} p_2)_\nu \hspace*{-1pt}+\hspace*{-1pt}\gamma_\nu(p_1 \hspace*{-1.8pt} + \hspace*{-1.5pt} p_2)_\mu\hspace*{-0.5pt}\Big)
\hspace*{-2.3pt} - \hspace*{-1.5pt} \eta_{\mu\nu} \hspace*{-0.7pt} \Bigg(\hspace*{-1pt}{1\over 2}(\! \not\!{p}_1 + \hspace*{-1pt} \!\not\!{p}_2)\hspace*{-2pt}- \hspace*{-1pt} m \hspace*{-1.5pt} \Bigg) \hspace*{-2pt} \Bigg]\nonumber\\
{}^{\frac{1}{2}}\tau_{\mu\nu,\rho\sigma}^{(2)}(p_2, p_1, m) \hspace*{-8pt} & = \hspace*{-7.5pt}& i\kappa^2\Bigg[
\hspace*{-1pt}-{1\over 2}\bigg({1\over 2}(\not \!{p}_1 \hspace*{1pt} + \! \not\!{p}_2)-m\bigg) \hspace*{1pt} P_{\mu\nu,\rho\sigma}\nonumber\\
&& \hspace*{20.3pt}{} - {1\over
16}\bigg[\eta_{\mu\nu} \Big(\gamma_\rho(p_1+p_2)_\sigma+\gamma_\sigma(p_1+p_2)_\rho\Big) \nonumber\\
&& \hspace*{43.5pt} {} + \hspace*{-1pt} \eta_{\rho\sigma}\Big(\gamma_\mu
(p_1+p_2)_\nu+\gamma_\nu(p_1+p_2)_\mu\Big)\bigg] \nonumber\\
&& \hspace*{20.3pt}{} + {3\over
16}(p_1+p_2)^{\epsilon}\rho^{\xi}(I_{\xi\phi,\mu\nu}{I^{\phi}}_{\epsilon,\rho\sigma}
+I_{\xi\phi,\rho\sigma}{I^{\phi}}_{\epsilon,\mu\nu}) \nonumber\\
&& \hspace*{20.3pt}{} + {i\over 16}\epsilon^{\epsilon\phi\eta\lambda}\gamma_\lambda \gamma_5
\Big(I_{\rho\sigma,\phi\xi} {I_{\mu\nu,\eta}}^\xi \hspace*{1pt} {k}_\epsilon
- I_{\mu\nu,\phi\xi} {I_{\rho\sigma,\eta}}^\xi \hspace*{1pt} (k+q)_\epsilon \Big)\Bigg]. \nonumber \\ \quad
\end{eqnarray}

The tree level transition amplitude from one-graviton exchange is then
\begin{eqnarray}
{}^{1\over 2}{\cal M}^{(1)}(q) \hspace*{-5pt} &= \hspace*{-5pt}&{- 16 \pi G m_a m_b \over
\sqrt{2E_12E_2E_3E_4}}\Bigg[-\frac{m_a m_b}{q^2} \hspace*{1pt} \bar{u}(p_4)u(p_3) \nonumber \\
&& \hspace*{76pt} {}+ \frac{s - m_a^2 - m_b^2 + \frac{1}{2} q^2}{q^2} \, {1\over m_a}\bar{u}(p_4)\!\not\!{p}_1u(p_3)\Bigg] . \quad
\end{eqnarray}
Defining the spin vector as
\begin{equation}
S_b^\mu={1\over 2}\bar{u}(p_4)\gamma_5\gamma^\mu u(p_3)
\end{equation}
we find the identity
\begin{equation}
\bar{u}(p_4)\gamma_\mu u(p_3)=\left({1\over 1-{q^2\over
4m_b^2}}\right)\left[{(p_3+p_4)_\mu\over
2m_b}\bar{u}(p_4)u(p_3)-{i\over
m_b^2}\epsilon_{\mu\beta\gamma\delta}q^\beta p_3^\gamma
S_b^\delta\right]\label{eq:id}
\end{equation}
whereupon the nonanalytic part of the transition amplitude in the thereshold
limit $s \rightarrow s_0 = (m_a + m_b)^2$ can be written in the form
\begin{equation}
{}^{1\over 2}{\cal M}^{(1)}(q) \simeq -{4\pi G m_a m_b\over
q^2}\left[\bar{u}(p_4)u(p_3)+{2 i\over
m_am_b^2}\epsilon_{\alpha\beta\gamma\delta}p_1^\alpha p_3^\beta
q^\gamma S_b^\delta \right].
\end{equation}
In order to define the potential we require the nonrelativistic
amplitude in the symmetric center of mass frame ($\vec p_1 = - \vec
p_3 = \vec p + \vec q /2$) where
\begin{equation}
S_b^\alpha\stackrel{NR}{\longrightarrow}(0,\vec{S}_b) \ \ \ \ \ {\rm
with} \ \ \ \ \ \vec{S}_b = {1\over
2}\chi_f^{b\dagger}\vec{\sigma}\chi_i^b,
\end{equation}
\begin{equation}
\bar{u}(p_4)u(p_3) \stackrel{NR}{\longrightarrow}
\chi_f^{b\dagger}\chi_i^b-{i\over 2m_b^2}\vec{S}_b\cdot \vec{p}
\times\vec{q} \label{eq:nr}
\end{equation}
and
\begin{equation} \label{eq:epsireduce}
\epsilon_{\alpha\beta\gamma\delta}p_1^\alpha p_3^\beta q^\gamma
S_b^\delta\stackrel{NR}{\longrightarrow}(m_a+m_b)\left(1+\frac{\vec
p^{\hspace*{1.4pt}2}}{2 m_a
m_b}\right)\vec{S}_b\cdot\vec{p}\times\vec{q},
\end{equation}
so that
\begin{equation}
{}^{1\over 2}{\cal M}^{(1)}(\vec q)\simeq {4\pi Gm_am_b\over
\vec{q}^{\hspace*{1.4pt} 2}}\left[\chi_f^{b\dagger}\chi_i^b + {i(3m_a + 4 m_b)\over
2m_am_b^2}\vec{S}_b\cdot\vec{p}\times\vec{q}+\ldots\right]
\end{equation}
and the lowest order potential becomes
\begin{eqnarray}
{}^{1\over 2}V_G^{(1)}(\vec r)&=&-\int {d^3q\over (2\pi)^3} \, \hspace*{1pt}
{}^{1\over 2}{\cal M}^{(1)}(\vec q \hspace*{1pt}) \, e^{-i\vec{q}\cdot\vec{r}}\nonumber\\
&=&-{G m_a m_b \over r}\chi_f^{b\dagger}\chi_i^b
-{3 m_a + 4 m_b\over
2 m_a m_b^2}\vec{S}_b\cdot\vec{p}\times\vec{\nabla}\left(-{G m_a m_b \over r}\right)\nonumber\\
&=&-{Gm_am_b\over r}\chi_f^{b\dagger}\chi_i^b+{G\over
r^3}{3m_a + 4m_b\over 2m_b}\vec{L}\cdot\vec{S}_b\label{eq:oj}
\end{eqnarray}
where $\vec{L}=\vec{r}\times\vec{p}$ is the angular momentum---the
modification of the leading spin-independent potential has a
spin-orbit character.

In order to determine the second order potential, we must evaluate
the one loop diagrams in Fig. \ref{fig_loops}. A subtlety that
arises in the calculation involving spin is that {\it two}
independent kinematic variables arise: the momentum transfer $q^2$
and $s - s_0$, which is to leading order proportional to $p_0^2$
(where $p_0^2 \equiv \vec p_i^{\hspace*{1.4pt}2},  \ i=1,2,3,4$) in
the center of mass frame.  We find that our results differ if we
perform an expansion first in $s - s_0$ and then in $q^2$ or vice
versa. This ordering issue occurs only for the box diagram, diagram
(d) of Fig. \ref{fig_loops}, where it stems from the reduction of
vector and tensor box integrals.  Their reduction in terms of scalar
integrals involves the inversion of a matrix whose Gram determinant
vanishes in the nonrelativistic threshold limit $q^2, s - s_0
\rightarrow 0$. More precisely, the denominators or the vector and
tensor box integrals (see Appendix A in \cite{hrem}) involve a
factor of $(4 p_0^2 - \vec q^{\hspace*{1.4pt} 2})$ when expanded in
the nonrelativistic limit. Since $q^{\hspace*{1.4pt} 2} = 4 p_0^2
\sin^2 \frac{\theta}{2}$ with $\theta$ the scattering angle, we
notice that $4 p_0^2 > \vec q^{\hspace*{1.4pt} 2}$ unless we
consider backward scattering where $\theta = \pi$ and where the
scattering amplitude diverges. And since $p_0^2$ originates from the
relativistic structure $s - s_0$, it is clear that we must first
expand our vector and tensor box integrals in $q^2$ and then in $s -
s_0$. With this procedure, the contributions of the loop diagrams in
Fig. \ref{fig_loops} to the spin-0 -- spin-1/2 scattering amplitude
read
\begin{eqnarray}
{}^{1\over 2}{\cal M}_{\ref{fig_loops}a}^{(2)}(q) \hspace*{-5pt} &=\hspace*{-5pt}&G^2 m_a m_b \bigg[\bar{u}(p_4) u(p_3) (-10 L) + {1\over
m_a}\bar{u}(p_4) \! \not \!{p}_1u(p_3) (-24L) \bigg]\nonumber\\
{}^{1\over 2}{\cal M}_{\ref{fig_loops}b}^{(2)}(q) \hspace*{-5pt} &=\hspace*{-5pt}&G^2 m_a m_b \bigg[\bar{u}(p_4) u(p_3) (- m_a S + 3 L) \nonumber\\
&& \hspace*{41pt} + {1\over m_a}\bar{u}(p_4) \! \not \!{p}_1u(p_3) (8 m_a S + 20 L ) \bigg]\nonumber\\
{}^{1\over 2}{\cal M}_{\ref{fig_loops}c}^{(2)}(q)  \hspace*{-5pt} &=\hspace*{-5pt}&G^2 m_a m_b \bigg[\bar{u}(p_4) u(p_3) (4 m_b S + 2 L) \nonumber\\
&& \hspace*{41pt} + {1\over m_a}\bar{u}(p_4) \! \not \!{p}_1u(p_3) (4 m_b S + 16 L ) \bigg]\nonumber\\
{}^{1\over 2}{\cal M}_{\ref{fig_loops}d}^{(2)}(q)  \hspace*{-5pt} &=\hspace*{-5pt}&G^2 m_a m_b \Bigg[\bar{u}(p_4)u(p_3) \hspace*{-1pt} \Bigg( \! - \frac{4 m_a m_b}{q^2} L- {m_am_b(3 m_a+4 m_b)\over s-s_0}S \nonumber\\
&&  \hspace*{98pt} - \hspace*{0.9pt} (6 m_a \hspace*{-2pt} + \hspace*{-2pt} 11 m_b) S  \hspace*{-1pt} -  \hspace*{-1pt} \frac{42 m_a^2 \hspace*{-2pt} + \hspace*{-2pt} 23m_am_b \hspace*{-2pt} - \hspace*{-2pt} 4 m_b^2}{6m_am_b} L \hspace*{-1pt}\Bigg)\nonumber\\
&& \hspace*{41pt} + {1\over
m_a}\bar{u}(p_4) \! \not \!{p}_1u(p_3) \hspace*{-1pt}\Bigg({8m_am_b\over q^2} L
+ {m_am_b(3 m_a + 4 m_b)\over s-s_0} S\nonumber\\
&& \hspace*{131pt} +  (9 m_a+ 13 m_b) S \nonumber\\
&& \hspace*{131pt} +  \frac{153 m_a^2 - 44 m_a m_b + 88 m_b^2}{12 m_a m_b} L \Bigg) \Bigg]\nonumber\\
&&- \ i 4\pi G^2 m_a^2 m_b^2 \hspace*{1pt} {L\over q^2}\sqrt{m_a m_b\over s-s_0} \left(2\,  {1\over m_a}\bar{u}(p_4) \! \not \!{p}_1u(p_3) - \bar{u}(p_4)u(p_3)\right)\nonumber\\
{}^{1\over 2}{\cal M}_{\ref{fig_loops}e}^{(2)}(q)  \hspace*{-5pt} &=\hspace*{-5pt}&G^2 m_a m_b \Bigg[\bar{u}(p_4)u(p_3) \hspace*{-1pt} \Bigg( \frac{4 m_a m_b}{q^2} L\nonumber\\
&&  \hspace*{98pt} + \frac{19 m_a}{4} S  +  \frac{42 m_a^2 \hspace*{-1pt} + \hspace*{-1pt} 21m_am_b \hspace*{-1pt} - \hspace*{-1pt} 4 m_b^2}{6m_am_b} L \hspace*{-1pt}\Bigg)\nonumber\\
&& \hspace*{41pt} + {1\over m_a}\bar{u}(p_4) \! \not \!{p}_1u(p_3)
\hspace*{-1pt}\Bigg(- {8m_am_b\over q^2} L - \frac{33 m_a + 16 m_b}{4} S \nonumber\\
&& \hspace*{131pt} -  \frac{153 m_a^2 + 268 m_a m_b + 88 m_b^2}{12 m_a m_b} L \Bigg) \Bigg]\nonumber\\
{}^{1\over 2}{\cal M}_{\ref{fig_loops}f}^{(2)}(q) \hspace*{-5pt} &=\hspace*{-5pt}&G^2 m_a m_b \bigg[\bar{u}(p_4) u(p_3) \bigg( \! - \frac{3 m_a}{2} S + 10 L\bigg) \nonumber\\
&& \hspace*{41pt} + {1\over m_a}\bar{u}(p_4) \! \not \!{p}_1u(p_3) \bigg( \! - \frac{m_a + 4 m_b}{2} S + 4 L \bigg) \bigg]\nonumber\\
{}^{1\over 2}{\cal M}_{\ref{fig_loops}g}^{(2)}(q) \hspace*{-5pt} &=\hspace*{-5pt}&G^2 m_a m_b \bigg[\bar{u}(p_4) u(p_3) \hspace*{-0.5pt} \bigg(\hspace*{-2.5pt} - \hspace*{-1pt}\frac{1}{15} L\bigg) \hspace*{-1.5pt} + \hspace*{-0.5pt}{1\over
m_a}\bar{u}(p_4) \! \not \!{p}_1u(p_3) \hspace*{-0.5pt} \bigg( \hspace*{-2.5pt} - \hspace*{-1pt} \frac{42}{15}L\bigg) \hspace*{-1pt}\bigg].
\end{eqnarray}
Again, we have calculated the form factor diagrams in Fig. \ref{fig_loops}(f) both directly and via the
results from \cite{Bjerrum-Bohr:2002ks, Holstein:FF}.
Summing, we find
\begin{eqnarray}
{}^{1\over 2}{\cal M}_{tot}^{(2)}(q) \hspace*{-5pt} &=\hspace*{-5pt}&G^2 m_a m_b \Bigg[L\left(\frac{23}{5}\bar{u}(p_4)u(p_3)
  -{64\over 5}{1\over m_a}\bar{u}(p_4) \! \not\!{p}_1u(p_3)\right) \nonumber\\
&&\hspace*{41pt}+ \hspace*{1.4pt}S \bigg(- \frac{15 m_a + 28 m_b}{4} \, \bar{u}(p_4)u(p_3) \nonumber\\
&&\hspace*{61pt} + \hspace*{1.4pt} \frac{11(3 m_a + 4 m_b)}{4} {1\over m_a}
\bar{u}(p_4) \! \not\!{p}_1u(p_3)  \bigg)\nonumber\\
&&\hspace*{41pt}- {(3 m_a\hspace*{-1.5pt}+\hspace*{-1pt}4m_b)m_am_bS\over
s-s_0}\hspace*{-1pt}\left(\bar{u}(p_4)u(p_3) \hspace*{-1pt}-\hspace*{-1pt}{1\over
m_a}\bar{u}(p_4) \! \not\!{p}_1u(p_3)\hspace*{-1.7pt}\right)\hspace*{-2pt}\Bigg]\nonumber\\
&&- \ i 4\pi G^2 m_a^2 m_b^2 \hspace*{1pt} {L\over q^2}\sqrt{m_a m_b\over s-s_0} \left(- \bar{u}(p_4)u(p_3) + 2\,  {1\over m_a}\bar{u}(p_4) \! \not \!{p}_1u(p_3) \right)\nonumber\\
\quad \label{eq:su}
\end{eqnarray}
Using the identity Eq. (\ref{eq:id}) and
$$p_1\cdot(p_3+p_4)=2m_am_b+s-s_0+{q^2\over 2}$$
Eq. (\ref{eq:su}) becomes
\begin{eqnarray}
{}^{1\over 2}{\cal M}_{tot}^{(2)}(q) \hspace*{-5pt} &=\hspace*{-5pt}&G^2 m_a m_b \Bigg[
\bar{u}(p_4)u(p_3) \bigg(6(m_a + m_b) S - \frac{41}{5} L\bigg) \nonumber \\
&& \hspace*{41pt} + \hspace*{0.5pt} \frac{i}{m_a m_b^2} \hspace*{1pt} \epsilon_{\alpha\beta\gamma\delta} \hspace*{1pt} p_1^\alpha p_3^\beta q^\gamma S_b^\delta \,
 \bigg(\frac{11(3 m_a + 4 m_b)}{4} S - \frac{64}{5} L\bigg) \nonumber\\
&& \hspace*{41pt}+ \hspace*{0.5pt} {i S (3 m_a + 4 m_b)\over
m_b (s-s_0)}\epsilon_{\alpha\beta\gamma\delta}
p_1^\alpha p_3^\beta q^\gamma S_b^\delta\Bigg]\nonumber\\
&&-i4\pi G^2 m_a^2 m_b^2 \hspace*{1pt} {L\over q^2}\sqrt{m_a m_b\over
s-s_0}\left(\bar{u}(p_4)u(p_3)+ {2 i\over
m_am_b^2}\epsilon_{\alpha\beta\gamma\delta}
p_1^\alpha p_3^\beta q^\gamma S_b^\delta\right) \nonumber\\
\quad \label{eq:ohla}
\end{eqnarray}
Finally, working in the center of mass frame and taking the
nonrelativistic limit, we find
\begin{eqnarray}
{}^{1\over 2}{\cal M}_{tot}^{(2)}(\vec q)\hspace*{-5pt} &\simeq\hspace*{-5pt}&\left[G^2 m_a m_b
\left( 6 (m_a + m_b) S - {41 \over 5}L\right) - i 4 \pi G^2 m_a^2 m_b^2 \hspace*{1pt} \frac{L}{q^2} \frac{m_r}{p_0}\right]
\chi_f^{b\dagger}\chi_i^b \nonumber\\
&+\hspace*{-5pt}&\Bigg[G^2  \hspace*{-1pt} \left(\frac{12m_a^3 \hspace*{-1pt} + \hspace*{-1pt} 45 m_a^2 m_b \hspace*{-1pt} + \hspace*{-1pt} 56 m_a m_b^2
 \hspace*{-1pt} + \hspace*{-1pt} 24 m_b^3}{2(m_a \hspace*{-1pt} + \hspace*{-1pt} m_b)}  \hspace*{1pt} S
- \frac{87 m_a \hspace*{-1pt} + \hspace*{-1pt} 128 m_b}{10}  \hspace*{1pt} L\right)\nonumber\\
&&+ \frac{G^2 m_a^2 m_b^2 (3 m_a + 4 m_b)}{(m_a + m_b)} \left(- i \frac{2 \pi
L}{p_0 q^2} + \frac{S}{p_0^2} \right)\Bigg] {i\over
m_b}\vec{S}_b\cdot\vec{p}\times\vec{q}\label{eq:yo}
\end{eqnarray}
We note from Eq. (\ref{eq:yo}) that the scattering amplitude
consists of two pieces---a spin-independent component proportional
to $\chi_f^{b\dagger}\chi_i^b$ whose functional form
\begin{equation}
 G^2 m_a m_b \left( 6 (m_a + m_b) S - {41 \over 5}L\right) - i 4 \pi G^2 m_a^2 m_b^2 \hspace*{1pt} \frac{L}{q^2} \frac{m_r}{p_0}
\end{equation}
is {\it identical} to that of spinless scattering---together with
a spin-orbit component proportional to
$${i\over m_b}\vec{S}_b\cdot\vec{p}\times\vec{q}$$
whose functional form is
\begin{eqnarray}
&&G^2 \left(\frac{12m_a^3 + 45 m_a^2 m_b + 56 m_a m_b^2 + 24 m_b^3}{2(m_a + m_b)}  \hspace*{1pt} S
- \frac{87 m_a + 128 m_b}{10}  \hspace*{1pt} L\right)\nonumber\\
& + & \frac{G^2 m_a^2 m_b^2 (3 m_a + 4 m_b)}{(m_a + m_b)} \left(- i \frac{2 \pi
L}{p_0 q^2} + \frac{S}{p_0^2} \right) \label{eq:im}
\end{eqnarray}
We note in Eq. (\ref{eq:im}) the presence in the spin-orbit
potential of an imaginary final state rescattering term proportional
to $i/p_0$, similar to that found in the case of spin-independent
scattering, together with a {\it completely new} type of kinematic
form, proportional to $1/p_0^2$ which diverges at threshold.  The
presence of {\it either} term would prevent us from writing down a
well defined second order potential.

The solution to this problem is, as before, to properly subtract the
iterated first order potential---
\begin{eqnarray}
{}^{1\over 2}{\rm Amp}_G^{(2)}(\vec q) &=&- \int{d^3\ell\over
(2\pi)^3} \, \frac{\left<\vec p_f \left| {}^{\frac{1}{2}} \hat
V^{(1)}_G \right| \vec \ell \, \right> \left<\vec \ell \left|
{}^{\frac{1}{2}} \hat V^{(1)}_G \right| \vec p_i
\right>}{{p_0^2\over 2m_r} - \frac{\ell^2}{2 m_r} + i \epsilon}
\end{eqnarray}
where we now use the one-graviton exchange potential ${}^{1\over
2}V_G^{(1)}(\vec r)$ given in Eq. (\ref{eq:oj}).  Splitting this
lowest order potential into spin-independent and spin-dependent
components---
\begin{equation}
\left<\vec p_f \left| {}^{\frac{1}{2}} \hat V^{(1)}_G \right| \vec
p_i \, \right> =
 \left<\vec p_f \left| {}^{\frac{1}{2}} \hat V^{(1)}_{S-I} \right| \vec p_i \, \right>
 + \left<\vec p_f \left| {}^{\frac{1}{2}} \hat V^{(1)}_{S-O} \right| \vec p_i \, \right>
\end{equation}
where
\begin{eqnarray}
 \left<\vec p_f \left| {}^{\frac{1}{2}} \hat V^{(1)}_{S-I} \right| \vec p_i \, \right>
 &=& - {4 \pi G m_a m_b \over \vec{q}^{\hspace*{1.4pt}2}} \, \chi_f^{b\dagger}\chi_i^b
  = - \frac{4 \pi G m_a m_b}{(\vec p_i - \vec p_f)^2} \, \chi_f^{b\dagger}\chi_i^b \nonumber\\
 \left<\vec p_f \left| {}^{\frac{1}{2}} \hat V^{(1)}_{S-O} \right| \vec p_i \, \right>
 &=&- {4 \pi G m_a m_b \over \vec{q}^{\hspace*{1.4pt}2}}{3 m_a+ 4 m_b\over 2m_am_b}\, \frac{i}{m_b}\vec{S}_b\cdot\vec{p}\times\vec{q} \nonumber\\
 &=&- {4 \pi G m_a m_b \over (\vec p_i - \vec p_f)^2}{3 m_a + 4 m_b\over 2m_am_b}\, \frac{i}{m_b}\vec{S}_b\cdot \frac{1}{2} (\vec p_i + \vec p_f) \times (\vec p_i - \vec p_f) \nonumber\\
\end{eqnarray}
we find that the iterated amplitude splits also into
spin-independent and spin-dependent pieces.  The leading
spin-independent amplitude is
\begin{eqnarray}
{}^{1\over 2}{\rm Amp}^{(2)}_{S-I}(\vec q) \hspace*{-3pt} &= \hspace*{-3pt} &-
\int{d^3\ell\over (2\pi)^3} \, \frac{\left<\vec p_f \left|
{}^{\frac{1}{2}} \hat V^{(1)}_{S-I} \right| \vec \ell \, \right>
\left<\vec \ell \left| {}^{\frac{1}{2}}
 \hat V^{(1)}_{S-I} \right| \vec p_i \right>}{\frac{p_0^2}{2 m_r} -
 \frac{\ell^2}{2 m_r} + i \epsilon} \nonumber\\
&= \hspace*{-3pt} &i \sum_{s_\ell} \int {d^3\ell\over (2\pi)^3} {c_G^2
\chi_f^{b\dagger}\chi_{s_\ell}^b \over
|\vec{\ell} - \vec{p}_f|^2+\lambda^2}{i\over {p_0^2\over
2m_r}-{\ell^2\over 2m_r}+i\epsilon}{c_G^2
\chi_{s_\ell}^{b\dagger}\chi_i^b \over
|\vec{p}_i - \vec{\ell}|^2+\lambda^2}\nonumber\\
&=  \hspace*{-3pt} &\chi_f^{b\dagger}\chi_i^b H =-i4\pi G^2 m_a^2
m_b^2 \hspace*{1pt} {L\over
q^2}\frac{m_r}{p_0}\chi_f^{b\dagger}\chi_i^b
\label{eq:iteration0hSI}
\end{eqnarray}
where we defined $c_G^2 \equiv - 4 \pi G m_a m_b$,
and the leading spin-dependent term is
\begin{eqnarray}
{}^{1\over 2}{\rm Amp}^{(2)}_{S-O}(\vec q) \hspace*{-6pt} &= \hspace*{-6pt}&-
\int{d^3\ell\over (2\pi)^3} \, \frac{\left<\vec p_f \left|
{}^{\frac{1}{2}} \hat V^{(1)}_{S-I}
 \right| \vec \ell \, \right> \left<\vec \ell \left| {}^{\frac{1}{2}}
 \hat V^{(1)}_{S-O} \right| \vec p_i \right>}{\frac{p_0^2}{2 m_r} -
 \frac{\ell^2}{2 m_r} + i \epsilon} \nonumber\\
&&- \int{d^3\ell\over (2\pi)^3} \, \frac{\left<\vec p_f \left|
{}^{\frac{1}{2}} \hat V^{(1)}_{S-O} \right| \vec \ell \, \right>
\left<\vec \ell \left| {}^{\frac{1}{2}} \hat V^{(1)}_{S-I} \right|
\vec p_i \right>}{\frac{p_0^2}{2 m_r} -
\frac{\ell^2}{2 m_r} + i \epsilon} \nonumber\\
& = \hspace*{-6pt} & {i(3 m_a+ 4 m_b)\over 2 m_am_b^2}
\vec{S}_b \cdot \nonumber\\
&& \left(\hspace*{-2pt}i \hspace*{-3.2pt} \int \hspace*{-3.2pt}
{d^3\ell\over (2\pi)^3} {c_G^2\over |\vec \ell \hspace*{-1.1pt} -
\hspace*{-1.1pt} \vec{p}_f \hspace*{1pt}|^2 \hspace*{-1.1pt} +
\hspace*{-1.2pt} \lambda^2}{i \over {p_0^2\over 2m_r}
\hspace*{-1.1pt} - \hspace*{-1.1pt} {\ell^2\over 2m_r}
\hspace*{-1.1pt} + \hspace*{-1.1pt} i\epsilon}{c_G^2 \,
\frac{1}{2}(\vec p_i \hspace*{-1.1pt} + \hspace*{-1.1pt} \vec \ell)
\hspace*{-2.5pt} \times \hspace*{-2.5pt} (\vec p_i \hspace*{-1.1pt}
- \hspace*{-1.1pt} \vec \ell)\over |\vec{p}_i \hspace*{-1.1pt} -
\hspace*{-1.1pt} \vec \ell|^2 \hspace*{-1.1pt}
 + \hspace*{-1.2pt} \lambda^2}\right.\nonumber\\
&&\left. \hspace*{-3.6pt}+ i \hspace*{-3.2pt} \int \hspace*{-3.2pt}
{d^3\ell\over (2\pi)^3} {c_G^2\, \frac{1}{2}(\vec \ell
\hspace*{-1.1pt} + \hspace*{-1.2pt} \vec p_f) \hspace*{-2.5pt}
\times \hspace*{-2.5pt} (\vec \ell \hspace*{-1.1pt} -
\hspace*{-1.2pt} \vec p_f)\over |\vec \ell \hspace*{-1.1pt} -
\hspace*{-1.1pt} \vec{p}_f \hspace*{1pt}|^2 \hspace*{-1.1pt} +
\hspace*{-1.2pt} \lambda^2}{i \over {p_0^2\over 2m_r}
\hspace*{-1.1pt} - \hspace*{-1.1pt} {\ell^2\over 2m_r}
\hspace*{-1.1pt} + \hspace*{-1.1pt} i\epsilon}{c_G^2\over |\vec{p}_i
\hspace*{-1.1pt} - \hspace*{-1.1pt} \vec \ell|^2 \hspace*{-1.1pt}
 + \hspace*{-1.2pt}\lambda^2}\hspace*{-3pt}\right)\nonumber\\
& = \hspace*{-6pt} &{i(3 m_a + 4 m_b)\over 2 m_a m_b^2} \,
\vec{S}_b\cdot\vec{H}\times
\vec{q}\nonumber\\
& = \hspace*{-6pt} &\frac{G^2 m_a^2 m_b^2 (3 m_a + 4 m_b)}{(m_a + m_b)} \left(- i \frac{2 \pi
L}{p_0 q^2} + \frac{S}{p_0^2} \right) {i\over
m_b}\vec{S}_b\cdot\vec{p}\times\vec{q} \label{eq:iteration0hSO}
\end{eqnarray}
(In principle we would also have to iterate the leading order
spin-orbit piece twice.  However this procedure yields only terms
higher order in $p_0^2$.)  We observe that when the amplitudes Eqs.
(\ref{eq:iteration0hSO}) and (\ref{eq:iteration0hSI}) are subtracted
from the full one loop scattering amplitude Eq. (\ref{eq:yo}) both
the terms involving $1/p_0^2$ and those proportional to $i/p_0$
disappear leaving behind a well-defined second order potential
\begin{eqnarray}
{}^{1\over 2}V^{(2)}_G(\vec r)&=&-\int{d^3q\over
(2\pi)^3}e^{-i\vec{q}\cdot\vec{r}}\left[{}^{1\over 2}{\cal
M}_{tot}^{(2)}(\vec{q})-{}^{1\over 2}{\rm
Amp}_G^{(2)}(\vec{q})\right]\nonumber\\
&=& \int{d^3q\over
(2\pi)^3}e^{-i\vec{q}\cdot\vec{r}}
\Bigg[G^2 m_a m_b \left(- 6(m_a+m_b)S+{41\over 5}L\right)
\chi_f^{b\dagger}\chi_i^b\nonumber\\
&& {}\hspace*{63pt} + G^2 \Bigg( \hspace*{-2pt} - \frac{12m_a^3 + 45 m_a^2 m_b + 56 m_a m_b^2 + 24 m_b^3}{2(m_a + m_b)}  \hspace*{1pt} S \nonumber\\
&& {}\hspace*{91pt} + \frac{87 m_a + 128 m_b}{10}  \hspace*{1pt} L\Bigg) \hspace*{1pt}
{i\over m_b}\vec{S}_b\cdot\vec{p}\times\vec{q} \hspace*{1pt} \Bigg]\nonumber\\
&=&\left[-{3G^2m_am_b(m_a+m_b)\over r^2} - {41G^2m_am_b\hbar\over 10\pi r^3}\right]
\chi_f^{b\dagger}\chi_i^b\nonumber\\
&+&{1\over m_b}\vec{S}_b\cdot \vec{p}\times\vec{\nabla}\Bigg[
\frac{G^2(12m_a^3 + 45 m_a^2 m_b + 56 m_a m_b^2 + 24 m_b^3)}{4(m_a + m_b)r^2} \nonumber\\
&& \hspace*{66pt} + \frac{G^2(87 m_a + 128 m_b) \hbar}{20 \pi r^3} \Bigg]\nonumber\\
&=&\left[-{3G^2m_am_b(m_a+m_b)\over r^2} - {41G^2m_am_b\hbar\over 10\pi r^3}\right]
\chi_f^{b\dagger}\chi_i^b\nonumber\\
&+&\Bigg[{G^2(12 m_a^3 + 45 m_a^2 m_b + 56 m_a m_b^2 + 24 m_b^3)\over
2 m_b (m_a+m_b)r^4} \nonumber \\
&&+{3 G^2(87 m_a+ 128m_b)\hbar\over 20 \pi
m_b r^5}\Bigg] \vec{L}\cdot\vec{S}_b \label{eq:mn}
\end{eqnarray}
We observe then that the second order potential for long range
gravitational scattering of a spinless and spin-1/2 particle
consists of two components:  one which is independent of the spin of
particle b and is identical to the potential found for the case of
spinless scattering, accompanied by a spin-orbit interaction
involving a new form for its classical and quantum potentials.  It
is tempting to speculate that the form of this new spin-orbit
potential is also universal.  In order to check this hypothesis we
consider the case of spin-0 -- spin-1 scattering.

\subsection{Spin-0 -- Spin-1}

The dynamics of a neutral spin-1 field $\phi_\mu$ having mass $m$ is
described by the Proca Lagrangian which, when coupled to gravity via
minimal substitution, takes the form
\begin{equation}
\sqrt{-g}{\cal L}_m = \sqrt{-g} \left[- {1\over 4} \hspace*{1pt} U_{\mu\nu} U_{\rho \sigma} g^{\mu \rho} g^{\nu \sigma}
+ {1\over 2}m^2\phi_\mu \phi_\nu g^{\mu \nu} \right] \label{eqn:la}
\end{equation}
where
\begin{equation}
U_{\mu\nu} = D_\mu \phi_\nu - D_\nu\phi_\mu = \partial_\mu \phi_\nu - \partial_\nu\phi_\mu \label{eq:fieldtens}
\end{equation}
is the spin-1 field tensor. The last equality in Eq.
(\ref{eq:fieldtens}) follows from the symmetry of the connection
coefficients $\Gamma^\alpha_{\mu \nu} = \Gamma^\alpha_{\nu \mu}$. 
Expanded in terms of the graviton field, the matter Lagrangian then has the form
\begin{eqnarray}
\sqrt{-g}{\cal L}_m^{(0)} \hspace*{-5pt} &= \hspace*{-5pt} &
- {1\over 2}\partial_\mu\phi_\nu\partial^\mu\phi^\nu
+ {1\over 2}\partial_\mu\phi_\nu\partial^\nu\phi^\mu
+ {1\over 2}m^2\phi_\mu\phi^\mu\nonumber\\
\sqrt{-g}{\cal L}_m^{(1)} \hspace*{-5pt} &= \hspace*{-5pt} &\frac{\kappa}{2} h \left(
- {1\over 2} \partial_\mu \phi_\nu \partial^\mu \phi^\nu
+ {1\over 2} \partial_\mu \phi_\nu \partial^\nu \phi^\mu
+ {1\over 2} m^2 \phi_\mu \phi^\mu \right)\nonumber\\
& - \hspace*{-5pt} &\kappa h^{\mu \nu} \left(
- \frac{1}{2} \partial_\mu \phi_\alpha \partial_\nu \phi^\alpha
- \frac{1}{2} \partial_\alpha \phi_\mu \partial^\alpha \phi_\nu
+ \partial_\mu \phi^\alpha \partial_\alpha \phi_\nu
+ \frac{1}{2} m^2 \phi_\mu \phi_\nu\right)\nonumber\\
\sqrt{-g}{\cal L}_m^{(2)} \hspace*{-5pt} &= \hspace*{-5pt} &\frac{\kappa^2}{8} \left(h^2 - 2 h_{\alpha \beta} h^{\alpha \beta}\right) \! \left(
- {1\over 2} \partial_\mu \phi_\nu  \partial^\mu \phi^\nu
+ {1\over 2} \partial_\mu \phi_\nu \partial^\nu \phi^\mu
+ {1\over 2} m^2 \phi_\mu \phi^\mu \right)\nonumber\\
& + \hspace*{-5pt} &\frac{\kappa^2}{2} \left(2 h^{\mu \rho} h^\nu_\rho - h h^{\mu \nu}\right) \nonumber \\
&& \hspace*{15pt} \times  \left(
- \frac{1}{2} \partial_\mu \phi_\alpha \partial_\nu \phi^\alpha
- \frac{1}{2} \partial_\alpha \phi_\mu \partial^\alpha \phi_\nu
+ \partial_\mu \phi^\alpha \partial_\alpha \phi_\nu
+ \frac{1}{2} m^2 \phi_\mu \phi_\nu\right)\nonumber\\
& + \hspace*{-5pt} & \kappa^2 \hspace*{1pt} h^{\mu \rho} h^{\nu \sigma} \left(
- \frac{1}{2} \partial_\mu \phi_\nu \partial_\rho \phi_\sigma
+ \frac{1}{2} \partial_\mu \phi_\nu \partial_\sigma \phi_\rho \right)
\end{eqnarray}
and the one- and two-graviton vertices are
\begin{figure}[h]
  \centering
  \includegraphics{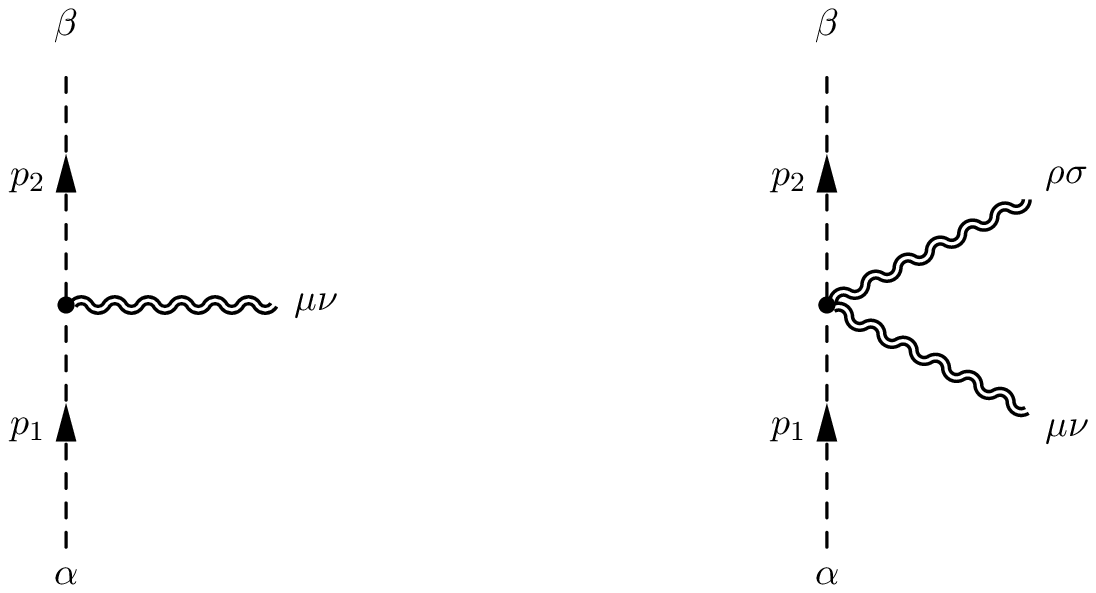}
\end{figure}
\vspace*{-15pt}
\begin{eqnarray}
{}^1 \tau^{(1)}_{\beta,\alpha,\mu\nu}(p_2,p_1, m)\hspace*{-7.5pt} &= \hspace*{-7.5pt} & - {i \kappa \over 2} \hspace*{1pt} \eta_{\mu \nu}
\Big[(p_1 \cdot p_2 - m^2) \eta_{\alpha \beta} - p_{1 \beta} p_{2 \alpha} \Big] \nonumber \\
& + \hspace*{-7.5pt} & i \kappa \hspace*{1pt} I_{\mu \nu, \kappa \lambda} \bigg[(p_1 \cdot p_2 - m^2) {I_{\alpha \beta,}}^{\kappa \lambda}
+ \frac{1}{2} \big(p_1^\kappa p_2^\lambda + p_1^\lambda p_2^\kappa\big) \eta_{\alpha \beta} \nonumber \\
&& \hspace*{39pt} - \big(p_1^\kappa p_{2 \alpha} \delta_\beta^\lambda + p_2^\kappa p_{1 \beta} \delta_\alpha^\lambda \big)\bigg] \nonumber \\
{}^1 \tau^{(2)}_{\beta,\alpha,\mu\nu, \rho \sigma}(p_2,p_1, m)\hspace*{-7.5pt} &= \hspace*{-7.5pt} & {i \kappa^2 \over 2} \hspace*{1pt} P_{\mu \nu, \rho \sigma}
\Big[(p_1 \cdot p_2 - m^2) \eta_{\alpha \beta} - p_{1 \beta} p_{2 \alpha} \Big] \nonumber \\
& - \hspace*{-7.5pt} & i \kappa^2 \hspace*{-2.6pt} \left( \hspace*{-1.8pt} {I_{\mu \nu,}}^{\kappa \delta} {I_{\rho \sigma, \delta}}^{\lambda}
\hspace*{-1.9pt} + \hspace*{-1.9pt} {I_{\rho \sigma,}}^{\kappa \delta} {I_{\mu \nu, \delta}}^{\lambda} \hspace*{-1pt} - \hspace*{-1pt} \frac{1}{2} \Big(\eta_{\mu \nu} {I_{\rho \sigma,}}^{\kappa \lambda} \hspace*{-1pt} + \hspace*{-1pt} \eta_{\rho \sigma} {I_{\mu \nu,}}^{\kappa \lambda}\Big) \! \hspace*{-1pt} \right)
\nonumber \\
&& \hspace*{30pt} \times \bigg[(p_1 \cdot p_2 - m^2) {I_{\alpha \beta,}}^{\kappa \lambda}
 + \frac{1}{2} \big(p_1^\kappa p_2^\lambda + p_1^\lambda p_2^\kappa\big) \eta_{\alpha \beta} \nonumber \\
&& \hspace*{39pt} - \big(p_1^\kappa p_{2 \alpha} \delta_\beta^\lambda + p_2^\kappa p_{1 \beta} \delta_\alpha^\lambda \big)\bigg] \nonumber \\
& - \hspace*{-7.5pt} & \frac{i \kappa^2 }{2} \hspace*{-1pt} \left({I_{\mu \nu,}}^{\eta \theta} {I_{\rho \sigma,}}^{\kappa \lambda} + {I_{\rho \sigma,}}^{\eta \theta} {I_{\mu \nu,}}^{\kappa \lambda} \right) \nonumber \\
&& \hspace*{16pt} \times \hspace*{-0.5pt} \bigg[p_{1 \eta} \eta_{\alpha \kappa} (p_{2 \theta} \eta_{\beta \lambda} \hspace*{-2.4pt} - \hspace*{-2pt} p_{2 \lambda} \eta_{\beta \theta}\hspace*{-1.3pt} ) \hspace*{-1.7pt}
+ \hspace*{-1.7pt} p_{2 \eta} \eta_{\beta \kappa} (p_{1 \theta} \eta_{\alpha \lambda} \hspace*{-2.4pt} - \hspace*{-2pt} p_{1 \lambda} \eta_{\alpha \theta}\hspace*{-1.3pt} ) \hspace*{-1pt} \bigg]. \nonumber \\
\end{eqnarray}

If we take the incoming spin-1 particle to have polarization
vector $\epsilon_i^b$ satisfying $\epsilon_i^b\cdot p_3=0$ and the
outgoing particle to have polarization $\epsilon_f^b$ satisfying
$\epsilon_f^b\cdot p_4=0$, then the one-graviton exchange
amplitude can then be written as
\begin{eqnarray}
{}^1{\cal M}^{(1)}(q)&=& -{8\pi G\over
\sqrt{2E_12E_22E_32E_4}} \nonumber \\
&&\times
\Bigg[- \epsilon_f^{b*} \hspace*{-0.2pt} \cdot \hspace*{-0.2pt} \epsilon_i^b \hspace*{-0.3pt} \left({(s-m_a^2-m_b^2+{1\over
2} q^2)^2 - 2m_a^2m_b^2 + m_a^2 q^2 - \frac{1}{4 }q^4 \over q^2}\right) \nonumber\\
&& \hspace*{17.8pt} - \left(\epsilon_f^{b*}\cdot q\epsilon_i^b\cdot p_1-\epsilon_f^{b*}\cdot p_1\epsilon_i^b\cdot q \right)
{2(s-m_a^2-m_b^2+{1\over 2}q^2)\over q^2} \nonumber\\
&& \hspace*{17.8pt} - \left(\epsilon_f^{b*}\cdot q\epsilon_i^b\cdot p_1 +\epsilon_f^{b*}\cdot p_1\epsilon_i^b\cdot q\right) \nonumber \\
&& \hspace*{17.8pt} + \hspace*{2.5pt} 2 \, \epsilon_f^{b*}\cdot q\epsilon_i^b\cdot q \, {m_a^2\over q^2}
+ 2 \, \epsilon_f^{b*}\cdot p_1\epsilon_i^b\cdot p_1\Bigg]\nonumber \\
& \simeq & - \frac{4 \pi G m_a m_b}{q^2}\Bigg[
- \epsilon_f^{b*}\cdot\epsilon_i^b
- \frac{2}{m_a m_b} \left(\epsilon_f^{b*}\cdot q\epsilon_i^b\cdot p_1-\epsilon_f^{b*}\cdot p_1\epsilon_i^b\cdot q \right) \nonumber \\
&&\hspace*{72pt} + \hspace*{0.5pt} \frac{1}{m_b^2} \hspace*{1pt} \epsilon_f^{b*}\cdot q\epsilon_i^b\cdot q \Bigg] \quad
\end{eqnarray}
where the first expression is exact while the second expression
contains only the nonanalytic part in the threshold limit $s
\rightarrow s_0 = (m_a + m_b)^2$. Now we rewrite this expression
using the identity
\begin{equation}
 \epsilon_{f\mu}^{b*} \, \epsilon_i^b\cdot q - \epsilon_{i\mu}^b \, \epsilon_f^{b*}\cdot q
 =\frac{1}{1-\frac{q^2}{4 m_b^2}}\left[\frac{i}{m_b} \hspace*{1pt} \epsilon_{\mu\beta\gamma\delta}
 \hspace*{1pt} p_3^\beta q^\gamma S_b^\delta  - \frac{(p_3+p_4)_\mu}{2 m_b^2} \hspace*{1pt}
 \epsilon_f^{b*}\cdot q \, \epsilon_i^b \cdot q\right]\label{eq:kj}
\end{equation}
where we have defined the spin vector
\begin{equation}
S_{b\mu}={i\over 2m_b} \hspace*{1pt}
\epsilon_{\mu\beta\gamma\delta} \hspace*{1pt}
\epsilon_f^{b*\beta}\epsilon_i^{b\gamma}(p_3+p_4)^\delta
\end{equation}
The leading one-graviton exchange amplitude can then be written as
\begin{equation}
{}^1{\cal M}^{(1)}(q)\simeq -{4 \pi G m_a m_b \over q^2}
\left[- \epsilon_f^{b*}\cdot\epsilon_i^b
+ {2i\over m_am_b^2}\epsilon_{\alpha\beta\gamma\delta}p_1^\alpha p_3^\beta q^\gamma S_b^\delta
- {1\over m_b^2}\epsilon_f^{b*}\cdot
q\epsilon_i^b\cdot q \right]
\end{equation}
In the nonrelativistic limit we have
\begin{equation}
\epsilon_i^{b0}\simeq {1\over m_b} \hspace*{1pt}
\hat{\epsilon}_i^b\cdot\vec{p_3}, \quad \quad \quad \epsilon_f^{b0}\simeq {1\over
m_b} \hspace*{1pt} \hat{\epsilon}_f^b\cdot\vec{p_4}
\end{equation}
so that
\begin{eqnarray}
\epsilon_f^{b*}\cdot\epsilon_i^b&\simeq&
-\hat{\epsilon}_f^{b*}\cdot\hat{\epsilon}_i^b+{1\over
m_b^2} \hspace*{1pt} \hat{\epsilon}_f^{b*}\cdot\vec{p}_4 \hspace*{1pt} \hat{\epsilon}_i^b\cdot\vec{p}_3\nonumber\\
&\simeq&-\hat{\epsilon}_f^{b*}\cdot\hat{\epsilon}_i^b+{1\over
2m_b^2}\hat{\epsilon}_f^{b*}\times\hat{\epsilon}_i^b\cdot\vec{p}_4\times\vec{p}_3\nonumber\\
&&+{1\over
2m_b^2}\left(\hat{\epsilon}_f^{b*}\cdot\vec{p}_4\hat{\epsilon}_i^b\cdot\vec{p}_3
+\hat{\epsilon}_f^{b*}\cdot\vec{p}_3\hat{\epsilon}_i^b\cdot\vec{p}_4\right)
\label{eq:ss}
\end{eqnarray}
Since
\begin{equation}
-i\hat{\epsilon}_f^{b*}\times\hat{\epsilon}_i^b= \left<1,m_f \left|
\vec{S}_b \right| 1,m_i\right>,
\end{equation}
Eq. (\ref{eq:ss}) becomes
\begin{eqnarray}
\epsilon_f^{b*}\cdot\epsilon_i^b \hspace*{-0.2pt}{}& \simeq&
-\hat{\epsilon}_f^{b*}\cdot\hat{\epsilon}_i^b-{i\over 2m_b^2}
\hspace*{1pt} \vec{S}_b\cdot\vec{p}_3\times\vec{p}_4 +{1\over
2m_b^2} \! \hspace*{-0.2pt}
\left(\hat{\epsilon}_f^{b*}\cdot\vec{p}_4 \hspace*{1pt}
\hat{\epsilon}_i^b\cdot\vec{p}_3
+\hat{\epsilon}_f^{b*}\cdot\vec{p}_3 \hspace*{1pt} \hat{\epsilon}_i^b\cdot\vec{p}_4\right) \nonumber \\
& \simeq& -\hat{\epsilon}_f^{b*}\cdot\hat{\epsilon}_i^b +{1\over
m_b^2} \hspace*{1pt} \hat{\epsilon}_f^{b*}\cdot\vec{p} \
\hat{\epsilon}_i^b\cdot\vec{p} +{i \over 2m_b^2} \hspace*{1pt}
\vec{S}_b\cdot\vec{p} \times\vec{q} -{1\over 4 m_b^2} \hspace*{1pt}
\hat{\epsilon}_f^{b*}\cdot\vec{q} \ \hat{\epsilon}_i^b\cdot\vec{q}
\nonumber \\
\end{eqnarray}
and the transition amplitude assumes the form
\begin{eqnarray}
{}^1{\cal M}^{(1)}(\vec{q}) \simeq {4\pi Gm_am_b\over
\vec{q}^{\hspace*{1.4pt} 2}}
 \!\!\!\!\!&\bigg[& \!\!\!\!\!
\hat{\epsilon}_f^{b*}\cdot\hat{\epsilon}_i^b -{1\over m_b^2}
\hspace*{1pt} \hat{\epsilon}_f^{b*}\cdot\vec{p} \
\hat{\epsilon}_i^b\cdot\vec{p}
+{i(3 m_a + 4m_b)\over 2m_am_b^2}\hspace*{1pt} \vec{S}_b\cdot\vec{p} \times\vec{q} \nonumber\\
&-& \!\!\!\! {3\over 4 m_b^2}\hspace*{1pt}
\hat{\epsilon}_f^{b*}\cdot \vec{q}\ \hat{\epsilon}_i^b\cdot
\vec{q}\bigg]
\end{eqnarray}
The spin-independent and spin-orbit terms here are identical in
form to those found in the spin-0 -- spin-1/2 case but now are
accompanied by new terms which are quadrupole in nature, as can be
seen from the identity
\begin{eqnarray}
T^b_{cd}&\equiv&{1\over 2}\left(\hat{\epsilon}_{fc}^{b*}
\hspace*{1pt}\hat{\epsilon}_{id}^b+ \hat{\epsilon}_{ic}^b
\hspace*{1pt}\hat{\epsilon}_{fd}^{b*}\right)-{1\over
3} \hspace*{1pt} \delta_{cd} \hspace*{1pt} \hat{\epsilon}_f^{b*}\cdot\hat{\epsilon}_i^b\nonumber\\
&=& - \left<1,m_f \left|{1\over 2} (S_cS_d+S_dS_c)-{2\over
3}\delta_{cd} \right|1,m_i \right>
\end{eqnarray}
The corresponding lowest order potential is then
\begin{eqnarray}
{}^1V^{(1)}_C(\vec r) &=&-\int {d^3q\over (2\pi)^3} \, \hspace*{1pt}
{}^1{\cal M}^{(1)}(\vec{q}) \, e^{-i\vec{q}\cdot\vec{r}} \nonumber\\
& \simeq &- {G m_a m_b \over
r}\left(\hat{\epsilon}_f^{b*}\cdot\hat{\epsilon}_i^b-{1\over
m_b^2}\hat{\epsilon}_f^{b*}\cdot\vec{p} \
\hat{\epsilon}_i^b\cdot\vec{p}\right) \nonumber \\
&& - {3 m_a + 4 m_b\over 2m_am_b^2}\, \vec{S}_b\cdot\vec{p} \times\vec{\nabla}\left(- {G m_a m_b \over r}\right)\nonumber\\
&&+ \, {3 \over 4 m_b^2} \hspace*{1pt} \hat{\epsilon}_f^{b*}\cdot\vec{\nabla} \ \hat{\epsilon}_i^b\cdot\vec{\nabla} \, \left(- {G m_a m_b \over r}\right) \nonumber\\
&\simeq&- \, {G m_a m_b \over r}
\left(\hat{\epsilon}_f^{b*}\cdot\hat{\epsilon}_i^b-{1\over m_b^2}\
\vec p : T^b : \vec p\right)
 + {G \over r^3}{3 m_a + 4 m_b\over 2 m_b}\vec{L}\cdot\vec{S}_b \nonumber \\
&& - \, \frac{G}{r^5} \frac{9 m_a}{4 m_b} \ \vec r : T^b : \vec r
\end{eqnarray}
where we have defined
$$\vec{w}:T^b:\vec{s}\equiv w_cT^b_{cd}s_d$$ and
which agrees precisely with its spin-1/2 analog in Eq.
(\ref{eq:oj}) up to quadrupole and tensor corrections.

The calculation of the one loop corrections proceeds as before,
but with increased complexity due to the unit spin. Evaluating
the diagrams in Fig. \ref{fig_loops}, we find then
\begin{eqnarray}
{}^1{\cal M}^{(2)}_{\ref{fig_loops}a}(q) \hspace*{-5pt} &=\hspace*{-5pt} & G^2 m_a m_b \Bigg[28 L \epsilon_f^{*b} \cdot \epsilon_i^b
- \frac{16 L}{m_a^2} \epsilon_f^{b*}\cdot p_1 \epsilon_i^b \cdot p_1 \nonumber\\
&&\hspace*{41pt} + \frac{16 L}{m_a m_b} (\epsilon_f^{b*}\cdot q \epsilon_i^b \cdot p_1 - \epsilon_f^{b*} \cdot p_1 \epsilon_i^b\cdot q) \nonumber\\
&&\hspace*{41pt} + \frac{8 L}{m_a^2} (\epsilon_f^{b*}\cdot q \epsilon_i^b \cdot p_1 + \epsilon_f^{b*} \cdot p_1 \epsilon_i^b\cdot q) \nonumber\\
&&\hspace*{41pt} + \frac{4 L}{m_b^2} \epsilon_f^{b*}\cdot q \epsilon_i^b \cdot q \Bigg] \nonumber\\
{}^1{\cal M}^{(2)}_{\ref{fig_loops}b}(q) \hspace*{-5pt} &=\hspace*{-5pt} & G^2 m_a m_b \Bigg[(- 4 m_a S - 16 L) \epsilon_f^{*b} \cdot \epsilon_i^b
+ \frac{8 m_a S + 16 L}{m_a^2} \epsilon_f^{b*}\cdot p_1 \epsilon_i^b \cdot p_1  \nonumber\\
&&\hspace*{41pt} - \frac{8 L}{m_a m_b} (\epsilon_f^{b*}\cdot q \epsilon_i^b \cdot p_1 - \epsilon_f^{b*} \cdot p_1 \epsilon_i^b\cdot q) \nonumber\\
&&\hspace*{41pt} - \frac{4 m_a S + 8 L}{m_a^2} (\epsilon_f^{b*}\cdot q \epsilon_i^b \cdot p_1 + \epsilon_f^{b*} \cdot p_1 \epsilon_i^b\cdot q) \nonumber\\
&&\hspace*{41pt} - \left(\left(2 m_a - \frac{3 m_b^2}{2 m_a}\right)S + \left(2 - \frac{2 m_b^2}{m_a^2}\right)L \right) \frac{1}{m_b^2} \epsilon_f^{b*}\cdot q \epsilon_i^b \cdot q \Bigg] \nonumber\\
{}^1{\cal M}^{(2)}_{\ref{fig_loops}c}(q) \hspace*{-5pt} &=\hspace*{-5pt} & G^2 m_a m_b \Bigg[(- 8 m_b S - 12 L) \epsilon_f^{*b} \cdot \epsilon_i^b
+ \frac{16 L}{m_a^2} \epsilon_f^{b*}\cdot p_1 \epsilon_i^b \cdot p_1  \nonumber\\
&&\hspace*{41pt} - \frac{4 m_b S + 8 L}{m_a m_b} (\epsilon_f^{b*}\cdot q \epsilon_i^b \cdot p_1 - \epsilon_f^{b*} \cdot p_1 \epsilon_i^b\cdot q) \nonumber\\
&&\hspace*{41pt} - \frac{8 L}{m_a^2} (\epsilon_f^{b*}\cdot q \epsilon_i^b \cdot p_1 + \epsilon_f^{b*} \cdot p_1 \epsilon_i^b\cdot q) \nonumber\\
&&\hspace*{41pt} - \frac{m_b S + 8 L}{m_b^2} \epsilon_f^{b*}\cdot q \epsilon_i^b \cdot q \Bigg] \nonumber\\
{}^1{\cal M}^{(2)}_{\ref{fig_loops}d}(q) \hspace*{-5pt} &=\hspace*{-5pt} & G^2 m_a m_b \Bigg[
\frac{4 m_a m_b} {q^2} L \bigg( \hspace*{-2.5pt} - \hspace*{-0.5pt} \epsilon_f^{*b} \hspace*{-1pt} \cdot \hspace*{-0.5pt} \epsilon_i^b  \hspace*{-0.5pt}
- \frac{2}{m_a m_b} (\epsilon_f^{b*} \hspace*{-1pt} \cdot \hspace*{-0.5pt} q \epsilon_i^b \hspace*{-1pt} \cdot \hspace*{-0.5pt} p_1 \hspace*{-1pt} - \epsilon_f^{b*} \hspace*{-1pt} \cdot \hspace*{-0.5pt} p_1 \epsilon_i^b\cdot \hspace*{-0.5pt} q) \nonumber \\
&& \hspace*{100pt} + \frac{1}{m_b^2} \epsilon_f^{b*}\cdot q \epsilon_i^b \cdot q\bigg) \nonumber \\
&&\hspace*{41pt} + \frac{S}{s-s_0} \bigg( \hspace*{-2.5pt} -(3m_a + 4 m_b) (\epsilon_f^{b*}\cdot q \epsilon_i^b \cdot p_1 - \epsilon_f^{b*} \cdot p_1 \epsilon_i^b\cdot q)\nonumber\\
&&\hspace*{88pt} + \frac{m_a (5 m_a + 7 m_b)}{2 m_b} \epsilon_f^{b*}\cdot q \epsilon_i^b \cdot q \bigg)\nonumber\\
&&\hspace*{41pt} - \bigg((6 m_a + 4 m_b) S + \frac{12 m_a^2 - 6 m_a m_b + 24 m_b^2}{3 m_a m_b} L \bigg) \epsilon_f^{*b} \cdot \epsilon_i^b \nonumber\\
&& \hspace*{41pt} + \bigg(- 4 m_a S + \frac{22 m_a - 24 m_b}{3 m_b} L \bigg) \frac{1}{m_a^2} \epsilon_f^{b*}\cdot p_1 \epsilon_i^b \cdot p_1 \nonumber \\
&& \hspace*{41pt} - \bigg(13 (m_a + m_b) S + \frac{25 m_a^2 + 7 m_a m_b + 22 m_b^2}{3 m_a m_b} L \bigg) \nonumber \\
&& \hspace*{59pt} \times \frac{1}{m_a m_b} (\epsilon_f^{b*}\cdot q \epsilon_i^b \cdot p_1 - \epsilon_f^{b*} \cdot p_1 \epsilon_i^b\cdot q) \nonumber \\
&& \hspace*{41pt} + \bigg(\hspace*{-1pt} 2 m_a S \hspace*{-1.5pt} - \hspace*{-0.8pt} \frac{11 m_a \hspace*{-1.9pt} - \hspace*{-1.4pt} 12 m_b}{3 m_b} L \hspace*{-1pt} \bigg) \frac{1}{m_a^2} (\epsilon_f^{b*} \hspace*{-1.5pt} \cdot \hspace*{-0.8pt} q \epsilon_i^b \hspace*{-1.5pt} \cdot \hspace*{-0.8pt} p_1 \hspace*{-1pt} + \hspace*{-0.7pt} \epsilon_f^{b*} \hspace*{-1.5pt} \cdot \hspace*{-0.8pt} p_1 \epsilon_i^b \hspace*{-1.5pt} \cdot \hspace*{-0.8pt} q)\nonumber \\
&& \hspace*{41pt} + \bigg(\!\bigg( \hspace*{-0.5pt} 7 m_a \hspace*{-1pt} + \hspace*{-1pt} 9 m_b \hspace*{-1pt} - \hspace*{-1pt} \frac{3 m_b^2}{4 m_a}\bigg)S \hspace*{-0.5pt}  + \hspace*{-0.5pt}  \bigg( \hspace*{-0.5pt} 5 \frac{m_a}{m_b} \hspace*{-1pt} + \hspace*{-1pt} 2 \hspace*{-1pt} + \hspace*{-1pt} 9 \frac{m_b}{m_a} \hspace*{-1pt} - \hspace*{-1pt} \frac{m_b^2}{m_a^2}\bigg) L\bigg) \nonumber \\
&& \hspace*{59pt} \times \frac{1}{m_b^2} \epsilon_f^{b*}\cdot q \epsilon_i^b \cdot q  \Bigg] \nonumber \\
&- \hspace*{-5pt} & i 4\pi G^2 m_a^2 m_b^2 \hspace*{1pt} {L\over q^2}\sqrt{m_a m_b\over s-s_0} \hspace*{-1pt} \Bigg( \hspace*{-5pt}
- \hspace*{-1.5pt} \epsilon_f^{*b} \hspace*{-2.5pt} \cdot \hspace*{-1.5pt} \epsilon_i^b \hspace*{-1.5pt}
- \hspace*{-1.5pt} \frac{2}{m_a m_b} (\epsilon_f^{b*} \hspace*{-2.5pt} \cdot \hspace*{-1.5pt} q \epsilon_i^b \hspace*{-2.5pt} \cdot \hspace*{-1.5pt} p_1 \hspace*{-1.5pt} - \hspace*{-1pt} \epsilon_f^{b*} \hspace*{-2.5pt} \cdot \hspace*{-1.5pt} p_1 \epsilon_i^b \hspace*{-2.5pt} \cdot \hspace*{-1.5pt} q) \nonumber \\
&& \hspace*{122pt} + \frac{1}{m_b^2} \epsilon_f^{b*} \hspace*{-1pt} \cdot \hspace*{-0.5pt} q \epsilon_i^b \hspace*{-1pt} \cdot \hspace*{-0.5pt} q \Bigg) \nonumber \\
{}^1{\cal M}^{(2)}_{\ref{fig_loops}e}(q) \hspace*{-5pt} &=\hspace*{-5pt} & G^2 m_a m_b \Bigg[
\frac{4 m_a m_b} {q^2} L \bigg( \epsilon_f^{*b} \cdot \epsilon_i^b
+ \frac{2}{m_a m_b} (\epsilon_f^{b*} \cdot q \epsilon_i^b \cdot p_1 - \epsilon_f^{b*} \cdot p_1 \epsilon_i^b \cdot q) \nonumber \\
&& \hspace*{100pt} - \frac{1}{m_b^2} \epsilon_f^{b*}\cdot q \epsilon_i^b \cdot q\bigg) \nonumber \\
&&\hspace*{41pt} + \bigg((2 m_a + 4 m_b) S + \frac{12 m_a^2 + 52 m_a m_b + 24 m_b^2}{3 m_a m_b} L \bigg) \epsilon_f^{*b} \cdot \epsilon_i^b \nonumber\\
&& \hspace*{41pt} - \bigg(4 m_a S + \frac{22 m_a + 24 m_b}{3 m_b} L \bigg) \frac{1}{m_a^2} \epsilon_f^{b*}\cdot p_1 \epsilon_i^b \cdot p_1 \nonumber \\
&& \hspace*{41pt} + \bigg(\! \bigg(\frac{17}{4}m_a + 4 m_b\bigg) S + \frac{25 m_a^2 + 49 m_a m_b + 22 m_b^2}{3 m_a m_b} L \bigg) \nonumber \\
&& \hspace*{59pt} \times \frac{1}{m_a m_b} (\epsilon_f^{b*}\cdot q \epsilon_i^b \cdot p_1 - \epsilon_f^{b*} \cdot p_1 \epsilon_i^b\cdot q) \nonumber \\
&& \hspace*{41pt} + \bigg(\hspace*{-1pt} 2 m_a S \hspace*{-1.5pt} + \hspace*{-0.8pt} \frac{11 m_a \hspace*{-1.9pt} + \hspace*{-1.4pt} 12 m_b}{3 m_b} L \hspace*{-1pt} \bigg) \frac{1}{m_a^2} (\epsilon_f^{b*} \hspace*{-1.5pt} \cdot \hspace*{-0.8pt} q \epsilon_i^b \hspace*{-1.5pt} \cdot \hspace*{-0.8pt} p_1 \hspace*{-1pt} + \hspace*{-0.7pt} \epsilon_f^{b*} \hspace*{-1.5pt} \cdot \hspace*{-0.8pt} p_1 \epsilon_i^b \hspace*{-1.5pt} \cdot \hspace*{-0.8pt} q)\nonumber \\
&& \hspace*{41pt} - \bigg(\! \hspace*{-1pt} \bigg( \hspace*{-0.8pt} \frac{13}{8} m_a \hspace*{-2.5pt} + \hspace*{-2pt} \frac{31}{8} m_b \hspace*{-2pt} + \hspace*{-2pt} \frac{3 m_b^2}{4 m_a} \hspace*{-0.8pt}\bigg) \hspace*{-0.8pt}S \hspace*{-2.2pt}
+ \hspace*{-2.5pt}  \bigg( \hspace*{-2pt} 5 \frac{m_a}{m_b} \hspace*{-2pt} + \hspace*{-2pt} \frac{16}{3} \hspace*{-2pt} + \hspace*{-2pt} 9 \frac{m_b}{m_a} \hspace*{-2pt} + \hspace*{-2pt} \frac{m_b^2}{m_a^2} \hspace*{-0.5pt}\bigg) \hspace*{-0.5pt}L \hspace*{-1.5pt}\bigg) \nonumber \\
&& \hspace*{59pt} \times \frac{1}{m_b^2} \epsilon_f^{b*}\cdot q \epsilon_i^b \cdot q  \Bigg] \nonumber \\
{}^1{\cal M}^{(2)}_{\ref{fig_loops}f}(q) \hspace*{-5pt} &=\hspace*{-5pt} & G^2 m_a m_b \Bigg[(2(m_a + m_b) S - 14 L) \epsilon_f^{*b} \cdot \epsilon_i^b \nonumber\\
&& \hspace*{41pt} + \bigg(\!\bigg(\frac{1}{2} m_a \hspace*{-2pt} + \hspace*{-1pt} 2 m_b\hspace*{-2pt}\bigg)  S \hspace*{-1.5pt}-\hspace*{-1.5pt} 4 L \hspace*{-2pt}\bigg)
\frac{1}{m_a m_b} (\epsilon_f^{b*} \hspace*{-2pt} \cdot \hspace*{-1pt} q \epsilon_i^b \hspace*{-1pt} \cdot \hspace*{-1pt} p_1 \hspace*{-1.5pt} - \hspace*{-1pt} \epsilon_f^{b*} \hspace*{-2pt} \cdot \hspace*{-1pt} p_1 \epsilon_i^b \hspace*{-1pt}\cdot \hspace*{-1pt} q) \nonumber\\
&& \hspace*{41pt} + \bigg( \! \bigg( \hspace*{-1.5pt} - \frac{1}{4} m_a + \frac{1}{2} m_b \bigg) S + 4 L \bigg) \frac{1}{m_b^2} \epsilon_f^{b*}\cdot q \epsilon_i^b \cdot q \Bigg] \nonumber\\
{}^1{\cal M}^{(2)}_{\ref{fig_loops}g}(q) \hspace*{-5pt} &=\hspace*{-5pt} & G^2 m_a m_b \Bigg[\frac{43 L}{15} \epsilon_f^{*b} \cdot \epsilon_i^b
+ \frac{14 L}{5 m_a m_b} (\epsilon_f^{b*} \cdot q \epsilon_i^b \cdot p_1 - \epsilon_f^{b*} \cdot p_1 \epsilon_i^b \cdot q) \nonumber\\
&&\hspace*{41pt} - \frac{7 L}{5 m_b^2} \epsilon_f^{b*}\cdot q \epsilon_i^b \cdot q \Bigg]
\end{eqnarray}
Combining, we find the full one loop amplitude
\begin{eqnarray}
{}^1{\cal M}_{tot}^{(2)}(q) \hspace*{-5pt} &=\hspace*{-5pt} & G^2 m_a m_b \Bigg[
\frac{S}{s-s_0} \bigg( \hspace*{-3pt} -(3m_a + 4 m_b) (\epsilon_f^{b*}\cdot q \epsilon_i^b \cdot p_1 - \epsilon_f^{b*} \cdot p_1 \epsilon_i^b\cdot q)\nonumber\\
&& \hspace*{88pt} + \frac{m_a (5 m_a + 7 m_b)}{2 m_b} \epsilon_f^{b*}\cdot q \epsilon_i^b \cdot q \bigg)\nonumber\\
&& \hspace*{41pt} + \bigg(6 (m_a + m_b) S - \frac{41}{5} L\bigg) \hspace*{-1.5pt} \left(- \epsilon_f^{b*}\cdot \epsilon_i^b\right) \nonumber\\
&& \hspace*{41pt} + \bigg(\hspace*{-3pt} - \frac{11(3 m_a + 4 m_b)}{4} S + \frac{64}{5}L\bigg) \nonumber \\
&& \hspace*{59pt} \times \frac{1}{m_a m_b} (\epsilon_f^{b*}\cdot q \epsilon_i^b \cdot p_1 - \epsilon_f^{b*} \cdot p_1 \epsilon_i^b\cdot q)\nonumber\\
&& \hspace*{41pt} + \bigg(\frac{25 m_a + 37 m_b}{8} S - \frac{101}{15} L \bigg) \frac{1}{m_b^2} \epsilon_f^{b*}\cdot q \epsilon_i^b \cdot q\Bigg]
\end{eqnarray}
which, using the identity Eq. (\ref{eq:kj}), becomes
\begin{eqnarray}
{}^{1}{\cal M}_{tot}^{(2)}(q) \hspace*{-5pt} &=\hspace*{-5pt}&G^2 m_a m_b \Bigg[
- \epsilon_f^{b*}\cdot \epsilon_i^b \bigg(6 (m_a + m_b) S - \frac{41}{5} L\bigg) \nonumber\\
&& \hspace*{41pt} + \hspace*{0.5pt} \frac{i}{m_a m_b^2} \hspace*{1pt} \epsilon_{\alpha\beta\gamma\delta} \hspace*{1pt} p_1^\alpha p_3^\beta q^\gamma S_b^\delta \,
 \bigg(\frac{11(3 m_a + 4 m_b)}{4} S - \frac{64}{5} L\bigg) \nonumber\\
&& \hspace*{41pt} + \hspace*{0.5pt} \frac{1}{m_b^2} \hspace*{1pt} \epsilon_f^{b*}\cdot q \epsilon_i^b \cdot q \hspace*{1pt}
 \bigg(\hspace*{-3pt}- \frac{53 m_a + 67 m_b}{8} S + \frac{91}{15} L\bigg) \nonumber\\
&& \hspace*{41pt}+ \hspace*{0.5pt} {i (3 m_a + 4 m_b) S\over
m_b (s-s_0)} \hspace*{1pt} \epsilon_{\alpha\beta\gamma\delta}\hspace*{1pt}
p_1^\alpha p_3^\beta q^\gamma S_b^\delta \nonumber\\
&& \hspace*{41pt} - \hspace*{0.5pt} \frac{m_a (m_a + m_b) S}{2 m_b (s - s_0)} \epsilon_f^{b*}\cdot q \epsilon_i^b \cdot q \Bigg]\nonumber\\
&- \hspace*{-5pt} & i 4\pi G^2 m_a^2 m_b^2 \hspace*{1pt} {L\over q^2}\sqrt{m_a m_b\over s-s_0} \Bigg( \hspace*{-3pt}
- \epsilon_f^{*b} \cdot \epsilon_i^b + \frac{2i}{m_a m_b}  \hspace*{1pt} \epsilon_{\alpha\beta\gamma\delta}\hspace*{1pt} p_1^\alpha p_3^\beta q^\gamma S_b^\delta \nonumber \\
&& \hspace*{125pt} - \hspace*{0.5pt} \frac{1}{m_b^2} \hspace*{1pt} \epsilon_f^{b*} \cdot q \epsilon_i^b \cdot q \Bigg)
\label{eq:M2spin1kir}
\end{eqnarray}
Notice here that without the $\epsilon_f^{b*}\cdot
q\epsilon_i^b\cdot q$ terms, Eq. (\ref{eq:M2spin1kir}) has an
identical structure to that of the case of spin-0 -- spin-1/2
scattering---Eq. (\ref{eq:ohla})---provided we substitute
$\bar{u}(p_4)u(p_3)\longrightarrow -
\epsilon_f^{b*}\cdot\epsilon_i^b$.

Finally, taking the nonrelativistic limit we find
\begin{eqnarray}
{}^{1}{\cal M}_{tot}^{(2)}(\vec q)\hspace*{-6pt} &\simeq\hspace*{-6pt}&\left[G^2 m_a m_b
\left( 6 (m_a + m_b) S - {41 \over 5}L\right) - i 4 \pi G^2 m_a^2 m_b^2 \hspace*{1pt} \frac{L}{q^2} \frac{m_r}{p_0}\right] \nonumber \\
&& \hspace*{15pt} \times
\left(\hat{\epsilon}_f^{b*} \cdot \hat{\epsilon}_i^b - {1\over m_b^2}\, \vec p \hspace*{-0.5pt}: \hspace*{-1pt} T^b \hspace*{-2pt}: \hspace*{-1pt}\vec p\right) \nonumber\\
&+\hspace*{-6pt}&\Bigg[G^2  \hspace*{-1pt} \left(\frac{12m_a^3 \hspace*{-1pt} + \hspace*{-1pt} 45 m_a^2 m_b \hspace*{-1pt} + \hspace*{-1pt} 56 m_a m_b^2
 \hspace*{-1pt} + \hspace*{-1pt} 24 m_b^3}{2(m_a \hspace*{-1pt} + \hspace*{-1pt} m_b)}  \hspace*{1pt} S
- \frac{87 m_a \hspace*{-1pt} + \hspace*{-1pt} 128 m_b}{10}  \hspace*{1pt} L\right)\nonumber\\
&&+ \frac{G^2 m_a^2 m_b^2 (3 m_a + 4 m_b)}{(m_a + m_b)} \left(- i \frac{2 \pi
L}{p_0 q^2} + \frac{S}{p_0^2} \right)\Bigg] {i\over
m_b}\vec{S}_b\cdot\vec{p}\times\vec{q} \nonumber \\
&+\hspace*{-6pt}&\Bigg[G^2 m_a m_b \left(- \frac{21 m_a^2 + 47 m_a m_b + 28 m_b^2}{4 (m_a + m_b)} \, S + \frac{241}{60} \hspace*{1pt} L \right) \nonumber \\
&&+ \frac{G^2 m_a^3 m_b^3}{2(m_a + m_b)} \left(i \frac{6 \pi L}{p_0
q^2} - \frac{S}{p_0^2}\right) \Bigg] \frac{1}{m_b^2} \, \vec q : T^b
: \vec q
\label{eq:yo2}
\end{eqnarray}
As found in the earlier calculations, there exist terms involving
both $i/p_0$ and $1/p_0^2$ which prevent the defining of a simple
second order potential.  The solution now is well
known---subtraction of the iterated first order potential.  Since
the form of the
spin-independent---$\hat{\epsilon}_B^*\cdot\hat{\epsilon}_A$---and
spin-orbit---$\vec{S}_b\cdot\vec{p} \times\vec{q}$---terms is
identical to that found for the case of spin-1/2, it is clear that
the subtraction goes through as before and that the corresponding
pieces of the second order potential have the {\it same} form as
found for spin-1/2.  In addition, there are now two new pieces of
the amplitude, the quadrupole structure $\vec{q}:T^b:\vec{q}$ which
multiplies terms involving both $i/p_0$ and $1/p_0^2$ and the tensor
structure $\vec{p}:T^b:\vec{p}$ multiplying only $i/p_0$. In order
to remove these we must iterate the full first order potential
including these quadrupole and tensor components. However, we find
that our simple nonrelativistic iteration fails to remove them! We
suspect the reason to be the presence of the tensor structure
$\vec{p}:T^b:\vec{p}$ in the lowest order potential which is in some
sense a relativistic correction but which when iterated yields also
quadrupole pieces $\vec{q}:T^b:\vec{q}$. A fully relativistic
iteration should thus be performed which is under study but is
beyond the scope of this paper. It would be interesting to
investigate if the requirement of canceling all $i/p_0$ and
$1/p_0^2$ forms in the quadrupole and tensor pieces could clarify
the ambiguity in the iteration of the leading order potential as
discussed in \cite{js, hrem}.

Since we did not perform the proper iteration of the quadrupole and
tensor pieces we include only the spin independent and spin-orbit
pieces in the resulting second order potential
\begin{eqnarray}
{}^{1}V^{(2)}_G(\vec r)&=&-\int{d^3q\over
(2\pi)^3}e^{-i\vec{q}\cdot\vec{r}}\left[{}^{1}{\cal
M}_{tot}^{(2)}(\vec{q})-{}^{1}{\rm
Amp}_G^{(2)}(\vec{q})\right]\nonumber\\
&=&\left[-{3G^2m_am_b(m_a+m_b)\over r^2} - {41G^2m_am_b\hbar\over 10\pi r^3}\right]
\hat{\epsilon}_f^{b*} \cdot \hat{\epsilon}_i^b \nonumber\\
&+&{1\over m_b}\vec{S}_b\cdot \vec{p}\times\vec{\nabla}\Bigg[
\frac{G^2(12m_a^3 + 45 m_a^2 m_b + 56 m_a m_b^2 + 24 m_b^3)}{4(m_a + m_b)r^2} \nonumber\\
&& \hspace*{66pt} + \frac{G^2(87 m_a + 128 m_b) \hbar}{20 \pi r^3} \Bigg] + {}^{1}V^{(2)}_T(\vec r) \nonumber\\
&=&\left[-{3G^2m_am_b(m_a+m_b)\over r^2} - {41G^2m_am_b\hbar\over 10\pi r^3}\right]
\hat{\epsilon}_f^{b*} \cdot \hat{\epsilon}_i^b \nonumber\\
&+&\Bigg[{G^2(12 m_a^3 + 45 m_a^2 m_b + 56 m_a m_b^2 + 24 m_b^3)\over
2 m_b (m_a+m_b)r^4} \nonumber \\
&&+{3 G^2(87 m_a+ 128m_b)\hbar\over 20 \pi
m_b r^5}\Bigg] \vec{L}\cdot\vec{S}_b + {}^{1}V^{(2)}_T(\vec r) \label{eq:NLOpot01}
\end{eqnarray}
where ${}^{1}V^{(2)}_T(\vec r)$ denotes the tensor pieces not
explicitly shown. Comparison with the corresponding form of
${}^{1\over 2}V^{(1)}(\vec r)$ given in Eq. (\ref{eq:mn}) confirms the
universality which we have suggested---the spin-independent and
spin-orbit terms have identical forms. The next task is to see
whether this universality applies when both scattered particles
carry spin. For this purpose we consider the case of spin-1/2 --
spin-1/2 scattering.

\section{Spin-Dependent Scattering: Spin-Spin Interaction}

\subsection{Spin-1/2 -- Spin-1/2}

In the case of scattering of a pair of spin-1/2 particles, the basic
vertices have already been developed in the spin-0 -- spin-1/2 scattering
section, so we can proceed directly to our calculation. The one-graviton
exchange amplitude at tree level reads
\begin{eqnarray}
{}^{{1\over 2}{1\over 2}}{\cal M}^{(1)}(q) \hspace*{-5pt} & =
\hspace*{-5pt} & - \frac{4 \pi G m_a m_b}{q^2}
\Bigg[\hspace*{-1pt} \frac{s - m_a^2 - m_b^2 + \frac{1}{2} q^2}{2 m_a m_b} \bar u(p_1) \gamma_\alpha u(p_2) \, \bar u(p_3) \gamma^\alpha u(p_4) \nonumber \\
&& \hspace*{60pt} {} + \frac{1}{m_a m_b} \hspace*{1pt} \bar u(p_1) \hspace*{-1pt} \! \not \! p_3  \hspace*{1pt} u(p_2) \, \bar u(p_3) \hspace*{-1pt} \! \not \! p_1  \hspace*{0.5pt} u(p_4) \nonumber \\
&& \hspace*{60pt} {} - \bar u(p_1) u(p_2) \, \bar u(p_3) u(p_4)
\Bigg] \times \sqrt{m_a^2m_b^2\over E_1E_2E_3E_4} \quad
\quad\label{eq:kh}
\end{eqnarray}
and with the spin identities Eq. (\ref{eq:id}) for system $b$ and
\begin{equation}
\bar{u}(p_2)\gamma_\mu u(p_1)=\left({1\over 1-{q^2\over
4m_a^2}}\right)\left[{(p_1+p_2)_\mu\over
2m_a}\bar{u}(p_2)u(p_1)+{i\over
m_a^2}\epsilon_{\mu\beta\gamma\delta}q^\beta p_1^\gamma
S_a^\delta\right]\label{eq:ida}
\end{equation}
for system $a$, Eq. (\ref{eq:kh}) can be written as
\begin{eqnarray}
{}^{{1\over 2}{1\over 2}}{\cal M}^{(1)}(q)\hspace*{-2pt}
\hspace*{-5pt} &=\hspace*{-5pt}&- {4\pi G m_a m_b \over q^2}
\Bigg[\bar{u}(p_2)u(p_1)\bar{u}(p_4)u(p_3) \nonumber \\
&&\hspace*{62pt}+\hspace*{1pt}{2 i\over
m_am_b^2}\bar{u}(p_2)u(p_1)\epsilon_{\alpha\beta\gamma\delta}p_1^\alpha
p_3^\beta q^\gamma S_b^\delta \nonumber\\
&&\hspace*{62pt}+\hspace*{1pt}{2 i\over
m_a^2 m_b}\bar{u}(p_4)u(p_3)\epsilon_{\alpha\beta\gamma\delta}p_1^\alpha
p_3^\beta q^\gamma S_a^\delta \nonumber\\
&&\hspace*{62pt}+{1\over m_am_b}(S_a\cdot qS_b\cdot q-q^2S_b\cdot
S_a)\Bigg].
\end{eqnarray}
In the symmetric center of mass frame we take the nonrelativistic
limit yielding
\begin{eqnarray}
{}^{{1\over 2}{1\over 2}}{\cal M}^{(1)}(\vec{q})&\simeq&{4\pi G m_a
m_b\over \vec{q}^{\hspace*{1.4pt} 2}}\Bigg[\chi_f^{a\dagger}
\chi_i^a \, \chi_f^{b\dagger}\chi_i^b
+{i(3 m_a + 4 m_b)\over 2m_am_b^2}\vec{S}_b\cdot\vec{p} \times\vec{q} \, \chi_f^{a\dagger}\chi_i^a \nonumber\\
&&\hspace*{54pt}+ {i(4 m_a + 3 m_b)\over 2m_a^2m_b}\vec{S}_a\cdot\vec{p} \times\vec{q} \, \chi_f^{b\dagger}\chi_i^b\nonumber\\
&&\hspace*{54pt}+{1\over m_am_b}\left(\vec{S}_a\cdot\vec{q} \,
\vec{S}_b\cdot\vec{q}-\vec{q}^{\hspace*{1.4pt} 2}
\vec{S}_a\cdot\vec{S}_b\right) \hspace*{-1pt}\Bigg]
\label{eq:LOMnonrelhh_gr}
\end{eqnarray}
whereby the lowest order potential for spin-1/2 -- spin-1/2
scattering becomes
\begin{eqnarray}
{}^{{1\over 2}{1\over 2}}V_G^{(1)}(\vec
r)\hspace*{-5pt}&=\hspace*{-5pt}& -\int {d^3q\over (2\pi)^3} \,
\hspace*{1pt}
{}^{{1\over 2}{1\over 2}}{\cal M}^{(1)}(\vec{q}) \, e^{-i\vec{q}\cdot\vec{r}} \nonumber\\
&\simeq \hspace*{-5pt}& - {G m_a m_b \over
r}\chi_f^{a\dagger}\chi_i^a \chi_f^{b\dagger}\chi_i^b \nonumber \\
&& - {(3 m_a + 4 m_b)\over 2m_am_b^2}\vec{S}_b\cdot\vec{p} \times\vec{\nabla}\bigg(\!
\! - {G m_a m_b \over r}\bigg) \, \chi_f^{a\dagger}\chi_i^a\nonumber\\
&& -{(4m_a + 3 m_b)\over 2m_a^2m_b}\vec{S}_a\cdot\vec{p}
\times\vec{\nabla} \bigg(\! \! - {G m_a m_b \over r}\bigg) \, \chi_f^{b\dagger}\chi_i^b \nonumber \\
&& - {1\over m_am_b}\vec{S}_a\cdot\vec{\nabla}\vec{S}_b\cdot\vec{\nabla} \bigg(\! \! - {G m_a m_b \over r}\bigg) \nonumber\\
&\simeq \hspace*{-5pt}& - {G m_a m_b \over r}\chi_f^{a\dagger}\chi_i^a\chi_f^{b\dagger}\chi_i^b
+ \frac{G}{r^3}{(3 m_a + 4 m_b)\over 2 m_b} \vec L \cdot \vec S_b\, \chi_f^{a\dagger}\chi_i^a\nonumber\\
&&+ \frac{G}{r^3} {(4m_a + 3 m_b)\over 2m_a} \vec L \cdot \vec{S}_a
\, \chi_f^{b\dagger}\chi_i^b + \frac{G}{r^5} \hspace*{-1pt} \left(3
\vec{S}_a\cdot\vec{r} \, \vec{S}_b\cdot\vec{r} - r^2 \vec S_a \cdot
\vec S_b \right). \nonumber \\ \label{eq:mj}
\end{eqnarray}
Note that since the piece proportional to $\vec{S}_a\cdot\vec{S}_b$
in Eq. (\ref{eq:LOMnonrelhh_gr}) is analytic in $\vec
q^{\hspace*{1.4pt} 2}$ it yields only  a short distance contribution
which is omitted in the potential Eq. (\ref{eq:mj}).

In this case when we evaluate the loop diagrams in Fig.
\ref{fig_loops}, we notice that part of the spin-spin structure
piece contains the form $q^2 S_a \cdot S_b$ multiplying the
nonanalytic structures $L$ and $S$. Due to the extra factor of $q^2$
in this form, we must expand all loop integrals to {\it one order
higher} in $q^2$ than before in order to be consistent. This has
been done and does make a difference in our results for the
spin-spin interaction piece. The results for the individual
diagrams are then found to be
\begin{eqnarray}
{}^{{1\over 2}{1\over 2}}{\cal
M}_{\ref{fig_loops}a}^{(2)}(q)\hspace*{-5pt}&=&\hspace*{-5pt}G^2
m_a m_b \Bigg[{\mathcal U}_a {\mathcal U}_b \left(- 26L\right)
 +  i \frac{{\mathcal E}_a {\mathcal U}_b}{m_a^2 m_b} \left(- 17 L\right)
 +  i \frac{{\mathcal U}_a {\mathcal E}_b}{m_a m_b^2} \left(- 17 L\right) \nonumber \\
 && \hspace*{37pt}+ \frac{S_a\cdot qS_b\cdot q}{m_a m_b} \left(- 9 L\right)
 - \frac{q^2 S_a \cdot S_b}{m_a m_b} \left(- \frac{15}{2} L\right) \hspace*{-1.5pt}\Bigg]\nonumber\\
{}^{{1\over 2}{1\over 2}}{\cal M}_{\ref{fig_loops}b}^{(2)}(q)\hspace*{-5pt}&=&\hspace*{-5pt} G^2 m_a m_b \Bigg[{\mathcal U}_a {\mathcal U}_b \left(15 L + 7 m_a S\right) \nonumber\\
 && \hspace*{37pt} +  i \frac{{\mathcal E}_a {\mathcal U}_b}{m_a^2 m_b} \left(\frac{21}{2} L + 4 m_a S\right)
 +  i \frac{{\mathcal U}_a {\mathcal E}_b}{m_a m_b^2} \left(13 L + 8 m_a S\right) \nonumber \\
 && \hspace*{37pt} + \frac{S_a\cdot qS_b\cdot q}{m_a m_b} \left(\frac{17}{3} L + \frac{5}{2} m_a S\right) \nonumber\\
 && \hspace*{37pt} - \frac{q^2 S_a \cdot S_b}{m_a m_b} \left(\frac{11}{3} L + \frac{3}{2} m_a S\right) \hspace*{-1.5pt}\Bigg]\nonumber\\
{}^{{1\over 2}{1\over 2}}{\cal M}_{\ref{fig_loops}c}^{(2)}(q)\hspace*{-5pt}&=&\hspace*{-5pt} G^2 m_a m_b \Bigg[{\mathcal U}_a {\mathcal U}_b \left(15 L + 7 m_b S\right) \nonumber\\
 && \hspace*{37pt} +  i \frac{{\mathcal E}_a {\mathcal U}_b}{m_a^2 m_b} \left(13 L + 8 m_b S\right)
 +  i \frac{{\mathcal U}_a {\mathcal E}_b}{m_a m_b^2} \left(\frac{21}{2} L + 4 m_b S\right) \nonumber \\
 && \hspace*{37pt} + \frac{S_a\cdot qS_b\cdot q}{m_a m_b} \left(\frac{17}{3} L + \frac{5}{2} m_b S\right) \nonumber\\
 && \hspace*{37pt} - \frac{q^2 S_a \cdot S_b}{m_a m_b} \left(\frac{11}{3} L + \frac{3}{2} m_b S\right) \hspace*{-1.5pt}\Bigg]\nonumber\\
{}^{{1\over 2}{1\over 2}}{\cal
M}_{\ref{fig_loops}d}^{(2)}(q)\hspace*{-5pt}&=&\hspace*{-5pt} G^2
m_am_b \Bigg[{\mathcal U}_a {\mathcal U}_b
\hspace*{-1.5pt}\Bigg(\hspace*{-1.5pt}L
\hspace*{-1pt}\left(\hspace*{-1.5pt} {4m_am_b\over
q^2}\hspace*{-1pt}+\hspace*{-1pt}\frac{31 m_a^2\hspace*{-1pt} - \hspace*{-1pt}16 m_a m_b\hspace*{-1pt} + \hspace*{-1pt} 31 m_b^2}{4m_am_b}\hspace*{-0.5pt}\right) \nonumber \\
 && \hspace*{68pt} + S \hspace*{2pt} {9\over 2}(m_a + m_b)\hspace*{-1pt} \Bigg) \nonumber\\
&&\hspace*{37pt}+  i \frac{{\mathcal E}_a {\mathcal U}_b}{m_a^2
m_b}\Bigg(\hspace*{-1pt} L \hspace*{-1pt} \left(\hspace*{-1pt}
\frac{8 m_a m_b}{q^2} \hspace*{-1pt}
  + \hspace*{-1pt} \frac{124 m_a^2 - 11 m_a m_b + 177 m_b^2}{12 m_a m_b}\right) \nonumber\\
&& \hspace*{82pt} + S \hspace*{-1pt} \left( \hspace*{-1pt} \frac{m_a
m_b(4m_a \hspace*{-1pt} + \hspace*{-0.5pt} 3 m_b)}{s-s_0}
\hspace*{-1pt}
+ \hspace*{-1pt} (13 m_a \hspace*{-1pt} + \hspace*{-0.5pt} 9 m_b) \hspace*{-1pt}\right) \! \hspace*{-1pt} \Bigg)\nonumber\\
&&\hspace*{37pt}+  i \frac{{\mathcal U}_a {\mathcal E}_b}{m_a m_b^2}\Bigg(\hspace*{-1pt} L \hspace*{-1pt} \left(\hspace*{-1pt} \frac{8 m_a m_b}{q^2}\hspace*{-1pt} +\hspace*{-1pt} \frac{177 m_a^2 - 11 m_a m_b + 124 m_b^2}{12 m_a m_b}\right) \nonumber\\
&& \hspace*{82pt} +S \hspace*{-1pt} \left(\hspace*{-1pt} \frac{m_a m_b(3 m_a \hspace*{-1pt} \hspace*{-1pt} + \hspace*{-0.5pt} 4 m_b)}{s-s_0} \hspace*{-1pt} + \hspace*{-1pt} (9 m_a \hspace*{-1pt} + \hspace*{-0.5pt} 13 m_b) \hspace*{-1pt}\right) \! \hspace*{-1pt} \Bigg)\nonumber\\
&&\hspace*{37pt} + \frac{S_a\cdot qS_b\cdot q}{m_a m_b} \Bigg( \hspace*{-1pt} L \hspace*{-1pt} \left(\frac{4 m_a m_b}{q^2} + \frac{39 m_a^2 \hspace*{-1pt} + \hspace*{-1pt} 25 m_a m_b \hspace*{-1pt} + \hspace*{-1pt} 39 m_b^2}{6 m_a m_b} \right) \nonumber\\
&& \hspace*{99pt} + S \hspace*{-1pt} \left(\frac{m_a m_b (m_a \hspace*{-1pt} + \hspace*{-1pt} m_b)}{s - s_0} \hspace*{-1pt} + \hspace*{-1pt} \frac{39}{4} \left(m_a \hspace*{-1pt} + \hspace*{-1pt} m_b \right) \hspace*{-1pt} \right) \hspace*{-3pt} \Bigg) \nonumber \\
&&\hspace*{37pt} - \frac{q^2 S_a \cdot S_b}{m_a m_b} \Bigg( \hspace*{-1.5pt} L \hspace*{-2pt} \left(\hspace*{-1pt} \frac{2 m_a m_b}{q^2} \hspace*{-1.5pt} + \hspace*{-1.5pt} \frac{153 m_a^2 \hspace*{-1.2pt} + \hspace*{-1.2pt} 218 m_a m_b \hspace*{-1.2pt} + \hspace*{-1.2pt} 153 m_b^2}{24 m_a m_b} \hspace*{-1pt}\right)\nonumber\\
&& \hspace*{92pt} + S \hspace*{-1pt} \left(\frac{m_a m_b (m_a \hspace*{-1pt} + \hspace*{-1pt} m_b)}{s - s_0} \hspace*{-1pt} + \hspace*{-1pt} \frac{41}{4} \left(m_a \hspace*{-1pt} + \hspace*{-1pt} m_b \right) \hspace*{-1pt} \right) \hspace*{-3pt} \Bigg) \nonumber \\
&&\hspace*{37pt} + \Big(2 S_a \hspace*{-0.5pt} \cdot \hspace*{-0.5pt} p_3 S_b \hspace*{-0.5pt} \cdot \hspace*{-0.5pt} p_1 \hspace*{-0.5pt} + \hspace*{-0.5pt} S_a \hspace*{-0.5pt} \cdot \hspace*{-0.5pt} q S_b \hspace*{-0.5pt} \cdot \hspace*{-0.5pt} p_1 \hspace*{-0.5pt} - \hspace*{-0.5pt} S_a \hspace*{-0.5pt} \cdot \hspace*{-0.5pt} p_3 S_b \hspace*{-0.5pt} \cdot \hspace*{-0.5pt} q\Big) \frac{22 L}{3 m_a m_b}\Bigg] \nonumber\\
&-& i 4\pi G^2 m_a^2 m_b^2 \hspace*{1pt} {L\over q^2}\sqrt{m_a m_b
\over s-s_0} \,
\Bigg({\mathcal U}_a {\mathcal U}_b + 2 \hspace*{0.5pt} i \frac{{\mathcal E}_a {\mathcal U}_b}{m_a^2 m_b} + 2 \hspace*{0.5pt} i \frac{{\mathcal U}_a {\mathcal E}_b}{m_a m_b^2}\nonumber\\
&&\hspace*{123pt}  + \frac{S_a \cdot qS_b \cdot q - \frac{1}{2}q^2 S_a \cdot S_b}{m_a m_b}\Bigg) \nonumber\\
{}^{{1\over 2}{1\over 2}}{\cal
M}_{\ref{fig_loops}e}^{(2)}(q)\hspace*{-5pt}&=&\hspace*{-5pt} G^2
m_am_b \Bigg[{\mathcal U}_a {\mathcal U}_b
 \hspace*{-1.5pt}\Bigg(\hspace*{-1.5pt}L \hspace*{-1pt}\left(\hspace*{-1.5pt} - {4m_am_b\over q^2}
 \hspace*{-1pt} - \hspace*{-1pt}\frac{93 m_a^2\hspace*{-1pt} + \hspace*{-1pt}232 m_a m_b\hspace*{-1pt} + \hspace*{-1pt} 93 m_b^2}{12 m_am_b}\hspace*{-0.5pt}\right) \nonumber \\
 && \hspace*{68pt} + S \left(- {7\over 2}(m_a + m_b) \right)\hspace*{-1pt} \Bigg) \nonumber\\
&&\hspace*{37pt}+  i \frac{{\mathcal E}_a {\mathcal U}_b}{m_a^2
m_b}\Bigg(\hspace*{-1pt} L \hspace*{-1.5pt} \left(\hspace*{-1.5pt} -
\frac{8 m_a m_b}{q^2} \hspace*{-1.5pt}
  - \hspace*{-1.5pt} \frac{124 m_a^2 \hspace*{-1.5pt} + \hspace*{-1.5pt} 235 m_a m_b \hspace*{-1.5pt} + \hspace*{-1.5pt} 177 m_b^2}{12 m_a m_b} \hspace*{-1.4pt} \right) \nonumber\\
&& \hspace*{82pt} + S \hspace*{-1pt} \left(  - 4 m_a - \frac{33}{4} m_b \right) \! \hspace*{-1pt} \Bigg)\nonumber\\
&&\hspace*{37pt}+  i \frac{{\mathcal U}_a {\mathcal E}_b}{m_a
m_b^2}\Bigg(\hspace*{-1pt} L \hspace*{-1.5pt} \left(\hspace*{-1.5pt}
- \frac{8 m_a m_b}{q^2} \hspace*{-1.5pt}
  - \hspace*{-1.5pt} \frac{177 m_a^2 \hspace*{-1.5pt} + \hspace*{-1.5pt} 235 m_a m_b \hspace*{-1.5pt} + \hspace*{-1.5pt} 124 m_b^2}{12 m_a m_b} \hspace*{-1.4pt} \right) \nonumber\\
&& \hspace*{82pt} + S \hspace*{-1pt} \left( - \frac{33}{4} m_a - 4 m_b \right) \! \hspace*{-1pt} \Bigg)\nonumber\\
&&\hspace*{37pt} + \frac{S_a\cdot qS_b\cdot q}{m_a m_b} \Bigg(
\hspace*{-1pt} L \hspace*{-1.5pt} \left(\hspace*{-1.5pt} - \frac{4
m_a m_b}{q^2} \hspace*{-1.5pt}
  - \hspace*{-1.5pt}  \frac{39 m_a^2 \hspace*{-1.7pt} + \hspace*{-1.5pt} 23 m_a m_b \hspace*{-1.7pt} + \hspace*{-1.5pt} 39 m_b^2}{6 m_a m_b} \hspace*{-1.4pt} \right) \nonumber\\
&& \hspace*{99pt} + S \hspace*{-1pt} \left( \frac{39}{4} \left(m_a \hspace*{-1pt} + \hspace*{-1pt} m_b \right) \hspace*{-1pt} \right) \hspace*{-3pt} \Bigg) \nonumber \\
&&\hspace*{37pt} - \frac{q^2 S_a \hspace*{-1.7pt} \cdot
\hspace*{-1.6pt} S_b}{m_a m_b} \hspace*{-1pt} \Bigg( \hspace*{-2pt}
L \hspace*{-2pt} \left(\hspace*{-2.3pt} - \frac{2 m_a m_b}{q^2}
\hspace*{-1.5pt}
  - \hspace*{-1.5pt} \frac{153 m_a^2 \hspace*{-1.8pt} + \hspace*{-2.2pt} 158 m_a m_b \hspace*{-1.8pt} + \hspace*{-2.2pt} 153 m_b^2}{24 m_a m_b} \hspace*{-1.4pt}\right)\nonumber\\
&& \hspace*{87pt} + S \left( - \frac{3}{2} \left(m_a \hspace*{-1pt} + \hspace*{-1pt} m_b \right) \hspace*{-1pt} \right) \hspace*{-3pt} \Bigg) \nonumber \\
&&\hspace*{37pt} + \Big(2 S_a \hspace*{-0.5pt} \cdot \hspace*{-0.5pt} p_3 S_b \hspace*{-0.5pt} \cdot \hspace*{-0.5pt} p_1 \hspace*{-0.5pt} + \hspace*{-0.5pt} S_a \hspace*{-0.5pt} \cdot \hspace*{-0.5pt} q S_b \hspace*{-0.5pt} \cdot \hspace*{-0.5pt} p_1 \hspace*{-0.5pt} - \hspace*{-0.5pt} S_a \hspace*{-0.5pt} \cdot \hspace*{-0.5pt} p_3 S_b \hspace*{-0.5pt} \cdot \hspace*{-0.5pt} q\Big) \frac{- 22 L}{3 m_a m_b}\Bigg] \nonumber\\
{}^{{1\over 2}{1\over 2}}{\cal M}_{\ref{fig_loops}f}^{(2)}(q)\hspace*{-5pt}&=&\hspace*{-5pt} G^2 m_a m_b \Bigg[{\mathcal U}_a {\mathcal U}_b \Big(14 L - 2 (m_a + m_b) S\Big) \nonumber\\
 && \hspace*{37pt} +  i \frac{{\mathcal E}_a {\mathcal U}_b}{m_a^2 m_b} \Bigg(4 L + \left(- 2 m_a - \frac{1}{2} m_b \right) \! S \Bigg) \nonumber \\
 && \hspace*{37pt} +  i \frac{{\mathcal U}_a {\mathcal E}_b}{m_a m_b^2} \Bigg(4 L + \left(- \frac{1}{2} m_a - 2 m_b \right) \! S \Bigg) \nonumber \\
 && \hspace*{37pt} + \frac{S_a\cdot qS_b\cdot q}{m_a m_b} \left(- 2 L - (m_a + m_b) S\right) \nonumber\\
 && \hspace*{37pt} - \frac{q^2 S_a \cdot S_b}{m_a m_b} \left(- 2 L - (m_a + m_b) S\right) \hspace*{-1.5pt}\Bigg]\nonumber\\
{}^{{1\over 2}{1\over 2}}{\cal
M}_{\ref{fig_loops}g}^{(2)}(q)\hspace*{-5pt}&=&\hspace*{-5pt}G^2
m_a m_b \Bigg[{\mathcal U}_a {\mathcal U}_b  \hspace*{-1pt}
\left(\hspace*{-1pt} - \frac{43}{15}L \hspace*{-0.5pt} \right)
\hspace*{-2pt}
 +  i \frac{{\mathcal E}_a {\mathcal U}_b}{m_a^2 m_b} \hspace*{-1pt} \left(\hspace*{-1pt} - \frac{14}{5} L \hspace*{-0.5pt} \right) \hspace*{-2pt}
 +  i \frac{{\mathcal U}_a {\mathcal E}_b}{m_a m_b^2} \hspace*{-1pt} \left(\hspace*{-1pt} - \frac{14}{5} L \hspace*{-0.5pt} \right) \nonumber \\
 && \hspace*{37pt}+ \frac{S_a\cdot qS_b\cdot q}{m_a m_b} \left(- \frac{7}{5} L\right)
 - \frac{q^2 S_a \cdot S_b}{m_a m_b} \left(- \frac{7}{5} L\right) \hspace*{-1.5pt}\Bigg].
\end{eqnarray}
where we have defined
\begin{equation}
{\mathcal U}_a=\bar{u}(p_2)u(p_1) \qquad \qquad {\mathcal
U}_b=\bar{u}(p_4)u(p_3)
\end{equation}
and
\begin{equation}
{\mathcal E}_i = \epsilon_{\alpha\beta\gamma\delta}p_1^\alpha
p_3^\beta q^\gamma S_i^\delta
\end{equation}
with $i = a, b$ to keep our notation compact.  The sum of all
diagrams is found to be
\begin{eqnarray}
{}^{{1\over 2}{1\over 2}}{\cal M}_{tot}^{(2)}(q) \hspace*{-8.5pt}
&=\hspace*{-7.5pt}&G^2 m_a m_b \Bigg[
{\mathcal U}_a {\mathcal U}_b \bigg(6(m_a + m_b) S - \frac{41}{5} L\bigg) \nonumber \\
&& \hspace*{41pt} + \hspace*{0.5pt} i \frac{{\mathcal E}_a {\mathcal U}_b}{m_a^2 m_b} \hspace*{-1pt} \bigg(\hspace*{-1.3pt} \frac{11 \hspace*{-0.7pt} (\hspace*{-0.9pt} 4 m_a \hspace*{-3.2pt} + \hspace*{-2.6pt} 3 m_b \hspace*{-0.7pt} )}{4} S \hspace*{-1.5pt} - \hspace*{-1.5pt} \frac{64}{5} L \hspace*{-1.5pt} + \hspace*{-1.5pt} \frac{m_a m_b(4 m_a \hspace*{-3pt} + \hspace*{-2.4pt} 3 m_b)}{s - s_0} S \hspace*{-1pt}\bigg) \nonumber\\
&& \hspace*{41pt} + \hspace*{0.5pt} i \frac{{\mathcal U}_a {\mathcal E}_b}{m_a m_b^2} \hspace*{-1pt} \bigg(\hspace*{-1.3pt} \frac{11 \hspace*{-0.7pt} (\hspace*{-0.9pt} 3 m_a \hspace*{-3.2pt} + \hspace*{-2.6pt} 4 m_b \hspace*{-0.7pt} )}{4} S \hspace*{-1.5pt} - \hspace*{-1.5pt} \frac{64}{5} L \hspace*{-1.5pt} + \hspace*{-1.5pt} \frac{m_a m_b(3 m_a \hspace*{-3pt} + \hspace*{-2.4pt} 4 m_b)}{s - s_0} S \hspace*{-1pt}\bigg) \nonumber\\
&& \hspace*{41pt} + S (m_a + m_b) \, \frac{S_a \cdot qS_b \cdot q -
q^2 S_a \cdot S_b}{m_a m_b}
\left(\frac{37}{4} + {m_a m_b\over s-s_0}\right)\nonumber\\
&&\hspace*{41pt} - L \, \frac{11 S_a \cdot qS_b \cdot q - 16 \hspace*{1pt} q^2 S_a \cdot S_b}{15 m_a m_b} \Bigg]\nonumber\\
&- \hspace*{-7.5pt}& i 4\pi G^2 m_a^2 m_b^2 {L\over q^2}\sqrt{m_a m_b
\over s-s_0} \,
\Bigg({\mathcal U}_a {\mathcal U}_b + 2 i \frac{{\mathcal E}_a {\mathcal U}_b}{m_a^2 m_b} + 2 i \frac{{\mathcal U}_a {\mathcal E}_b}{m_a m_b^2}\nonumber\\
&&\hspace*{124pt}  + \frac{S_a \cdot qS_b \cdot q - \frac{1}{2}q^2
S_a \cdot S_b}{m_a m_b}\Bigg). \quad \label{eq:ohla2}
\end{eqnarray}
Comparing with our finding for spin-0 -- spin-1/2 scattering in Eq.
(\ref{eq:ohla}), we notice the by now familiar universality of the
amplitude: The form of the component proportional to $\bar u(p_4)
u(p_3)$ of Eq. (\ref{eq:ohla}) is found here in the component
proportional to ${\mathcal U}_a {\mathcal U}_b$, and the form of the
structure proportional to
$\epsilon_{\alpha\beta\gamma\delta}p_1^\alpha p_3^\beta q^\gamma
S_b^\delta$ of Eq. (\ref{eq:ohla}) is now found in the component
proportional to ${\mathcal U}_a {\mathcal E}_b$ in Eq.
(\ref{eq:ohla2}). Moreover, the amplitude is symmetric in $a
\leftrightarrow b$ and new gravitational spin-spin interaction
corrections arise. The quantum part of the spin-spin component has
been calculated previously by Kirilin \cite{ggk} whose result
disagrees with our result for the numerical prefactors. For the quantum
terms in the spin-orbit components however, we fully agree with Kirilin's results in \cite{ggk}.
In the nonrelativistic limit we obtain the expression
\begin{eqnarray}
{}^{{1\over 2}{1\over 2}}{\cal M}_{tot}^{(2)}(\vec q)\hspace*{-8pt}
&\simeq\hspace*{-7.5pt}&\left[\hspace*{-0.8pt} G^2 m_a m_b \hspace*{-2.6pt} \left( \hspace*{-1.8pt} 6 (m_a \hspace*{-2.8pt} + \hspace*{-2.2pt} m_b) S \hspace*{-1.5pt} - \hspace*{-1.5pt} {41 \over 5}L \hspace*{-2pt} \right) \hspace*{-2.4pt} - \hspace*{-0.8pt} i 4 \pi G^2 m_a^2 m_b^2 \hspace*{1pt}
\frac{L}{q^2} \frac{m_r}{p_0} \hspace*{-0.8pt} \right] \hspace*{-1.8pt} \chi_f^{a\dagger}\chi_i^a \hspace*{1.3pt}  \chi_f^{b\dagger}\chi_i^b \nonumber\\
&+\hspace*{-7.5pt}&\Bigg[\hspace*{-0.8pt} G^2  \hspace*{-1pt} \left(\frac{24m_a^3
\hspace*{-1pt} + \hspace*{-1pt} 56 m_a^2 m_b \hspace*{-1pt} +
\hspace*{-1pt} 45 m_a m_b^2
 \hspace*{-1pt} + \hspace*{-1pt} 12 m_b^3}{2(m_a \hspace*{-1pt} + \hspace*{-1pt} m_b)}  \hspace*{1pt} S
- \frac{128 m_a \hspace*{-1pt} + \hspace*{-1pt} 87 m_b}{10}  \hspace*{1pt} L\right)\nonumber\\
&&+ \frac{G^2 m_a^2 m_b^2 (4 m_a + 3 m_b)}{(m_a + m_b)} \left(- i
\frac{2 \pi L}{p_0 q^2} + \frac{S}{p_0^2} \right)\Bigg] {i\over
m_a}\vec{S}_a\cdot\vec{p}\times\vec{q} \, \chi_f^{b\dagger}\chi_i^b \nonumber \\
&+\hspace*{-7.5pt}&\Bigg[\hspace*{-0.8pt} G^2  \hspace*{-1pt} \left(\frac{12m_a^3
\hspace*{-1pt} + \hspace*{-1pt} 45 m_a^2 m_b \hspace*{-1pt} +
\hspace*{-1pt} 56 m_a m_b^2
 \hspace*{-1pt} + \hspace*{-1pt} 24 m_b^3}{2(m_a \hspace*{-1pt} + \hspace*{-1pt} m_b)}  \hspace*{1pt} S
- \frac{87 m_a \hspace*{-1pt} + \hspace*{-1pt} 128 m_b}{10}  \hspace*{1pt} L\right)\nonumber\\
&&+ \frac{G^2 m_a^2 m_b^2 (3 m_a + 4 m_b)}{(m_a + m_b)} \left(- i
\frac{2 \pi L}{p_0 q^2} + \frac{S}{p_0^2} \right)\Bigg]
\chi_f^{a\dagger}\chi_i^a \, {i\over
m_b}\vec{S}_b\cdot\vec{p}\times\vec{q} \nonumber \\
&+\hspace*{-7.5pt}& G^2 m_a m_b \, \frac{19 m_a^2 + 36 m_a m_b + 19 m_b^2}{2(m_a + m_b)}\, S \, \frac{\vec S_a \cdot \vec q \, \vec S_b \cdot \vec q - \vec q^{\hspace*{1.4pt}2} \vec S_a \cdot \vec S_b}{m_a m_b}\nonumber\\
&-\hspace*{-7.5pt}& G^2 m_a m_b \,  L \, \frac{11 \vec S_a \cdot \vec q \, \vec S_b \cdot \vec q - 16 \hspace*{1pt} \vec q^{\hspace*{1.4pt}2} \vec S_a \cdot \vec S_b}{15 m_a m_b}\nonumber\\
&+\hspace*{-7.5pt}& \frac{G^2 m_a^3 m_b^3}{m_a + m_b} \, \frac{S}{p_0^2} \,  \frac{\vec S_a \cdot \vec q \, \vec S_b \cdot \vec q - \vec q^{\hspace*{1.4pt}2} \vec S_a \cdot \vec S_b}{m_a m_b}\nonumber\\
&+\hspace*{-7.5pt}& \frac{G^2 m_a^3 m_b^3}{m_a + m_b} \left(-i \frac{4 \pi L}{p_0
q^2}\right) \frac{\vec S_a \cdot \vec q \, \vec S_b \cdot \vec q -
\frac{1}{2}\vec q^{\hspace*{1.4pt}2} \vec S_a \cdot \vec S_b}{m_a
m_b}
 \label{eq:yohh}
\end{eqnarray}
which obviously has to exhibit the universalities of the
spin-independent and the spin-orbit pieces. Therefore, we know that
the subtraction of the second Born iteration successfully removes
the unwanted $i/p_0$ and $1/p_0^2$ structures for the
spin-independent and the spin-orbit components.

The leading spin-spin term of the second Born iteration amplitude is
new, however, and we compute
\begin{eqnarray}
{}^{{1\over 2}{1\over 2}} {\rm Amp}^{(2)}_{S-S}(\vec q)
\hspace*{-10pt} &= \hspace*{-10pt}&- \int{d^3\ell\over
(2\pi)^3} \, \frac{\left<\vec p_f \left| {}^{{1\over 2}{1\over 2}} \hat V^{(1)}_{S-I} \right| \vec \ell \, \right> \left<\vec \ell \left| {}^{{1\over 2}{1\over 2}} \hat V^{(1)}_{S-S} \right| \vec p_i \right>}{\frac{p_0^2}{2 m_r} - \frac{\ell^2}{2 m_r} + i \epsilon} \nonumber\\
&&- \int{d^3\ell\over
(2\pi)^3} \, \frac{\left<\vec p_f \left| {}^{{1\over 2}{1\over 2}} \hat V^{(1)}_{S-S} \right| \vec \ell \, \right> \left<\vec \ell \left| {}^{{1\over 2}{1\over 2}} \hat V^{(1)}_{S-I} \right| \vec p_i \right>}{\frac{p_0^2}{2 m_r} - \frac{\ell^2}{2 m_r} + i \epsilon} \nonumber\\
&=\hspace*{-10pt}&{1\over m_am_b} \, S_a^r \, S_b^s \nonumber\\
&& \left(\hspace*{-2pt}i \hspace*{-3.2pt} \int \hspace*{-3.2pt}
{d^3\ell\over (2\pi)^3} {c_G^2 \over |\vec \ell
\hspace*{-1.1pt} - \hspace*{-1.1pt} \vec{p}_f \hspace*{1pt}|^2
\hspace*{-1.1pt} + \hspace*{-1.2pt} \lambda^2}{i \over {p_0^2\over
2m_r} \hspace*{-1.1pt} - \hspace*{-1.1pt} {\ell^2\over 2m_r}
\hspace*{-1.1pt} + \hspace*{-1.1pt} i\epsilon}{c_G^2 (p_i
\hspace*{-1.1pt} - \hspace*{-1.1pt} \ell)^r (p_i \hspace*{-1.1pt} -
\hspace*{-1.1pt} \ell)^s\over
|\vec{p}_i \hspace*{-1.1pt} - \hspace*{-1.1pt} \vec{\ell} \hspace*{1pt}|^2
\hspace*{-1.1pt} + \hspace*{-1.2pt} \lambda^2}\right.\nonumber\\
&&\left. \hspace*{-3.6pt}+ i \hspace*{-3.2pt} \int \hspace*{-3.2pt}
{d^3\ell\over (2\pi)^3} {c_G^2 (\ell \hspace*{-1.1pt} -
\hspace*{-1.2pt} p_f)^r (\ell \hspace*{-1.1pt} - \hspace*{-1.2pt}
p_f)^s \over |\vec{\ell} \hspace*{-1.1pt} - \hspace*{-1.1pt}
\vec{p}_f \hspace*{1pt}|^2 \hspace*{-1.1pt} + \hspace*{-1.2pt}
\lambda^2}{i \over {p_0^2\over 2m_r} \hspace*{-1.1pt} -
\hspace*{-1.1pt} {\ell^2\over 2m_r} \hspace*{-1.1pt} +
\hspace*{-1.1pt} i\epsilon}{c_G^2 \over
|\vec{p}_i \hspace*{-1.1pt} - \hspace*{-1.1pt} \vec{\ell} \hspace*{1pt}|^2 \hspace*{-1.1pt} + \hspace*{-1.2pt}\lambda^2}\hspace*{-3pt}\right)\nonumber\\
&\stackrel{\lambda\rightarrow 0}{\longrightarrow}\hspace*{-8pt}& {}\, {1\over m_a m_b}
\Bigg[\left(\vec{S}_a\cdot\vec{p_i} \, \vec{S}_b\cdot\vec{p_i} + \vec{S}_a\cdot\vec{p_f} \, \vec{S}_b\cdot\vec{p_f} \right) H\nonumber\\
&& \hspace*{34pt} - \vec S_a \cdot (\vec p_i + p_f) \vec S_b \cdot \vec H - \vec S_a \cdot \vec H \, \vec S_b \cdot (\vec p_i + p_f) \nonumber\\
&& \hspace*{34pt} + 2 \, S_a^r S_b^s \, H^{rs}\Bigg]\nonumber\\
&=\hspace*{-10pt}& \frac{G^2m_a^2m_b^2}{m_a + m_b} \, \frac{S}{p_0^2} \,
(\vec S_a \cdot \vec q \, \vec S_b \cdot \vec q - \vec q^{\hspace*{1.4pt}2} \vec S_a \cdot \vec S_b)\nonumber\\
&+\hspace*{-10pt}& \frac{G^2m_a^2m_b^2}{m_a + m_b} \left(-i \frac{4
\pi L}{p_0 q^2}\right) (\vec S_a \cdot \vec q \, \vec S_b \cdot \vec
q - \frac{1}{2}\vec q^{\hspace*{1.4pt}2} \vec S_a \cdot \vec S_b).
\end{eqnarray}
where we again defined $c_G^2 \equiv - 4 \pi G m_a m_b$.
With this, the full second Born iteration amplitude becomes
\begin{eqnarray}
{}^{{1\over 2}{1\over 2}} {\rm Amp}_G^{(2)}(\vec q) \hspace*{-3pt}{}
&=&{}^{{1\over 2}{1\over 2}} {\rm Amp}^{(2)}_{S-I}(\vec q) +
{}^{{1\over 2}{1\over 2}} {\rm Amp}^{(2)}_{S-O}(\vec q) +
{}^{{1\over 2}{1\over 2}} {\rm Amp}^{(2)}_{S-S}(\vec q) \nonumber \\
&=&- i 4 \pi G^2m_a^2m_b^2
\frac{L}{q^2} \frac{m_r}{p_0} \,  \chi_f^{a\dagger}\chi_i^a \, \chi_f^{b\dagger}\chi_i^b \nonumber\\
&+& \frac{G^2m_a^2m_b^2 (2 m_a + m_b)}{m_a + m_b} \left(- i
\frac{2 \pi L}{p_0 q^2} + \frac{S}{p_0^2} \right) \, {i\over
m_a}\vec{S}_a\cdot\vec{p}\times\vec{q} \ \chi_f^{b\dagger}\chi_i^b\nonumber\\
&+& \frac{G^2m_a^2m_b^2 (m_a + 2 m_b)}{m_a + m_b} \left(- i
\frac{2 \pi L}{p_0 q^2} + \frac{S}{p_0^2} \right)
\chi_f^{a\dagger}\chi_i^a \ {i\over
m_b}\vec{S}_b\cdot\vec{p}\times\vec{q} \nonumber\\
&+& \frac{G^2m_a^2m_b^2}{m_a + m_b} \, \frac{S}{p_0^2} \,
(\vec S_a \cdot \vec q \, \vec S_b \cdot \vec q - \vec q^{\hspace*{1.4pt}2} \vec S_a \cdot \vec S_b)\nonumber\\
&+& \frac{G^2m_a^2m_b^2}{m_a + m_b} \left(-i \frac{4 \pi L}{p_0
q^2}\right) (\vec S_a \cdot \vec q \, \vec S_b \cdot \vec q -
\frac{1}{2}\vec q^{\hspace*{1.4pt}2} \vec S_a \cdot \vec S_b)
\end{eqnarray}
and we observe that when this amplitude is subtracted from the full
one loop scattering amplitude Eq. (\ref{eq:yohh}), {\it all} terms
involving $1/p_0^2$ and $i/p_0$ disappear leaving behind a
well-defined second order potential
\begin{eqnarray}
{}^{{1\over 2}{1\over 2}} V^{(2)}_G(\vec r) \hspace*{-5pt} & = \hspace*{-5pt} &-\int{d^3q\over
(2\pi)^3}e^{-i\vec{q}\cdot\vec{r}}\left[{}^{{1\over 2}{1\over 2}}
{\cal M}_{tot}^{(2)}(\vec{q})- {}^{{1\over 2}{1\over 2}} {\rm
Amp}_G^{(2)}(\vec{q})\right]\nonumber\\
& \simeq \hspace*{-5pt} &\left[-{3G^2m_am_b(m_a+m_b)\over r^2} -
{41G^2m_am_b\hbar\over 10\pi r^3}\right]
\chi_f^{a\dagger}\chi_i^a \, \chi_f^{b\dagger}\chi_i^b \nonumber\\
& + \hspace*{-5pt} &{1\over m_a}\vec{S}_a\cdot \vec{p}\times\vec{\nabla}\Bigg[
\frac{G^2(24m_a^3 + 56 m_a^2 m_b + 45 m_a m_b^2 + 12 m_b^3)}{4(m_a + m_b)r^2} \nonumber\\
&& \hspace*{68pt} + \frac{G^2(128 m_a + 87 m_b) \hbar}{20 \pi r^3} \Bigg] \chi_f^{a\dagger}\chi_i^a \nonumber\\
& + \hspace*{-5pt} &\chi_f^{a\dagger}\chi_i^a \, {1\over m_b}\vec{S}_b\cdot
\vec{p}\times\vec{\nabla}\Bigg[
\frac{G^2(12m_a^3 + 45 m_a^2 m_b + 56 m_a m_b^2 + 24 m_b^3)}{4(m_a + m_b)r^2} \nonumber\\
&& \hspace*{97pt} + \frac{G^2(87 m_a + 128 m_b) \hbar}{20 \pi r^3} \Bigg]\nonumber\\
& + \hspace*{-5pt} & \frac{\vec S_a \cdot \vec{\nabla} \vec S_b \cdot \vec{\nabla} -
\vec{\nabla}^2 \vec S_a \cdot \vec S_b}{m_a m_b}
\left[\frac{G^2 m_a m_b(19 m_a^2 + 36 m_a m_b + 19 m_b^2)}{4(m_a + m_b)r^2}\right]\nonumber\\
& + \hspace*{-5pt} & \frac{11 \vec S_a \cdot \vec{\nabla} \vec S_b \cdot \vec{\nabla}
- 16 \vec{\nabla}^2 \vec S_a \cdot \vec S_b}{15 m_a m_b}
\left[\frac{G^2 m_a m_b \hbar}{2 \pi r^3}\right]\nonumber\\
& \simeq \hspace*{-5pt} & \left[-{3G^2m_am_b(m_a+m_b)\over r^2} -
{41G^2m_am_b\hbar\over 10\pi r^3}\right]
\chi_f^{a\dagger}\chi_i^a \, \chi_f^{b\dagger}\chi_i^b\nonumber\\
& + \hspace*{-5pt} &\Bigg[{G^2(24 m_a^3 + 56 m_a^2 m_b + 45 m_a m_b^2 + 12
m_b^3)\over
2 m_a (m_a+m_b)r^4} \nonumber \\
&&+{3 G^2(128 m_a + 87 m_b)\hbar\over 20 \pi m_a r^5}\Bigg]
 \vec{L}\cdot\vec{S}_a \, \chi_f^{b\dagger}\chi_i^b  \nonumber \\
& + \hspace*{-5pt} &\Bigg[{G^2(12 m_a^3 + 45 m_a^2 m_b + 56 m_a m_b^2 + 24
m_b^3)\over
2 m_b (m_a+m_b)r^4} \nonumber \\
&&+{3 G^2(87 m_a+ 128m_b)\hbar\over 20 \pi m_b r^5}\Bigg]
 \chi_f^{a\dagger}\chi_i^a \, \vec{L}\cdot\vec{S}_b \nonumber \\
& + \hspace*{-5pt} & \left[\frac{2 G^2(19 m_a^2 + 36 m_a m_b + 19 m_b^2)}{(m_a + m_b) r^4}\right] \! \left(\vec S_a \cdot \vec r \, \vec S_b \cdot \vec r \, / r^2 - \frac{1}{2} \hspace*{1pt} \vec S_a \cdot \vec S_b\right)\nonumber\\
& + \hspace*{-5pt} & \frac{11 G^2 \hbar}{2 \pi r^5} \left(\vec S_a \cdot \vec r \,
\vec S_b \cdot \vec r \, / r^2 - \frac{43}{55} \vec S_a \cdot \vec
S_b\right)
\end{eqnarray}
which besides the universal spin-independent and spin-orbit
components displays a new (and presumably universal) gravitational
spin-spin interaction.

\section{Conclusions}

Above we have analyzed the gravitational scattering of two particles
having nonzero mass.  In lowest order the interaction arises from
one-graviton exchange and leads at threshold to the well known
Newtonian interaction $V(r)=-Gm_am_b/r$.  Inclusion of two-graviton
exchange effects means adding the contribution from box, cross-box,
triangle, and bubble diagrams, which have a rather complex form. The
calculation can be simplified, however, by using ideas from
effective field theory.  The point is that if one is interested only
in the leading long-range behavior of the interaction, then one need
retain only the leading nonanalytic small momentum-transfer piece of
the scattering amplitude. Specifically, the terms which one retains
are those which are nonanalytic and behave as either
$1/\sqrt{-q^2}$ or $\log -q^2$. When Fourier transformed, the
former leads to classical ($\hbar$-independent) terms in the
potential of order $G^2 M^3/r^2$ while the latter generates quantum
mechanical ($\hbar$-dependent) corrections of order
$G^2 M^2\hbar/r^3$. (Of course, there are also shorter range
nonanalytic contributions than these that are generated by
scattering terms of order $q^{2n}/\sqrt{-q^2}$ or $q^{2n}\log
-q^2$. However, these pieces are higher order in momentum transfer
and thus lead to shorter distance effects than those considered
above and are therefore neglected in our discussion.)

Specific calculations were done for particles with spin $0-0$,
$0-1/2$, $0-1$, and $1/2-1/2$ and various universalities were found.
In particular, we found that in each case there was a
spin-independent contribution of the form
\begin{eqnarray}
{}^{S_aS_b}{\cal M}^{(2)}_{tot}(q) \hspace*{-7.5pt} & = \hspace*{-7.5pt} & \left[G^2m_am_b\left(6(m_a+m_b)S - {41\over 5}L \right) - i4\pi G^2m_a^2m_b^2 \hspace*{1pt} {L\over
q^2}\sqrt{m_am_b\over s-s_0} \, \hspace*{1pt} \right] \nonumber \\
&& \times \left<S_a,m_{af}|S_a,m_{ai}\right> \left<S_b,m_{bf}|S_b,m_{bi}\right>
\end{eqnarray}
where $L=\log -q^2$ and $S=\pi^2/\sqrt{-q^2}$ and with $S_a$ the
spin of particle $a$ and $S_b$ the spin of particle $b$ with
projections $m_a$ and $m_b$ on the quantization axis. The imaginary
component of the amplitude, which would not, when
Fourier-transformed lead to a real potential, is eliminated when the
iterated lowest order potential contribution is subtracted, leading
to a well defined spin-independent second order potential of
universal form
\begin{eqnarray}
{}^{S_aS_b}V_{S-I}^{(2)}(\vec r)&=&\left[-{3 G^2m_am_b(m_a+m_b)\over r^2} - {41 G^2 m_a m_b \hbar \over 10\pi r^3}\right] \nonumber\\
&& \times \left<S_a,m_{af}|S_a,m_{ai}\right>
\left<S_b,m_{bf}|S_b,m_{bi}\right>
\end{eqnarray}
whose classical component depends on the way the iteration
of the leading order potential is performed. This ambiguity
shows that the second order potential itself is not an observable,
but we use it as a nice way to display the long distance components
of the scattering amplitude in coordinate space.

If either scattering particle carries spin then there exists an
additional spin-orbit contribution, whose form is also universal
\begin{eqnarray}
{}^{S_aS_b}V_{S-O}^{(2)}(\vec r)
& = \hspace*{-5pt} &\Bigg[{G^2(24 m_a^3 + 56 m_a^2 m_b + 45 m_a m_b^2 + 12
m_b^3)\over
2 m_a (m_a+m_b)r^4} \nonumber \\
&&+{3 G^2(128 m_a + 87 m_b)\hbar\over 20 \pi m_a r^5}\Bigg]
\times \vec{L}\cdot\vec{S}_a \left<S_b,m_{bf}|S_b,m_{bi}\right>  \nonumber \\
& + \hspace*{-5pt} &\Bigg[{G^2(12 m_a^3 + 45 m_a^2 m_b + 56 m_a m_b^2 + 24
m_b^3)\over
2 m_b (m_a+m_b)r^4} \nonumber \\
&&+{3 G^2(87 m_a+ 128m_b)\hbar\over 20 \pi m_b r^5}\Bigg]
\times \left<S_a,m_{af}|S_a,m_{ai}\right> \vec{L}\cdot\vec{S}_b \quad \quad
\end{eqnarray}
where we have defined
$$\vec{S}_a= \left<S_a,m_{af} \left|{}\, \vec{S} \, \right| S_a,m_{ai}\right>\quad{\rm and}
\quad \vec{S}_b= \left<S_b,m_{bf} \left|{}\, \vec{S} \, \right|
S_b,m_{bi}\right>. $$  
In this case a well defined second order
potential required the subtraction of infrared singular terms
behaving as both $i/p_0$ {\it and} $1/p_0^2$ which arise from the
iterated lowest order potential.

In the calculation of spin-0 -- spin-1 scattering we encountered new
tensor structures including a quadrupole interaction. Unfortunately,
the subtraction of the $i/p_0$ and $1/p_0^2$ tensor pieces in the
two-graviton exchange amplitude was not successful with our simple
nonrelativistic iteration of the leading order potential so that we
cannot at this time give the form of the quadrupole component of the
potential. Further work is needed to clarify this issue. The
corrections to the spin-spin interaction have only been calculated
in spin-1/2 -- spin-1/2 scattering where we found their
contributions to the scattering amplitude and to the potential.
Since we verified these forms only for a single spin configuration
we have not confirmed its universality which we, however, strongly
suspect. Of course, for higher spin configurations, there also exist
quadrupole-quadrupole interactions, spin-quadrupole interactions,
etc.  However, the calculation of such forms becomes increasingly
cumbersome as the spin increases, and the phenomenological
importance becomes smaller.  Thus we end our calculations here.

One point of view to interpret the universalities of the long
distance components of the scattering amplitudes and the resulting
potentials is that if we increase the spins of our scattered
particles, all we do is to add additional multipole moments. The
spin-independent component can then be viewed as a monopole-monopole
interaction, the spin-orbit piece as a dipole-monopole interaction
etc. As long as we do not change the quantum numbers that
characterize the lower multipoles, an increase in spin of the
scattered particles merely adds new interactions that are less
important at long distances.
The same kind of universalities were also found in long distance effects
in electromagnetic scattering \cite{hrem} and in the long range
components of mixed electromagnetic-gravitational scattering \cite{hrmix}.

\begin{center}
{\bf Acknowledgements}
\end{center}

We would like to thank John Donoghue for many clarifying discussions
and Walter Goldberger for important comments. This work was supported
in part by the National Science Foundation under award PHY05-53304
(BRH and AR) and by the the US Department of Energy under grant
DE-FG-02-92ER40704 (AR).

\appendix

\section{Iteration Integrals} \label{app_iter}

In this appendix we give the integrals
\begin{equation}
[H;H_r;H_{rs}] = i\int {d^3\ell\over (2\pi)^3}
{- 4\pi Gm_am_b\over |\vec{\ell} - \vec{p}_f|^2+\lambda^2}
{i[1;\ell_r;\ell_r\ell_s]\over {p_0^2\over 2m_r}-{\ell^2\over 2m_r}+i\epsilon}
{- 4\pi Gm_am_b\over |\vec{p}_i - \vec{\ell}|^2+\lambda^2}
\end{equation}
which are needed in order to perform the iteration of the lowest
order Newton potentials. Here we list only the results; for a
more detailed derivation, albeit with a different prefactor, see
\cite{hrem}. The leading expressions for the iteration integrals read
\begin{eqnarray}
H & \simeq &-i \hspace*{0.7pt} 4\pi G^2m_a^2m_b^2 \, \frac{L}{q^2} \frac{m_r}{p_0}\nonumber\\
H_r & \simeq &(p_i+p_f)_r \ G^2 m_a^2 m_b^2 \left(-i \hspace*{0.8pt} 2 \pi \hspace*{1pt} \frac{L}{q^2} {m_r\over p_0} + S {m_r\over p_0^2} \right)\nonumber\\
H_{rs} &\simeq & \delta_{rs} \ \vec q^{\hspace*{1.4pt}2} \ G^2 m_a^2 m_b^2
\left(i \hspace*{0.7pt} \pi \frac{L}{q^2}{m_r\over p_0} - \frac{1}{2} S {m_r\over p_0^2} \right)\nonumber\\
&+&(p_i+p_f)_r(p_i+p_f)_s \ G^2 m_a^2 m_b^2 \left(- i \hspace*{0.7pt} \pi \frac{L}{q^2}{m_r\over p_0} + S {m_r\over p_0^2} \right)\nonumber\\
&+& (p_i-p_f)_r (p_i-p_f)_s \ G^2 m_a^2 m_b^2 \left(-i \hspace*{0.7pt} \pi \frac{L}{q^2}{m_r\over p_0}+ \frac{1}{2} S {m_r\over p_0^2}\right).
\end{eqnarray}

\section{Classical Equations of Motion} \label{app_eom}

Above we have argued that the scattering amplitude which is
ultimately related to observables in quantum field theory is a
physical result while the potential we have given is not an
observable and depends both on the gauge, {\it i.e.}, the choice of
coordinates used, and on the way we perform the iteration, {\it
i.e.}, on the way we perform the matching. While the classical
component of our potential is in fact plagued by these ambiguities,
the quantum part is unique since it is unaffected by how we perform
the matching and since a quantum field theory calculation in any
gauge would result in the same result \cite{bdh}.

In this appendix we will demonstrate how we can recover the classical equations of motion
from our scattering amplitudes by setting up the Einstein-Infeld-Hoffmann (EIH) Lagrangian \cite{Einstein:1938yz}.
The EIH Lagrangian is itself dependent on the choice of coordinates,
but can be expressed in the center of mass frame
($\vec P \equiv \vec p_a = - \vec p_b$, $\vec r \equiv \vec r_a - \vec r_b$) in a general way as
\cite{Hiida:1972xs, Barker:1975ae}
\begin{equation}
 L_{EIH} = T - V
\end{equation}
where the kinetic energy to NLO in the nonrelativistic expansion reads
\begin{equation}
 T = \frac{\vec P^{\hspace*{1.4pt} 2}}{2 m_a} + \frac{\vec P^{\hspace*{1.4pt} 2}}{2 m_b} - \frac{\vec P^{\hspace*{1.4pt} 4}}{8 m_a^3} - \frac{\vec P^{\hspace*{1.4pt} 4}}{8 m_b^3}
\end{equation}
and the potential is
\begin{equation}
 V = V^{(1)} + V^{(2)}
\end{equation}
with
\begin{eqnarray}
 V^{(1)} & = & - \frac{G m_a m_b}{r} \left\{1 + \left[\frac{1}{2} + \left(\frac{3}{2} - \alpha\right) \frac{(m_a + m_b)^2}{m_a m_b}\right] \frac{\vec P^{\hspace*{1.4pt} 2}}{m_a m_b} \right. \nonumber \\
  && \hspace*{69.5pt} \left.+ \left[\frac{1}{2} + \alpha \, \frac{(m_a + m_b)^2}{m_a m_b} \right] \frac{\left(\vec P \cdot \hat r\right)^2}{m_a m_b}\right\} \label{eq_eihV1} \\
 V^{(2)} & = & \left(1 - 2 \alpha\right) \frac{G^2 m_a m_b (m_a + m_b)}{2 r^2}. \label{eq_eihV2}
\end{eqnarray}
The parameter $\alpha$ parameterizes the choice of coordinates used, where $\alpha = 0$ was the gauge of the original EIH result.
The coordinate change
\begin{equation}
 \vec r \rightarrow \vec r \hspace*{1pt} \left(1 - \alpha \hspace*{1pt} \frac{G (m_a + m_b)}{r}\right)
\end{equation}
which implies
\begin{equation}
 \vec P \rightarrow \vec P + \alpha \hspace*{1pt} \frac{G (m_a + m_b)}{r} \left[\vec P - \left(\vec P \cdot \hat r \right) \hat r \right]
\end{equation}
brings the original EIH Lagrangian into the form above, which is the
most general result.

Since we perform our matching on-shell, {\it i.e.}, we use the
on-shell one-graviton exchange amplitude to define the leading order
$\mathcal O(G)$ potential, terms proportional to $\vec P \cdot \hat
r$ would never arise. Clearly, our result must be in a gauge such
that the coefficient of the structure
$$\frac{\left(\vec P \cdot \hat r\right)^2}{m_a m_b}$$ in Eq. (\ref{eq_eihV1}) vanishes.
That is the case if and only if the gauge parameter is
\begin{equation}
 \alpha = - \frac{m_a m_b}{2(m_a + m_b)^2}
\end{equation}
whereby the EIH potential becomes
\begin{eqnarray}
 V^{(1)} & = & - \frac{G m_a m_b}{r} \left\{1 + \left[1 + \frac{3}{2} \hspace*{0.8pt} \frac{(m_a + m_b)^2}{m_a m_b}\right] \frac{\vec P^{\hspace*{1.4pt} 2}}{m_a m_b} \right\} \label{eq_eihHRV1} \\
 V^{(2)} & = & \left(1 + \frac{m_a m_b}{(m_a + m_b)^2}\right) \frac{G^2 m_a m_b (m_a + m_b)}{2 r^2}. \label{eq_eihHRV2}
\end{eqnarray}
Comparing the EIH potential $V^{(1)}$ in this gauge of Eq.
(\ref{eq_eihHRV1}) with the long distance component of the leading
order spin-independent potential in Eq. (\ref{eq:po}) we find full
agreement for the relativistic corrections to the $\mathcal O(G)$
potential. However, comparing the EIH potential $V^{(2)}$ in this
gauge of Eq. (\ref{eq_eihHRV2}) with the classical component of our
spin-independent potential in Eq. (\ref{eq:so}) we see that the two
do not agree!  The reason for this discrepancy is that we elected to
use a nonrelativistic iteration when we performed the second Born
iteration of the leading order potential in Eq.
(\ref{eq:iteration00b}).  This procedure, however, is not
self-consistent when we are interested in equations of motion at
NLO, for which we must account for the leading relativistic
corrections in the iteration.  In particular, we must use
expressions for the potential and the propagator in Eq.
(\ref{eq:iteration00a}) which include the leading relativistic
corrections\footnote{The subscript $NLO$ in this sections refers to
the iteration being performed at NLO in the relativistic expansion.}
\begin{equation}
\left<\vec p_f \left| {}^0 \hat V^{(1)}_{NLO} \right| \vec p_i
\right> \simeq -{4\pi Gm_am_b\over
\vec{q}^{\hspace*{1.4pt}2}}\left[1+{\vec p_i^{\hspace*{1.4pt}2} +
\vec p_f^{\hspace*{1.4pt}2} \over 2m_am_b}\left(1 + \frac{3(m_a + m_b)^2}{2 m_a m_b}\right)\right]
\end{equation}
\begin{equation}
 G_{NLO}^{(0)}(\ell)={i\over {p_0^2\over 2m_r}-{\ell^2\over 2m_r}+i\epsilon} \times \left[1 + \left({p_0^2\over 4m_r^2}+{\ell^2\over 4m_r^2}\right) \left(1-3{m_r^2\over m_am_b}\right)\right]
\end{equation}
which yields a second Born iteration amplitude
\begin{eqnarray}
{}^0{\rm Amp}^{(2)}_{NLO}(\vec{q}) \hspace*{-5pt} &\simeq \hspace*{-5pt} &-\int{d^3\ell\over
(2\pi)^3} {4\pi Gm_am_b\over |\vec{p}_f-\vec{\ell}|^2}{1\over
{p_0^2\over 2m_r}-{\ell^2\over 2m_r} + i \epsilon} {4\pi
Gm_am_b\over
|\vec{\ell}-\vec{p}_i|^2}\nonumber\\
&& \times \left[1+{(p_0^2+\ell^2)\over m_am_b}\left(\frac{1}{4} + \frac{7}{4} \frac{m_a m_b}{m_r^2}\right) \right]\nonumber\\
& \simeq \hspace*{-5pt} &H+{1\over
m_am_b}\left(p_0^2H+\delta_{rs}H_{rs}\right) \left(\frac{1}{4} + \frac{7}{4} \frac{m_a m_b}{m_r^2}\right) \nonumber\\
&\simeq \hspace*{-5pt} & - i \hspace*{0.6pt} 4 \pi G^2 m_a^2 m_b^2 \hspace*{1pt} {L\over q^2} {m_r\over p_0}
  + \frac{G^2 m_a^2 m_b^2}{m_a + m_b} \left(1 + \frac{7 (m_a + m_b)^2}{m_a m_b} \right)S . \nonumber \\ \label{eq:jk}
\end{eqnarray}
Subtracting this iterated amplitude which includes all corrections to NLO from the scattering
amplitude ${}^0 \! {\cal M}^{(2)}_{tot}(\vec{q})$ of Eq. (\ref{eq_ampLO_00})
we find then the second order potential
\begin{eqnarray}
{}^0V_{NLO}^{(2)}(\vec r) \hspace*{-5pt} &=\hspace*{-5pt}&-\int{d^3q\over
(2\pi)^3}e^{-i\vec{q}\cdot\vec{r}}\left[{}^0 \! {\cal
M}^{(2)}_{tot}(\vec q)-{}^0{\rm Amp}_{NLO}^{(2)}(\vec q)\right]\nonumber\\
&=\hspace*{-5pt}&\int{d^3q\over (2\pi)^3} e^{-i\vec{q}\cdot\vec{r}} \, G^2 m_a m_b
\left[ \left((m_a + m_b) + {m_am_b\over m_a+m_b}\right) S + {41\over 5}L\right]\nonumber\\
&=\hspace*{-5pt}& \left(1 + \frac{m_a m_b}{(m_a + m_b)^2}\right) \frac{G^2 m_a m_b (m_a + m_b)}{2 r^2}
  - {41 G^2 m_a m_b \hbar \over 10 \pi r^3} \label{eq:sf}
\end{eqnarray}
and observe that now the classical component agrees with the $\mathcal O(G^2)$ EIH potential
of Eq. (\ref{eq_eihHRV2}).

Thus we have shown that if we consistently take into account the
$v^2$ and $GM/r$ corrections beyond Newtonian physics we reproduce
the EIH Lagrangian in a certain gauge. From the resulting EIH
Lagrangian we could evaluate observables such as the precession of
the perihelion of Mercury which must clearly be independent of the
gauge used. The inclusion of the $v^2$ corrections is required since
the equations of motion can be used to describe bound states where
$v^2 \sim GM/r$ by the virial theorem.

However, our methods are clearly clumsy for the calculation of classical observables. Recently,
Goldberger and Rothstein have developed an effective field theory of gravity which is optimized
for calculating classical observables of bound states called NRGR
\cite{Goldberger:2004jt, Goldberger:2006bd, Goldberger:2007hy, Goldberger:2005cd, Kol:2007rx, Kol:2007bc}. 
Here the external particles are static sources so that no loops are to be calculated in their theory
when calculating classical observables since the only propagating particles present
are gravitons which are massless and thus the loop expansion in NRGR corresponds to an
expansion in $\hbar$. In the NRGR framework the spin-dependent classical equations of motion were
calculated recently to NLO by Porto and Rothstein \cite{Porto:2005ac, Porto:2006bt, Porto:2007qi, Porto:2007tt}
so that we will not continue here to evaluate the corresponding spin-dependent classical
potentials consistently taking into account all relativistic $\mathcal O(v^2)$ effects in the iteration.

\end{document}